\documentclass[times,final]{elsarticle}

\usepackage{geometry}
\usepackage{framed,multirow}

\usepackage{amssymb}
\usepackage{latexsym}
\usepackage{amsmath} 
\usepackage{booktabs}
\usepackage{url}
\usepackage{xcolor}
\usepackage{graphicx}
\usepackage{subfigure}
\usepackage{algorithm}
\usepackage{algorithmicx}
\usepackage{algpseudocode}
\usepackage{bm}
\usepackage{multicol}  
\usepackage{multirow}  
\usepackage{booktabs}  
\usepackage{threeparttable} 
\definecolor{newcolor}{rgb}{.8,.349,.1}

\begin{document}


\begin{frontmatter}

\title{Hybrid Finite Difference with the Physics-informed Neural Network for solving PDE in complex geometries}%

\author[1]{Zixue Xiang}
\author[2]{Wei Peng}
\author[2]{Weien Zhou}
\author[2]{Wen Yao \corref{cor1}}
\cortext[cor1]{Corresponding author}
\ead{wendy0782@126.com}
\address[1]{College of Aerospace Science and Engineering, National University of Defense Technology, Changsha 410073, China}
\address[2]{Defense Innovation Institute, Chinese Academy of Military Science, Beijing 100071, China}


\begin{abstract}
	The physics-informed neural network (PINN) is effective in solving the partial differential equation (PDE) by capturing the physics constraints as a part of the training loss function through the Automatic Differentiation (AD). This study proposes the hybrid finite difference with the physics-informed neural network (HFD-PINN) to fully use the domain knowledge. The main idea is to use the finite difference method (FDM) locally instead of AD in the framework of PINN. In particular, we use AD at complex boundaries and the FDM in other domains. The hybrid learning model shows promising results in experiments. To use the FDM locally in the complex boundary domain and avoid the generation of background mesh, we propose the HFD-PINN-sdf method, which locally uses the finite difference scheme at random points. In addition, the signed distance function is used to avoid the difference scheme from crossing the domain boundary. In this paper, we demonstrate the performance of our proposed methods and compare the results with the different number of collocation points for the Poisson equation, Burgers equation. We also chose several different finite difference schemes, including the compact finite difference method (CDM) and crank-nicolson method (CNM), to verify the robustness of HFD-PINN. We take the heat conduction problem and the heat transfer problem on the irregular domain as examples to demonstrate the efficacy of our framework. In summary, HFD-PINN, especially HFD-PINN-sdf, are more instructive and efficient, significantly when solving PDEs in complex geometries.
\end{abstract}
\end{frontmatter}

\section{Introduction}

Partial differential equations (PDEs) are an important mathematical concept in the real world to describe, for example, the motion of fluids, the propagation of waves, the evolution of stock markets and more. Much work has been devoted to proposing various solutions for the numerical approximation of PDEs. Common numerical methods including finite difference method (FDM) \cite{1965Finite}, finite element method (FEM) \cite{2003Finite} and finite volume method (FVM) \cite{2019Finite}, require spatial discretization and rely on the grid to discretize PDEs. However, establishing a grid is time-consuming and laborious, and the quality of the grid affects the accuracy of the method. Over the past several years, there has been an amount of work concerning meshless methods. The research work on the meshless methods includes Smooth Particle Hydrodynamics (SPH), distance basis function (RBF), Diffuse Element Method (DEM), Element Free Galerkin Method (EFG), Meshless weighted least-squares (MWLS), Least Square Element Differentiation Method (LSEDM) \cite{2020Local}, etc. However, these methods sometimes could not achieve the optimal balance of accuracy and computational complexity. 

Advances of machine learning (ML) based approaches have led to promising results that can cope with the high computational costs associated with classical numerical methods. Recent research on the theory of deep neural network (DNN) approximation shows that deep network approximation is a powerful tool for parameterization of mesh-free functions. Thus an burgeoning field called Scientific Machine Learning (SciML) \cite{lu2020deepxde, SML}, which combines techniques of machine learning into traditional scientific computing especially numerical methods of PDEs \cite{M1994Neural,2017DGM, Maziar2017Machine,2017Inferring,2017NumericalGP}, has received widespread attention. However, in many scientific or engineering practices, data acquisition may be difficult and time-consuming, so the available data for DNN training may be scarce. On the other hand, DNN may generate unreasonable or unrealistic predictions for specific scientific problems in the lack of understanding of domain knowledge such as scientific laws and practical theories. By incorporating prior knowledge as constraints to guide the training of ML models, Raissi et al. proposed the physics-informed neural network (PINN) \cite{raissi2019physics}. In PINN, the residual form of PDE and its boundary conditions are concentrated into an objective function as an unconstrained optimization problem. PINN was shown to behave well in solving both the forward and inverse problems of various kinds of PDE with very little data, such as phase field equation \cite{wight2020solving,liu2021novel,xiang2021selfadaptive,peng2021idrlnet}, stochastic PDE \cite{PIGAN,FokkerPlanck,2019Learning,2019Quantifying}, and fractional PDE \cite{fPINNs}. In addition, PINN is useful in several engineering applications, such as biomedical problems \cite{CardiacActivationMapping}, materials \cite{Material,chen2020physics,Magnetic}, or solid mechanics \cite{2021Asolidmechanics}. PINN also found rich applications in computational fluid mechanics \cite{2020NSFnets,0Surrogate,largesimulation,highspeedflows,2020Physics}. 

More work has been devoted to combining numerical methods and machine learning to solve PDE in recent years. In particular, deep network approximation can alleviate or overcome the dimensionality curse in some cases, making it an attractive tool for solving high-dimensional problems. Deep Galerkin method (DGM) \cite{sirignano2018dgm} was proposed as a natural merger of meshless deep learning algorithm and Galerkin method to solve high-dimensional PDE. It is similar to the Galerkin method, but its solution is approximated by a neural network trained to satisfy differential operators, initial conditions, and boundary conditions rather than a combination of basis functions. In addition, Ramabathiran et al. introduced SPINN \cite{SPINN}, using PINN to reinterpret the traditional meshless representation of PDE based on the distance basis function. The basic idea of SPINN is to transcribe the meshless approximation directly into a sparse DNN. The difference between the SPINN model and PINN is that it is more interpretable and accurate. During the training process of PINN, the dynamics of the physical process may not be enforced accurately via automatic differentiation. When we carry out relevant work, Chiu et al. proposed CAN-PINN \cite{chiu2021canpinn} based on the Coupled-Automatic-Numerical Differentiation Method to provide more robust and efficient training than PINN. In addition, Chen et al. used FDM to convert PDE into a form that DNN is easy to handle and then used projection to achieve hard-constrained optimization \cite{2021HCP}. The theory-guided hard constraint projection (HCP) can ensure that the model predictions strictly comply with the physical mechanisms based on rigorous mathematical and experimental proofs. 

In this work, we propose the hybrid finite difference with the Physics-informed neural network (HFD-PINN), which merges PINN and finite difference method to solve PDE. The PINN ensure that the prediction result is close to the physical mechanism by embedding PDE residual, initial, and boundary constraint in the loss function as regularization terms. When defining the PDE residual, it is necessary to calculate the derivative by the numerical differentiation of FDM or the Automatic Differentiation (AD) used by the basic PINN named AD-PINN here. We define the FD loss in the regular domain to guide the PINN training by discretizing PDEs into a finite difference equation. AD is suitable to deal with derivatives at complex boundaries. We propose HFD-PINN-sdf to avoid generating background mesh in irregular areas. In addition, the sign distance function is used to define the difference interval at each random collocation point. The performance of the HFD-PINN, especially HFD-PINN-sdf are verified by experiments based on classical PDEs, such as the Poisson equation, viscous Burgers equation, or the complex heat transfer problem. We further solve the heat conduction problem on the irregular domain with HFD-PINN-sdf. As we shall see later, numerical results show that the HFD-PINN is competitive with the AD-PINN, and it outperforms in the aspect of convergence speed.
Moreover, the self-adaptive finite difference method with sdf effectively improves the predictions of HFD-PINN. We investigate the performance of AD-PINN and HFD-PINN with the different number of collocation points. To further study the impact of the finite difference schemes on the learning performance of the newly proposed model, we compare seveal difference schemes with PINN, including the compact finite difference method and crank-nicolson method. Our findings indicate that HFD-PINN with different difference schemes outperforms AD-PINN in terms of accuracy. 

The outline of this paper is given as follows. Section 2 briefly summarizes the hybrid finite difference Method, the Physics-informed neural network. Details about the proposed approaches HFD-PINN and HFD-PINN-sdf are also presented. Numerical results showcasing the performance of the proposed approach are presented in Section 3. A summary together with an outlook is given in Section 4.

\section{Methods}

The FDM based on discretization requires mesh generation, which will lead to the curse of dimensionality, and the parameters increase exponentially with the input dimension. In particular, when dealing with regions of complex shape, it is more troublesome and difficult to generate a suitable mesh. The details of the hybrid finite difference Method are introduced in section 2.1. With the advancement of deep learning, neural network approximation is a powerful tool for mesh-free function parametrization, especially suitable for high-dimensional and complex geometric problems. The most notable method is that a physics-informed neural network called AD-PINN uses automatic differentiation to calculate the residual of the differential equation and use it as an AD loss term to facilitate training. This framework is introduced in detail in section 2.2. In addition, HFD-PINN, which combines the advantages of AD and FDM, is introduced in detail in section 2.3. We propose the Self-adaptive HFD-PINN with signed distance function to extend the proposed method to irregular regions in section 2.4.

\subsection{Hybrid finite difference Method}
This section provides a brief overview of the finite difference method, one of the popular and well-developed methods used to solve partial differential equations. In the FDM, the principle is to employ a Taylor series expansion to discretize the derivatives of the variables. Suppose we compute the first derivative of a scalar $u(x)$ at point $x_i$. we denote $u(x_i +\Delta x)$ and $u(x_i -\Delta x)$ as a Taylor series in $x$ as follow:
\begin{equation}
\begin{aligned}
u\left(x_{i}+\Delta x\right)&=u\left(x_{i}\right)+\left.\Delta x \frac{\partial u}{\partial x}\right|_{x_{i}}+\left.\frac{\Delta x^{2}}{2} \frac{\partial^{2} u}{\partial x^{2}}\right|_{x_{i}}+\cdots, \\
u\left(x_{i}-\Delta x\right)&=u\left(x_{i}\right)-\left.\Delta x \frac{\partial u}{\partial x}\right|_{x_{i}}+\left.\frac{\Delta x^{2}}{2} \frac{\partial^{2} u}{\partial x^{2}}\right|_{x_{i}}+\cdots.
\end{aligned}
\end{equation}

Thus the first and second derivatives of $u$ can be approximated as
\begin{equation}
\begin{aligned}
\left.\frac{\partial u}{\partial x}\right|_{x_{i}}&=\frac{u\left(x_{i}+\Delta x\right)-u\left(x_{i}\right)}{\Delta x}+\mathcal{O}(\Delta x), \\
\left.\frac{\partial^{2} u}{\partial x^{2}}\right|_{x_{i}}&=\frac{u\left(x_{i}-\Delta x\right)-2u\left(x_{i}\right) +u\left(x_{i}+\Delta x\right)}{\Delta x^2}+\mathcal{O}(\Delta x ^{2}).
\end{aligned}
\end{equation}

\begin{figure}[htbp]
	\centering
	\subfigure{
		\includegraphics[scale=0.5]{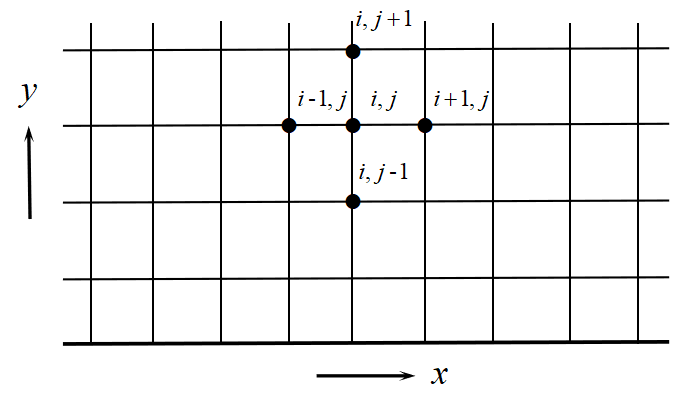}
	}
	\caption{Grids and nodes for a two-dimensional finite difference mesh.}
	\label{FDgird}
\end{figure}

The FDM replaces derivatives in the governing equations by difference quotients, which involve values of the solution at discrete mesh points in the domain. For two-dimensional problems, the grids are formulated with vertical and horizontal lines and define the mesh points as the intersection of the grid lines shown in Fig. \ref{FDgird}. The indexes $i$ and $j$ indicate the calculation space, directly corresponding to how variables are stored in computer memory. PDE describes the dependence between the partial derivatives of multivariate functions in the domain and usually need to be solved for the given initial conditions at the beginning of the simulation and the boundary conditions at the domain boundary. A simple and clear description of the FDM is given using the two-dimensional Poisson equation. 

\begin{equation}
\begin{aligned}
\frac{\partial^{2} u(x, y)}{\partial x^{2}}+\frac{\partial^{2} u(x, y)}{\partial y^{2}}&=q(x, y), \quad (x,y)\in \Omega, \\
u(x,y)&=g(x,y), \quad (x,y)\in \partial\Omega,
\label{2dheat}
\end{aligned}
\end{equation}
where $\Omega$ is the corresponding domain of the equations, $\partial \Omega$ denotes the boundary of the computational domain. The goal is thus to find the solution which satisfies the Eq. \eqref{2dheat} under the boundary condition $g(x,y)$. The equation at node $(x_i,y_j)$ is written in discrete form as:
\begin{equation}
\frac{u\left(x_{i+1}, y_{j}\right)-2 u\left(x_{i}, y_{j}\right)+u\left(x_{i-1}, y_{j}\right)}{(\Delta x)^{2}}+\frac{u\left(x_{i}, y_{j+1}\right)-2 u\left(x_{i}, y_{j}\right)+u\left(x_{i}, y_{j-1}\right)}{(\Delta y)^{2}}=q\left(x_{i}, y_{j}\right),
\label{fddiscretize}
\end{equation}
where $\Delta x$ and $\Delta y$ denote the difference interval along with the $x$ and $y$ directions, by applying FDM, the differential terms of the equation are converted into a finite-difference equation.
\begin{equation}
\begin{aligned}
&AU=q, \\
&U=\left[u(x+\Delta x,y), \quad u(x,y), \quad u(x-\Delta x,y), \quad u(x,y+\Delta y), \quad u(x,y-\Delta y)\right]^{T}, \\
&A=\left[\frac{1}{\Delta x^{2}}, \quad \frac{-2}{\Delta x^{2}}+\frac{-2}{\Delta y^{2}}, \quad\frac{1}{\Delta x^{2}},\quad \frac{1}{\Delta y^{2}}, \quad \frac{1}{\Delta y^{2}}\right],
\end{aligned}
\label{finitedifferenceequation}
\end{equation}
where $U$ is the prediction matrix at all nodes except the boundary values and $A$ is the physical constraint matrix, $q$ is a vector of known energy input to the system. An essential advantage of the FDM is uncomplicated to implement. FDM also has the possibility to achieve high-order accuracy of spatial discretization. Therefore, FDM is stable, of rapid convergence, accurate, and simple to solve PDEs.

On the basis of the research on the finite difference method, we pay attention to the high-precision finite difference format and its application. Two high-precision methods are mainly considered, including the compact finite difference method (CDM) and crank-nicolson method (CNM). For the convenience of calculation, we set the space step as $h=1/J$ and the time step as $\tau$, then the grid nodes $(x_j,t_n)$ can be expressed as $x_j=jh,t_n=n\tau,j=0,...,J$. The difference operators are given as Eq. \eqref{CDNM}. These methods use fewer grid points in the calculation area to improve the efficiency of the solution. Hence, the accuracy and stability of the difference scheme can be improved in theory and practice.
\begin{equation}
\begin{aligned}
&u_{j}^{n+\frac{1}{2}}=\frac{u_{j}^{n+1}+u_{j}^{n}}{2}, \delta_{x} u_{j}^{n}=\frac{u_{j+1}^{n}-u_{j-1}^{n}}{2 h} \\
&\delta_{t}^{*} u_{j}^{n}=\frac{u_{j}^{n+1}-u_{j}^{n}}{\tau}, \delta_{x}^{2} u_{j}^{n}=\frac{u_{j+1}^{n}-2 u_{j}^{n}+u_{j-1}^{n}}{h^{2}}
\end{aligned}
\label{CDNM}
\end{equation}

\subsection{physics-informed neural network (AD-PINN)}
This section briefly reviews the physics-informed neural network (PINN) for solving PDEs. The problem of two-dimensional Poisson equation (Eq. \eqref{2dheat}) is taken as an example to demonstrate the PINN. The idea of the PINN is to build a map from $\Omega$ to the solution by a feed-forward multi-layer neural network $\hat{u}(x,y;\theta)$. For simplicity, biases and weights of the neural network are concatenated in the parameter $\theta$. More precisely, the unknown parameters are found by solving an optimization problem that minimizes the following loss function.

Substitute $\hat{u}(x,y;\theta)$ into the PDE to define the residual as:
\begin{equation}
f(x, y;\theta):=\frac{\partial^{2} \hat{u}(x,y;\theta)}{\partial x^{2}}+\frac{\partial^{2} \hat{u}(x,y;\theta)}{\partial y^{2}} - q(x, y).
\label{residual}
\end{equation}

All partial derivatives are computed by applying the chain rule for the network through Automatic Differentiation (AD), which can be easily implemented in the deep learning framework, such as Pytorch \cite{DBLP}, TensorFlow \cite{abadi2016tensorflow}. The penalty terms of governing equation and boundary condition are defined as Eq. \eqref{ADloss} and Eq. \eqref{BCloss}, which are integrated into the loss function to enhance the prior knowledge of physics.
\begin{equation}
\begin{aligned}
\mathcal{L}_{\mathrm{AD}}(\theta)=& \frac{1}{N_{\mathrm{f}}} \sum_{i=1}^{N_{\mathrm{f}}}\left\|f(x_{f}^{i},y_{f}^{i};\theta)\right\|_{2}^{2},
\end{aligned}
\label{ADloss}
\end{equation}
\begin{equation}
\begin{aligned}
\mathcal{L}_{\mathrm{BC}}(\theta)=& \frac{1}{N_{\mathrm{b}}} \sum_{i=1}^{N_{\mathrm{b}}}\left\|\hat{u}(x_{b}^{i},y_{b}^{i}) - g(x_{b}^{i},y_{b}^{i})\right\|_{2}^{2},
\end{aligned}
\label{BCloss}
\end{equation}
where $N_f$ and $N_b$ denote the number of collocation points, boundary points. $\mathcal{L}_{\mathrm{AD}}(\theta)$ means the PDE residual evaluated at collocation points defined in domain $\left\{x_{f}^{i}, y_{f}^{i}\right\}^{N_f}_{i=1} \in \Omega$. $\mathcal{L}_{\mathrm{BC}}(\theta)$ enforces boundary conditions over boundary surface $\left\{x_{b}^{i}, y_{b}^{i}\right\}^{N_b}_{i=1} \in \partial\Omega$. It should be noted that if the PDE has the initial condition, we also need to define the IC loss $\mathcal{L}_{\mathrm{IC}}(\theta)$. Sometimes we define the Data loss $\mathcal{L}_{\mathrm{Data}}(\theta)$ according to the observations. Finally, the loss function of AD-PINN can be obtained as follows:
\begin{equation}
\begin{aligned}
\mathcal{L}^{\mathrm{AD-PINN}}(\theta)=& \omega_{f} \mathcal{L}_{\mathrm{AD}}(\theta)+\omega_{b}\mathcal{L}_{\mathrm{BC}}(\theta). \\
\end{aligned}
\end{equation}

\begin{figure}[htbp]
	\centering
	\subfigure{
		\includegraphics[scale=0.5]{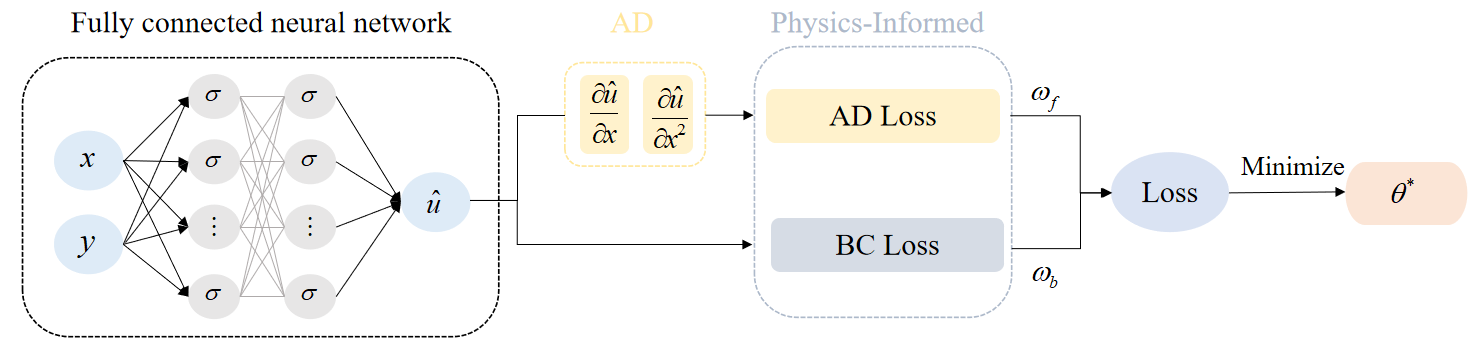}
	}
	\caption{Illustration of the physics-informed neural network (AD-PINN).}
	\label{PINN}
\end{figure}

We adjust the importance of each loss term through the weight parameter $\omega = \left\{\omega_{f},\omega_{b}\right\}$. The framework of AD-PINN is constructed shown in Fig. \ref{PINN}.

\subsection{Hybrid finite difference with physics-informed neural network (HFD-PINN)}
Derivatives, mostly in the form of gradients and Hessians, are ubiquitous in solving PDE residuals. Conventionally, methods of calculating derivatives include numerical differentiation using FDM, symbolic differentiation, and Automatic Differentiation. As mentioned above in section 2.1, the FDM is stable and simple to define the first and second derivatives. Therefore, we can combine FDM and PINN using the finite difference equation (Eq. \eqref{finitedifferenceequation}) to define the governing equation loss as follows. Matrix decomposition transforms the given constraint from the physical to the computational space, which is significant to generate predictions that obey the physical mechanisms. We need to generate background mesh and $N_f$ nodes through two-dimensional finite difference to obtain model predictions matrix $\hat{U}$. 
\begin{equation}
\begin{aligned}
&\hat{U}=\left[\hat{u}(x+\Delta x,y), \quad \hat{u}(x,y), \quad \hat{u}(x-\Delta x,y), \quad \hat{u}(x,y+\Delta y), \quad \hat{u}(x,y-\Delta y)\right]^{T}.
\end{aligned}
\label{FDPINN equation}
\end{equation}

The constraint matrix $A$ reflects the relationship determined by the physical mechanism at different positions in the computational space. The remaining part $q$ is determined based on PDE parameters and the point $(x,y)$. Thus the FD loss could be defined as:
\begin{equation}
\begin{aligned}
\mathcal{L}_{\mathrm{FD}}(\theta)=& \frac{1}{N_{\mathrm{f}}} \sum_{i=1}^{N_{\mathrm{f}}}\left\|\hat{U}A-q\right\|_{2}^{2},
\end{aligned}
\label{FDlosseq}
\end{equation}
where $N_{f}, N_{b}$ are the number of points used in discretizing the domain and boundary. Denote the positive weights $\omega_{f},\omega_{b}$ to define the loss function of HFD-PINN as follows:
\begin{equation}
\begin{aligned}
\mathcal{L}^{\mathrm{HFD-PINN}}(\theta)=& \omega_{f} \mathcal{L}_{\mathrm{FD}}(\theta)+\omega_{b}\mathcal{L}_{\mathrm{BC}}(\theta).
\end{aligned}
\label{FDPINNlosseq}
\end{equation}

\begin{figure}[htbp]
	\centering
	\subfigure{
		\includegraphics[scale=0.5]{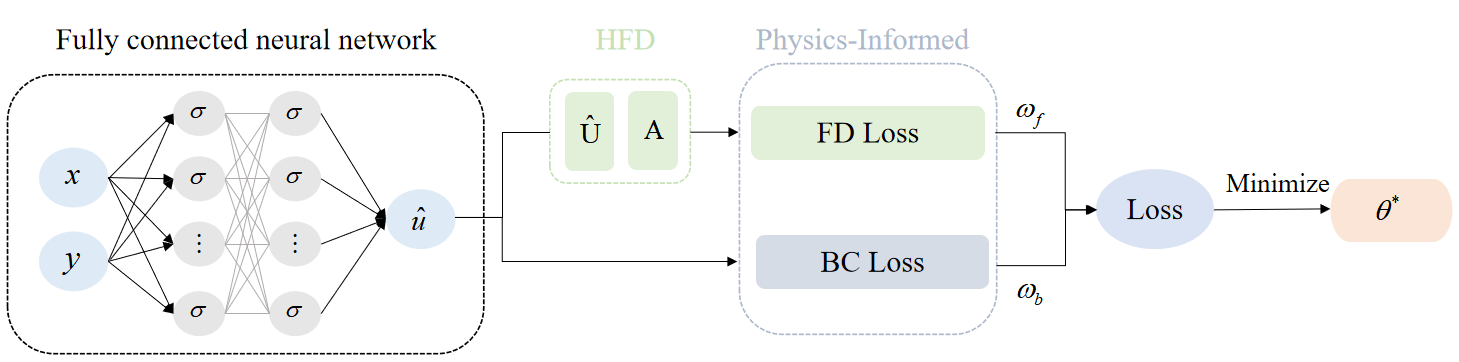}
	}
	\caption{Illustration of hybrid finite difference with the physics-informed neural network (HFD-PINN).}
	\label{FDPINN}
\end{figure}

On the other hand, because the method requires a structured background mesh, the significant difficulty of application lies in the inaccuracies in dealing with the complex domain. However, as a powerful method for calculating derivatives, automatic differentiation redefines the semantics of operators to propagate derivatives according to the chain rule of calculus. The AD is an indispensable tool for gradient-based machine learning, especially suitable for dealing with derivatives at the complex boundary. It is applied in computational fluid dynamics, atmospheric sciences, and engineering design optimization. Therefore, these two methods can be combined to complement each other to improve the prediction ability and efficiency of PINN. The algorithm of Finite difference with PINN (FD-PINN), compact finite difference with PINN (CD-PINN), and crank-nicolson with PINN (CD-PINN) are referred to as HFD-PINN. The illustration of HFD-PINN is shown in Fig. \ref{FDPINN}. The hybrid finite difference with the physics-informed neural network is summarized as the algorithm\ref{alg:FDPINN}. 

\begin{algorithm}[htb]
	\caption{Hybrid finite difference with physics-informed neural network (HFD-PINN)}
	\label{alg:FDPINN}
	
	\begin{algorithmic}
		\State \textbf{Step\hspace{0.5em}1}:Consider a fully connected neural network to define the PINN $\hat{u}(x,y;\theta)$.
		\State \textbf{Step\hspace{0.5em}2}:Generate a background mesh and collocation points $\left\{x_{f}^{i}, y_{f}^{i}\right\}^{N_f}_{i=1} \in \Omega$.
		\State \textbf{Step\hspace{0.5em}3}:Discretize the governing equation (Eq. \eqref{2dheat}) into the finite difference equation (Eq. \eqref{fddiscretize}).
		\State \textbf{Step\hspace{0.5em}4}:Obtain the constraint matrix $A$ (Eq. \eqref{finitedifferenceequation}) and the model predictions matrix $\hat{U}$ (Eq. \eqref{FDPINN equation}).
		\State \textbf{Step\hspace{0.5em}5}:Define the FD loss term (Eq. \eqref{FDlosseq}).
		\State \textbf{Step\hspace{0.5em}6}:Randomly sample boundary condition points $\left\{x_{b}^{i}, y_{b}^{i}\right\}^{N_b}_{i=1} \in \partial\Omega$.
		\State \textbf{Step\hspace{0.5em}7}:Choose AD or FD to deal with derivatives according to the complexity of the boundary.
		\State \textbf{Step\hspace{0.5em}8}:Define the BC loss term (Eq. \eqref{BCloss}).
		\State \textbf{Step\hspace{0.5em}9}:Denote the positive weights $\omega_{f},\omega_{b}$ and define the loss function of HFD-PINN (Eq. \eqref{FDPINNlosseq}).
		\State \textbf{Step\hspace{0.5em}10}:Use K gradient descent iterations to update the network parameters $\theta$ as:
		\For{$k = 1$ to $K$}
		\State  Update the parameters $\theta$ via Adam optimizer.\\
		\hspace{5em} $\theta_{k+1} \leftarrow$ Adam $ \mathcal{L}^{\mathrm{HFD-PINN}}(\theta_{k})$
		\EndFor
		\State \textbf{Return} \\
		The best parameters $\theta ^{\ast }$.
	\end{algorithmic}
\end{algorithm}

\subsection{Self-adaptive HFD-PINN with signed distance function (HFD-PINN-sdf)}
\begin{figure}[htbp]
	\centering
	\subfigure[HFD-PINN]{
		\includegraphics[scale=0.31]{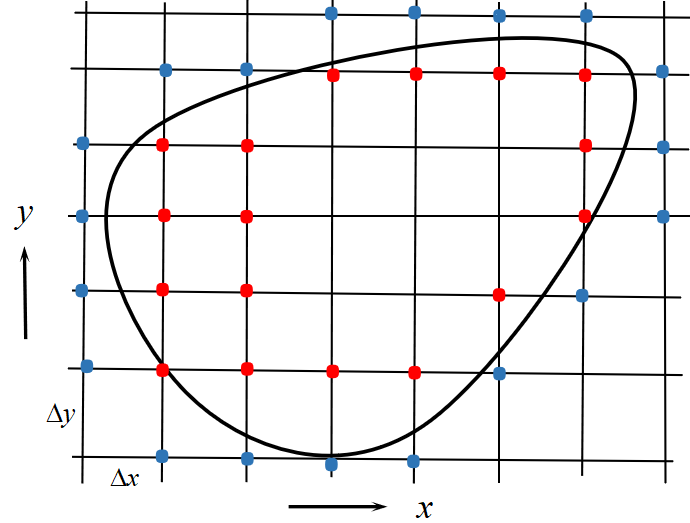}
	}%
	\subfigure[HFD-PINN-adapt]{
		\includegraphics[scale=0.35]{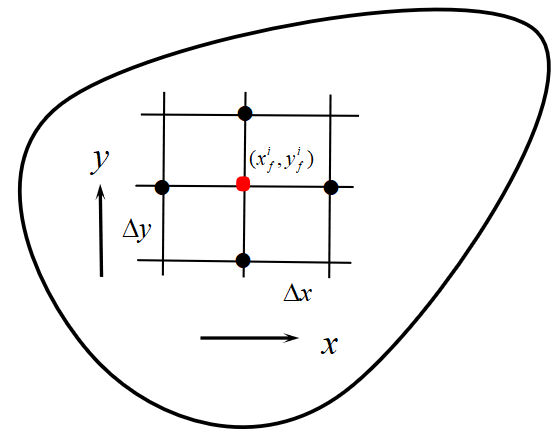}
	}%
	\subfigure[HFD-PINN-sdf]{
		\includegraphics[scale=0.36]{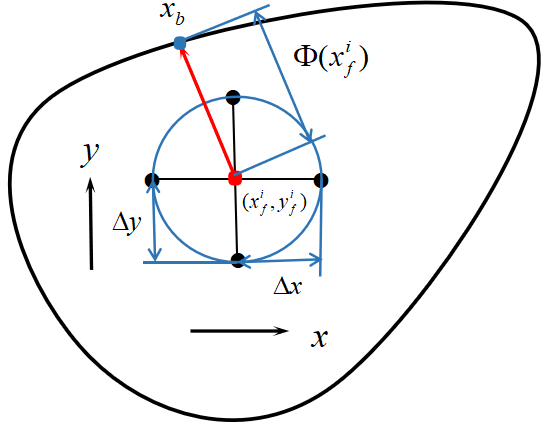}
	}%
	\centering
	\caption{The main idea of HFD-PINN (a), HFD-PINN-adapt (b), and HFD-PINN-sdf (c) are demonstrated. The red marks are configuration points. The blue marks are points out of boundary. The black marks are points to define the finite difference.}
	\label{Self}
\end{figure}

As for the irregular area shown in Fig. \ref{Self}(a), using a uniform mesh to obtain configuration points (red marks), including out of boundary points (blue marks) to define the derivatives with FDM, has low accuracy at the boundary. Hence, HFD-PINN deals with FDM derivatives in the regular domain while using the AD at complex boundaries. Further, it is hoped to avoid generating background mesh and randomly sample points in the area to define the Self-adaptive finite difference method. Firstly, we define the five-point difference scheme with a fixed difference interval at each random point to calculate the PDE residual as Fig. \ref{Self}(b). It is the main idea of Self-adaptive hybrid finite difference with physics-informed neural network (HFD-PINN-adapt). To avoid the difference points (black marks) out of bounds, we often define a tiny difference interval whose selection is often limited. Therefore, A signed distance function giving the distance for $x \in \Omega$ to $\partial \Omega$ is computed. The signed distance function (sdf) is a typical form of the level-set function that is defined as:
\begin{equation}
\Phi(x)=\left\{\begin{array}{lc}
d(x), & x \in \Omega \\
0, & x \in \partial \Omega \\
-d(x), & x \in \mathrm{D} / \bar{\Omega}
\end{array}\right.
\label{sdf}
\end{equation}

\begin{figure}[htbp]
	\centering
	\subfigure[Inward]{
		\includegraphics[scale=0.32]{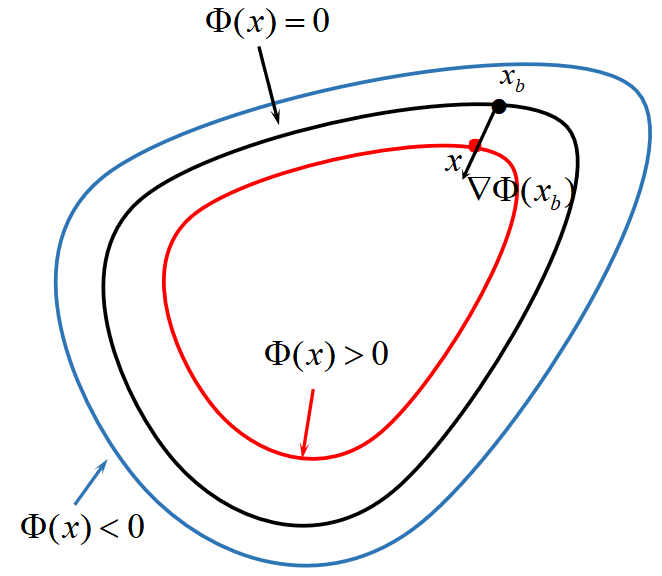}
	}%
	\subfigure[Outward]{
		\includegraphics[scale=0.32]{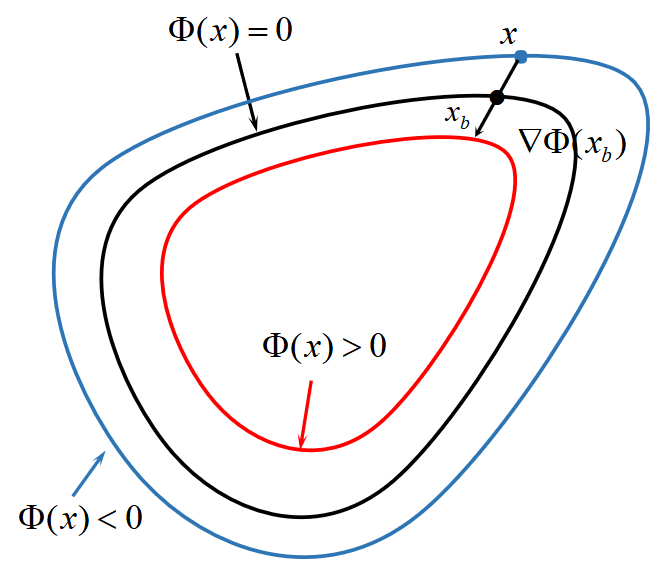}
	}%
	\centering
	\caption{Signed distance function: (a) $x$ is the offset of $x_{b}$ in the direction of inward normal. (b) $x$ is the offset of $x_{b}$ in the direction of outward normal. }
	\label{SDF}
\end{figure}

We define the signed distance function to be positive on the exterior, negative on the interior and zero on the boundary shown as Fig. \ref{SDF}. For each point, $x$, suppose that $x_{b}$ is the point on the interface closest to $x$. We define $d(x)$ as the minimum distance to a boundary point where a boundary condition should be imposed.
\begin{equation}
d(x)=\min _{x_{b} \in \partial \Omega}|| x-x_{b}||.
\end{equation}

The signed distance function has the property of the unit gradient module with $\left\|\nabla \Phi\right\|=1 $. The difference interval of each collocation point $x_{f}^{i}$ could be define as:
\begin{equation}
\Delta x_{f}^{i} =\alpha \Phi (x_{f}^{i}),
\label{step}
\end{equation} 
where $\alpha \in [0.01, 0.05]$. The main idea of HFD-PINN-sdf is shown in Fig. \ref{Self}(c). Thus Self-adaptive HFD-PINN with signed distance function (HFD-PINN-sdf) is summarized as the algorithm \ref{alg:SdfFDPINN}.

\begin{algorithm}[htb]
	\caption{Self-adaptive HFD-PINN with signed distance function (HFD-PINN-sdf)}
	\label{alg:SdfFDPINN}
	
	\begin{algorithmic}
		\State \textbf{Step\hspace{0.5em}1}:Consider a fully connected neural network to define the PINN $\hat{u}(x,y;\theta)$.
		\State \textbf{Step\hspace{0.5em}2}:Randomly generate collocation points $\left\{x_{f}^{i}, y_{f}^{i}\right\}^{N_f}_{i=1} \in \Omega$ in complex domain.
		\State \textbf{Step\hspace{0.5em}3}:Choose the differnence interval $\Delta x_{f}^{i}$ and $\Delta y_{f}^{i}$ of each collocation point to define the finite difference.
		\State \textbf{Step\hspace{0.5em}4}:Define the the difference interval (Eq. \eqref{step}) through signed distance function (Eq. \eqref{sdf}).
		\State \textbf{Step\hspace{0.5em}5}:Discretize the governing equation (Eq. \eqref{2dheat}) into the finite difference equation (Eq. \eqref{fddiscretize}).
		\State \textbf{Step\hspace{0.5em}6}:Obtain the constraint matrix $A$ (Eq. \eqref{finitedifferenceequation}) and the model predictions matrix $\hat{U}$ (Eq. \eqref{FDPINN equation}).
		\State \textbf{Step\hspace{0.5em}7}:Define the FD loss term (Eq. \eqref{FDlosseq}).
		\State \textbf{Step\hspace{0.5em}8}:Randomly sample boundary condition points $\left\{x_{b}^{i}, y_{b}^{i}\right\}^{N_b}_{i=1} \in \partial\Omega$.
		\State \textbf{Step\hspace{0.5em}9}:Choose AD or FD to deal with derivatives according to the complexity of the boundary.
		\State \textbf{Step\hspace{0.5em}10}:Define the BC loss term (Eq. \eqref{BCloss}).
		\State \textbf{Step\hspace{0.5em}11}:Denote the positive weights $\omega_{f},\omega_{b}$ and define the loss function of HFD-PINN-sdf (Eq. \eqref{FDPINNlosseq}).
		\State \textbf{Step\hspace{0.5em}12}:Use K gradient descent iterations to update the network parameters $\theta$ as:
		\For{$k = 1$ to $K$}
		\State Update the parameters $\theta$ via Adam optimizer.\\
		\hspace{5em} $\theta_{k+1} \leftarrow$ Adam $ \mathcal{L}^{\mathrm{HFD-PINN-sdf}}(\theta_{k})$
		\EndFor
		\State \textbf{Return} \\
		The best parameters $\theta ^{\ast }$.
	\end{algorithmic}
\end{algorithm}

\section{Results}
In this section, our goal is to systematically analyze the performance of the AD-PINN and HFD-PINN by setting a uniform criterion and quantifying their prediction results. Hence, multiple experiments are carried out. We first consider the viscous Burgers equation that is commonly used as testbeds. To further study the impact of the finite difference schemes on the learning performance, more challenging heat transfer problems of square plate with hole in the middle or corner are also considered here. We test the ability of HFD-PINN to solve the two-dimensional Poisson equation in the regular domain. We further solve the two-dimensional Poisson equation on the irregular domain with HFD-PINN-sdf. Furthermore, we describe the experimental conditions and results in detail. The performance and robustness of different models are evaluated by comparing AD-PINN and HFD-PINN in the case of using a different number of collocation points, different finite difference methods. The relative L2 error is the ratio of the 2-norm of the difference between the predictions and the 2-norm of the observations. The relative L2 error measures the accuracy of the estimation. The exact value $u(x_{i}, y_{i})$ and the trained approximation $\hat {u}\left(x_{i}, y_{i}\right)$ are inferred at the data points $\left\{x_{i}, y_{i}\right\}^{N}_{i=1}$.
\begin{equation}
\text { L2 error }=\frac{\sqrt{\sum_{i=1}^{N}\left|\hat {u}\left(x_{i},y_{i}\right)-u\left(x_{i},y_{i}\right)\right|^{2}}}{\sqrt{\sum_{i=1}^{N}\left|u\left(x_{i}, y_{i}\right)\right|^{2}}}.
\label{L2error}
\end{equation}

\begin{table}[tp]  
	
	\centering  
	\fontsize{8}{8}\selectfont  
	\caption{Comparing the relative error with the different number of collocation points when learning the Viscous Burgers equation.}  
	\label{tab1}  
	\begin{tabular}{|c|c|c|c|c|c|c|}  
		\hline  
		\multirow{2}{*}{collocation points}&  
		\multicolumn{3}{c|}{AD-PINN}&\multicolumn{3}{c|}{ FD-PINN}\cr\cline{2-7}  
		&AD loss&BC loss&Data loss&FD loss&BC loss&Data loss\cr  
		\hline  
		\hline  
		25&8.9e-02&7.3e-02&{\bf 6.0e-02}&1.9e+00&1.5e-03&{\bf 3.6e-02}\cr\hline  
		100&1.6e-02&5.0e-02&{\bf 2.5e-02}&1.4e+00&1.4e-03&{\bf 9.9e-03}\cr\hline  
		225&2.5e-03&1.4e-03&{\bf 2.2e-03}&1.2e+00&5.2e-04&{\bf 4.8e-04}\cr\hline  
		400&2.0e-03&3.0e-04&{\bf 4.1e-04}&1.1e+00&3.5e-05&{\bf 2.4e-04}\cr\hline  
	\end{tabular}  
\end{table} 

\begin{figure}[htbp]
	\centering
	\subfigure{
		\includegraphics[scale=0.45]{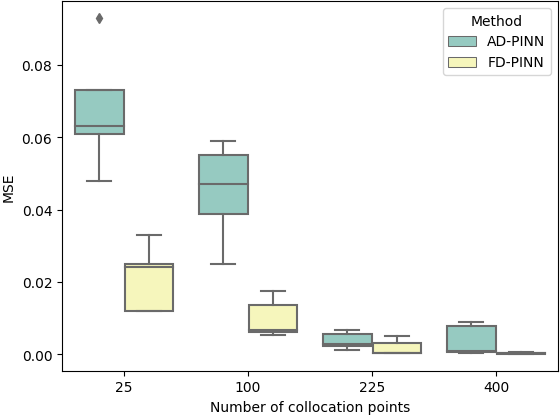}
	}
	\caption{Viscous Burgers equation: Boxplot of the relative L2 loss of AD-PINN and FD-PINN with different numbers of collocation points.}
	\label{BurFDPINNpa}
\end{figure} 

\subsection{Viscous Burgers Equation}
This section covers the viscous Burgers equation, which describes the propagation and reflection of shock waves. The equation forms the basis of many model PDEs, such as fluid mechanics and gas dynamics. Here, our focus is to analyze the ability of AD-PINN and HFD-PINN to solve nonlinear and time-dependent problems. Boundary value problems corresponding to the viscous Burgers equation is specified as follows:

\begin{equation}
\begin{array}{c}
u_{t}+uu_{x}- \nu  u_{x x}=0, \quad x \in[-1,1], \quad t \in[0,1], \\
u(0, x)=-\sin (\pi x), \\
u(t,-1)=u(t, 1)=0,
\end{array}
\label{Bureq}
\end{equation}
where the viscosity parameter is set as $\nu = 0.01 / \pi$. Similarly, the AD-PINN takes advantage of the automatic differentiation of neural networks to calculate the derivative of each feature at the $N_f$ collocation points. On the contrary, the FD-PINN uses the  to define the difference between the derivatives and the governing equation (Eq. \eqref{Bureq}). Specifically, given a set of $N_b =100$ randomly distributed boundary data, we learn the latent solution $u(t, x)$ by training all parameters of an 8-layer network using Adam with a learning rate of 0.001. Each hidden layer contains 20 neurons and a hyperbolic tangent activation function. The performance of AD-PINN and FD-PINN with the same numbers of collocation points and neural network architecture are compared to evaluate the effectiveness. We analyze the performance of AD-PINN and FD-PINN in different numbers of collocation points through several repeated and independent experiments. Table \ref{tab1} summarizes our results for the solution of the Burgers equation through AD-PINN and FD-PINN.
Moreover, a boxplot is drawn to reflect the relative error distribution in Fig. \ref{BurFDPINNpa}. As the number of collocation points increases, the model performance of AD-PINN and FD-PINN gradually converges. The ability of FD-PINN to extract information from collocation points is better than AD-PINN.

\begin{figure}[htbp]
	\centering
	\subfigure[Exact]{
		\includegraphics[scale=0.35]{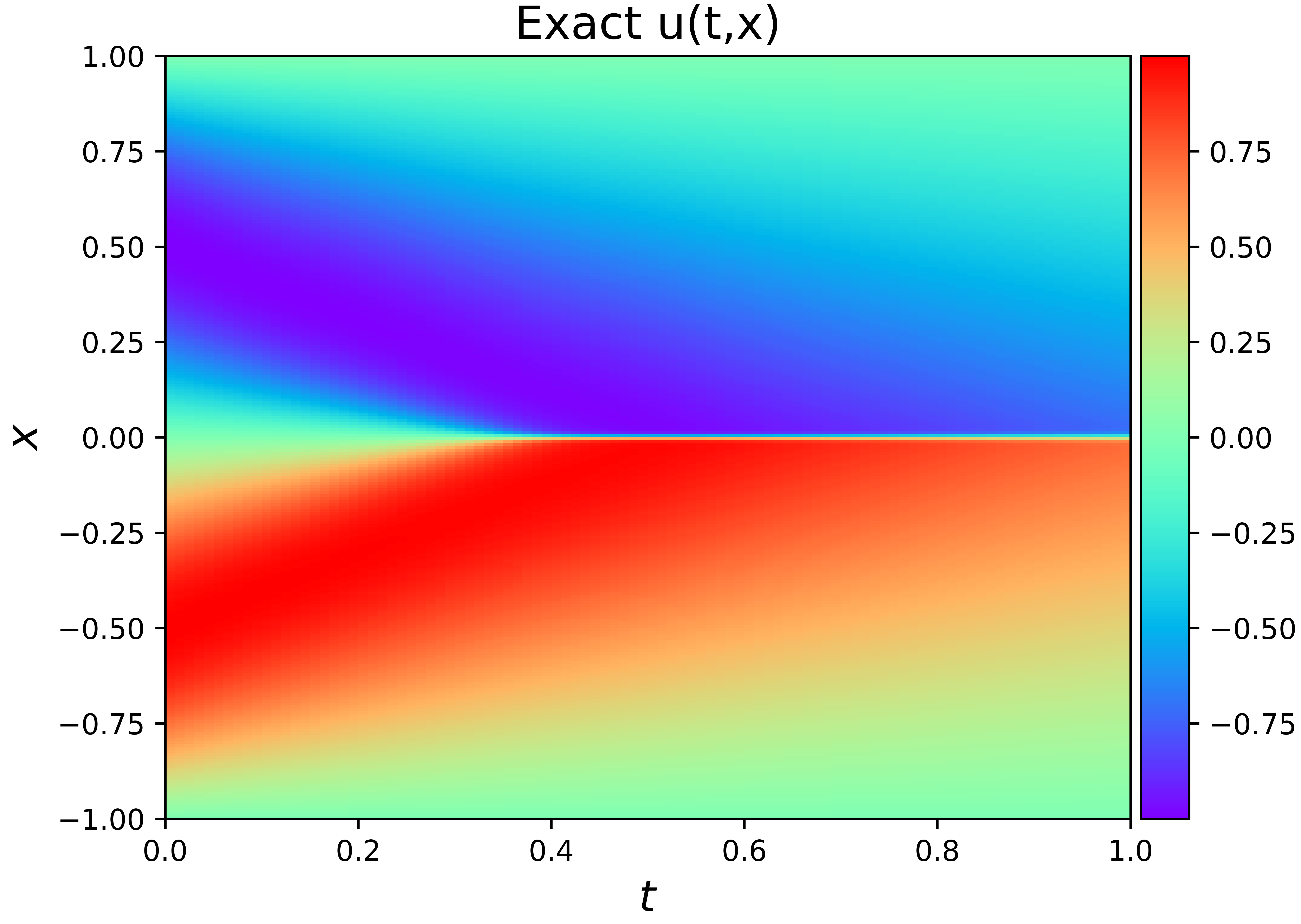}
	}%
	\subfigure[FD-PINN]{
		\includegraphics[scale=0.35]{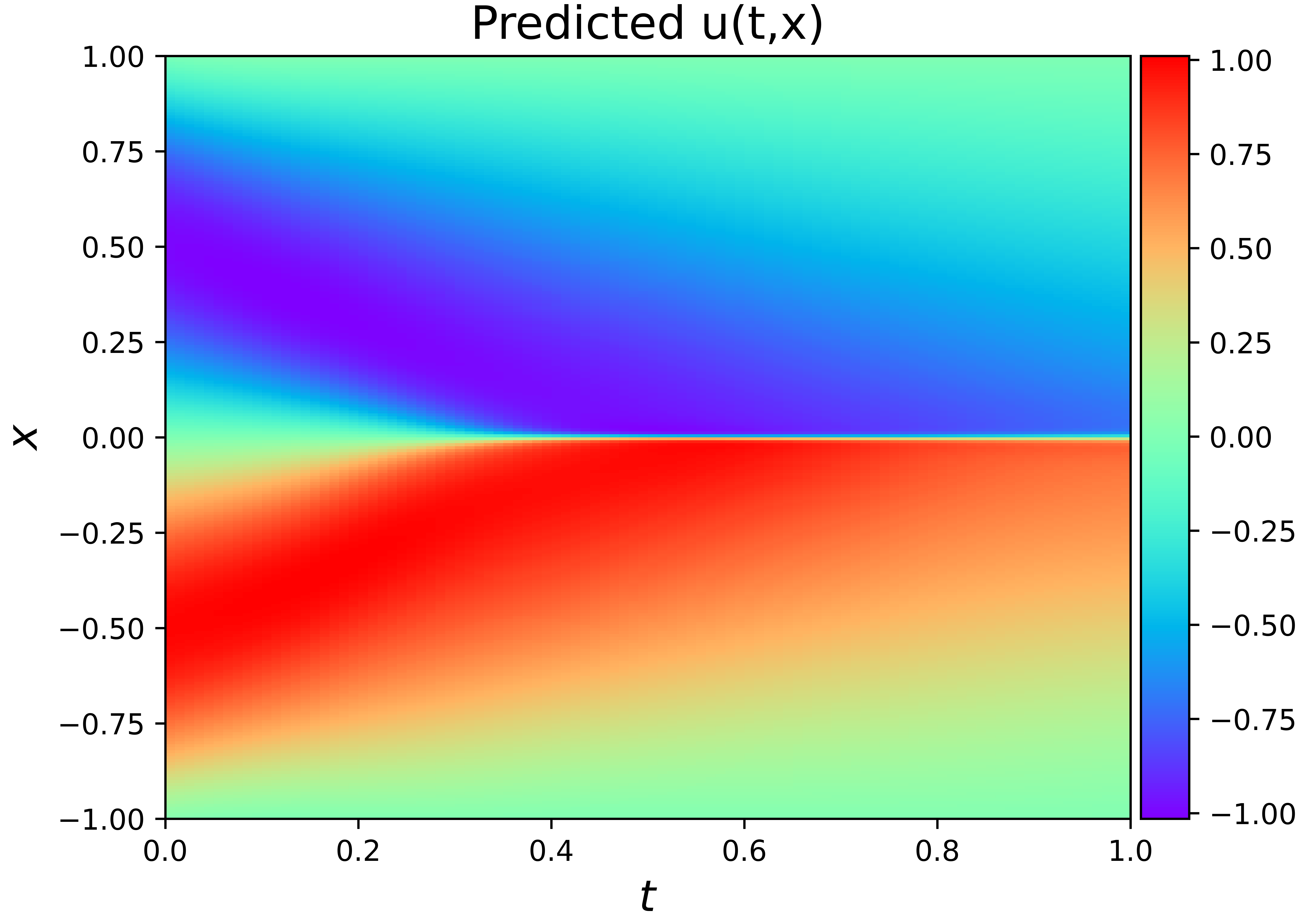}
	}%
	\centering
	\caption{Viscous Burgers equation:(a) The reference solution. (b) The prediction solution of FD-PINN.}
	\label{BurFDPINNpred}
\end{figure}

\begin{figure}[htbp]
	\centering
	\subfigure{
		\includegraphics[scale=0.35]{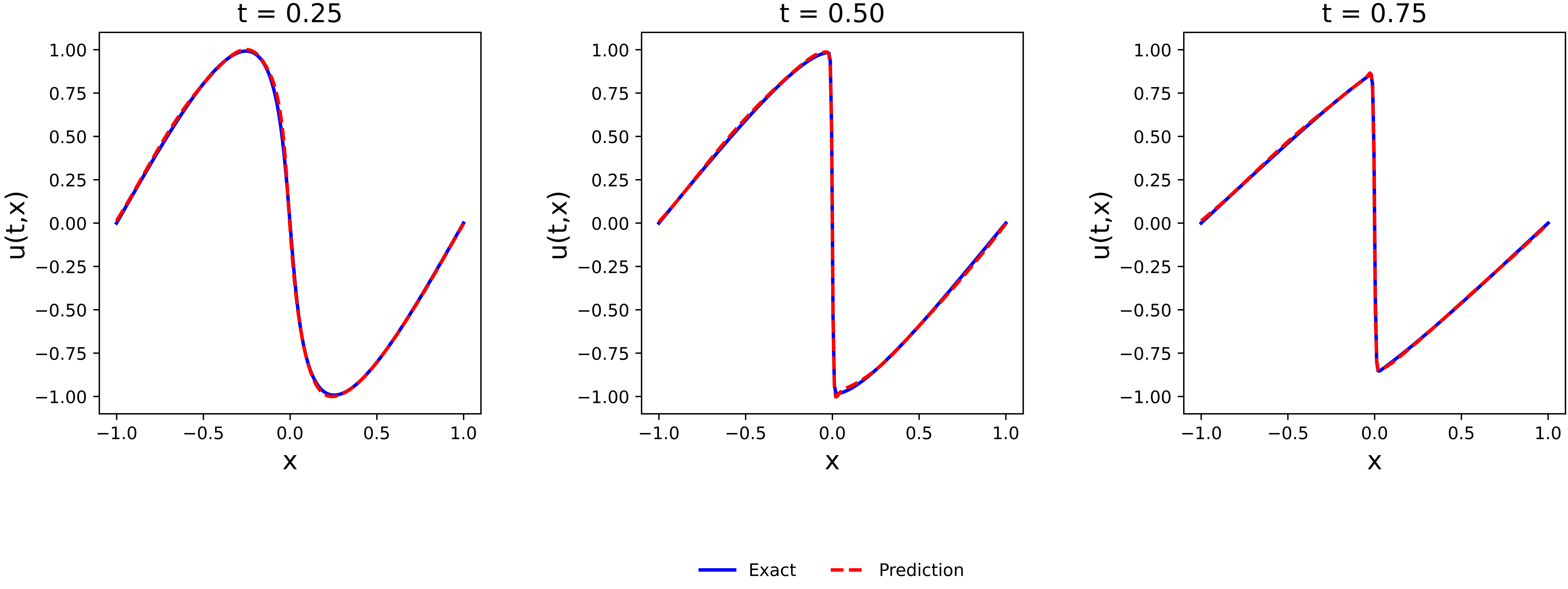}
	}
	\caption{Viscous Burgers equation: According to three temporal snapshots to compare the predicted and exact solutions with FD-PINN.}
	\label{BurFDPINNtest}
\end{figure}

\begin{figure}[htbp]
	\centering
	\subfigure[AD-PINN]{
		\includegraphics[scale=0.35]{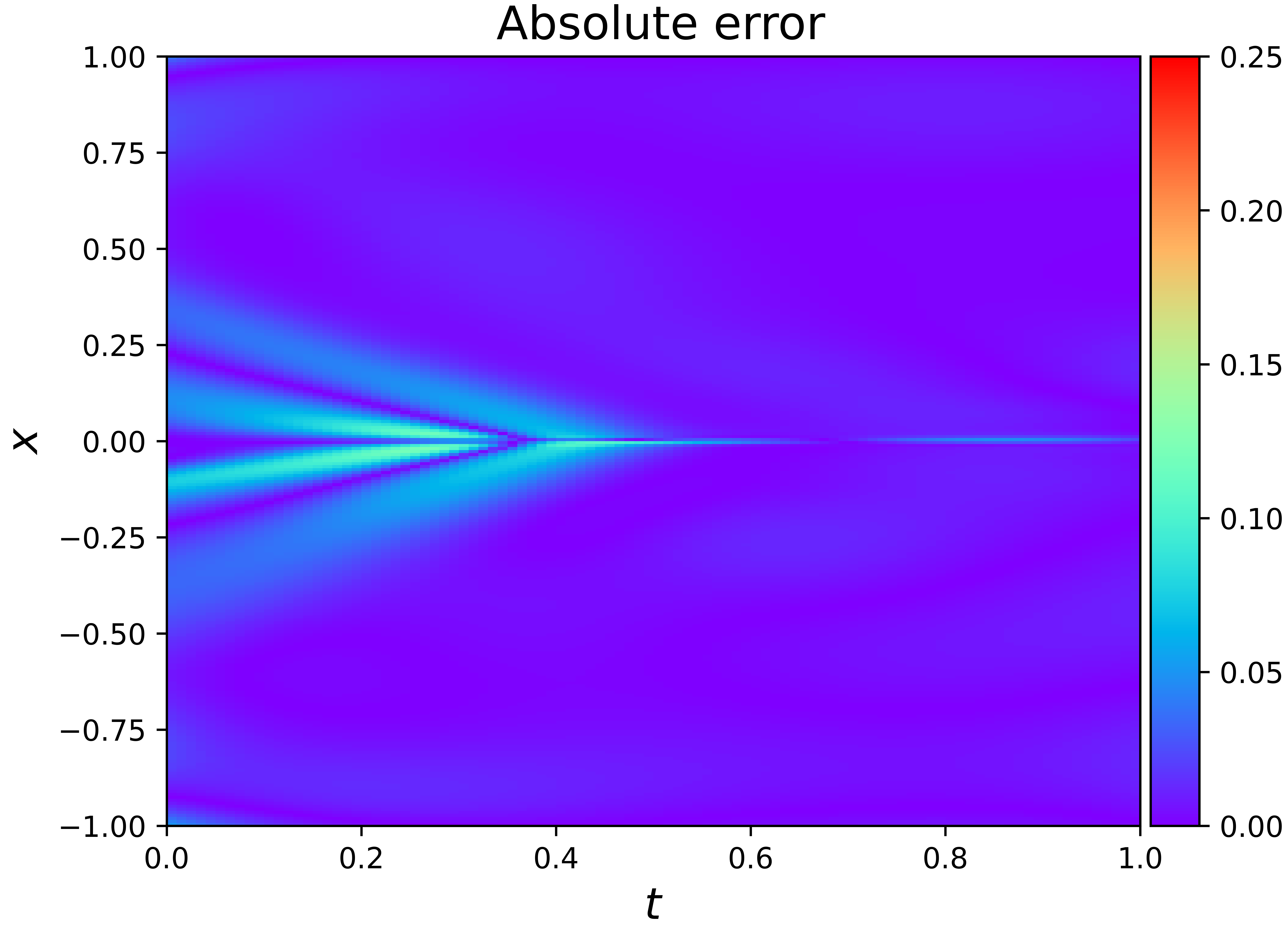}
	}%
	\subfigure[FD-PINN]{
		\includegraphics[scale=0.35]{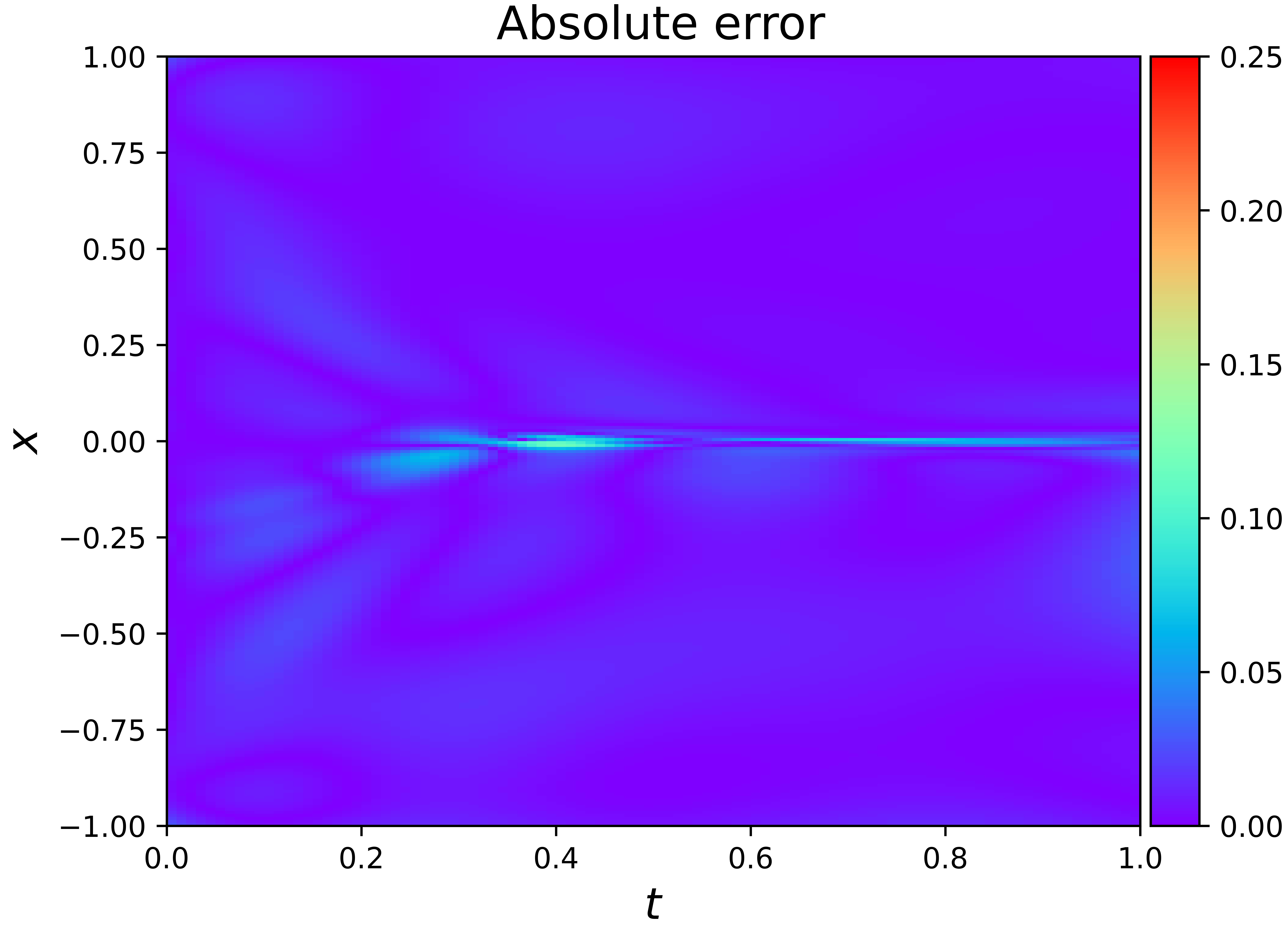}
	}%
	\centering
	\caption{Viscous Burgers equation:(a) The absolute error of AD-PINN. (b) The absolute error of FD-PINN}
	\label{BurADPINNerror}
\end{figure}

\begin{figure}[htbp]
	\centering
	\subfigure[AD-PINN]{
		\includegraphics[scale=0.35]{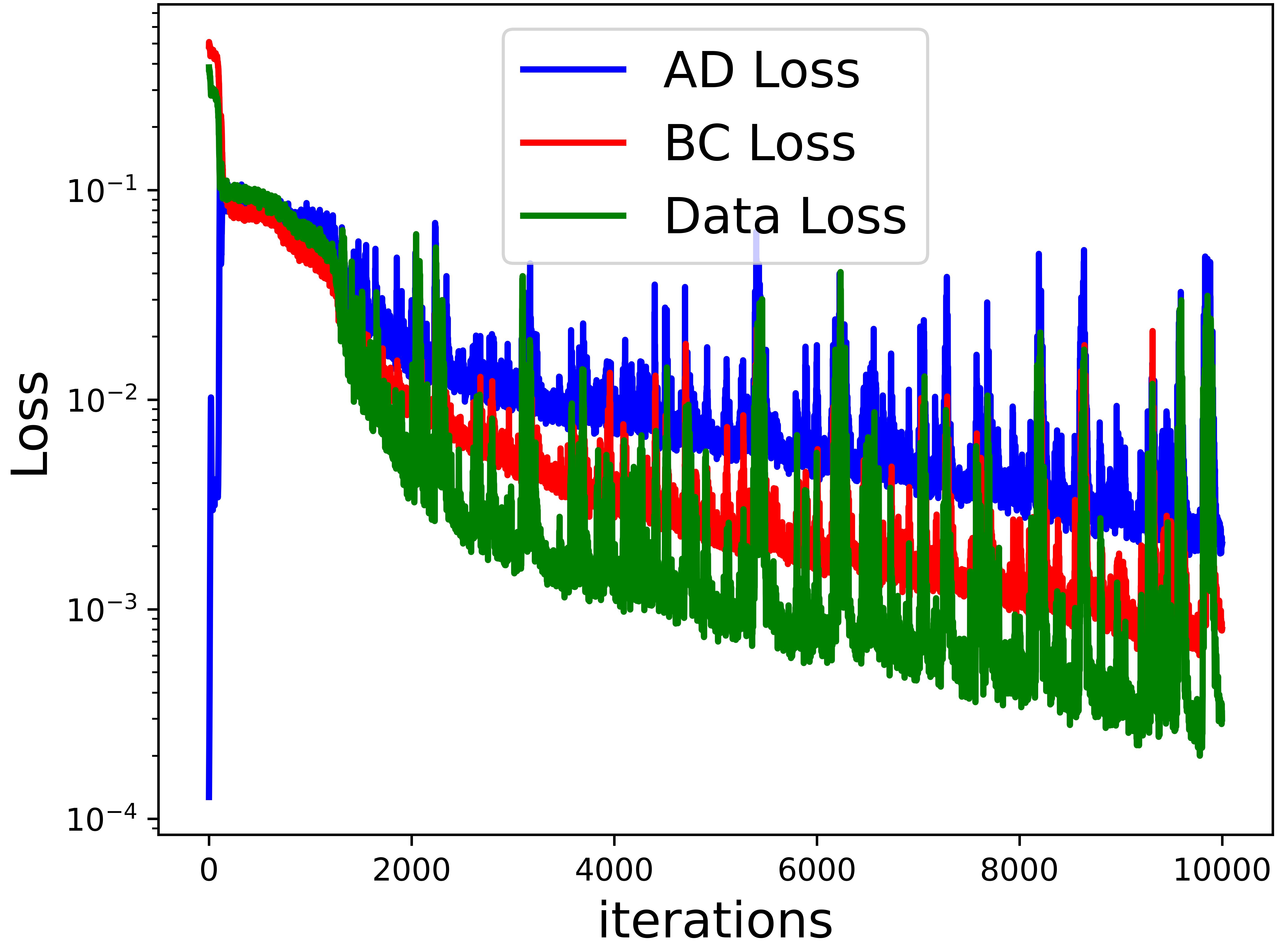}
	}%
	\subfigure[FD-PINN]{
		\includegraphics[scale=0.35]{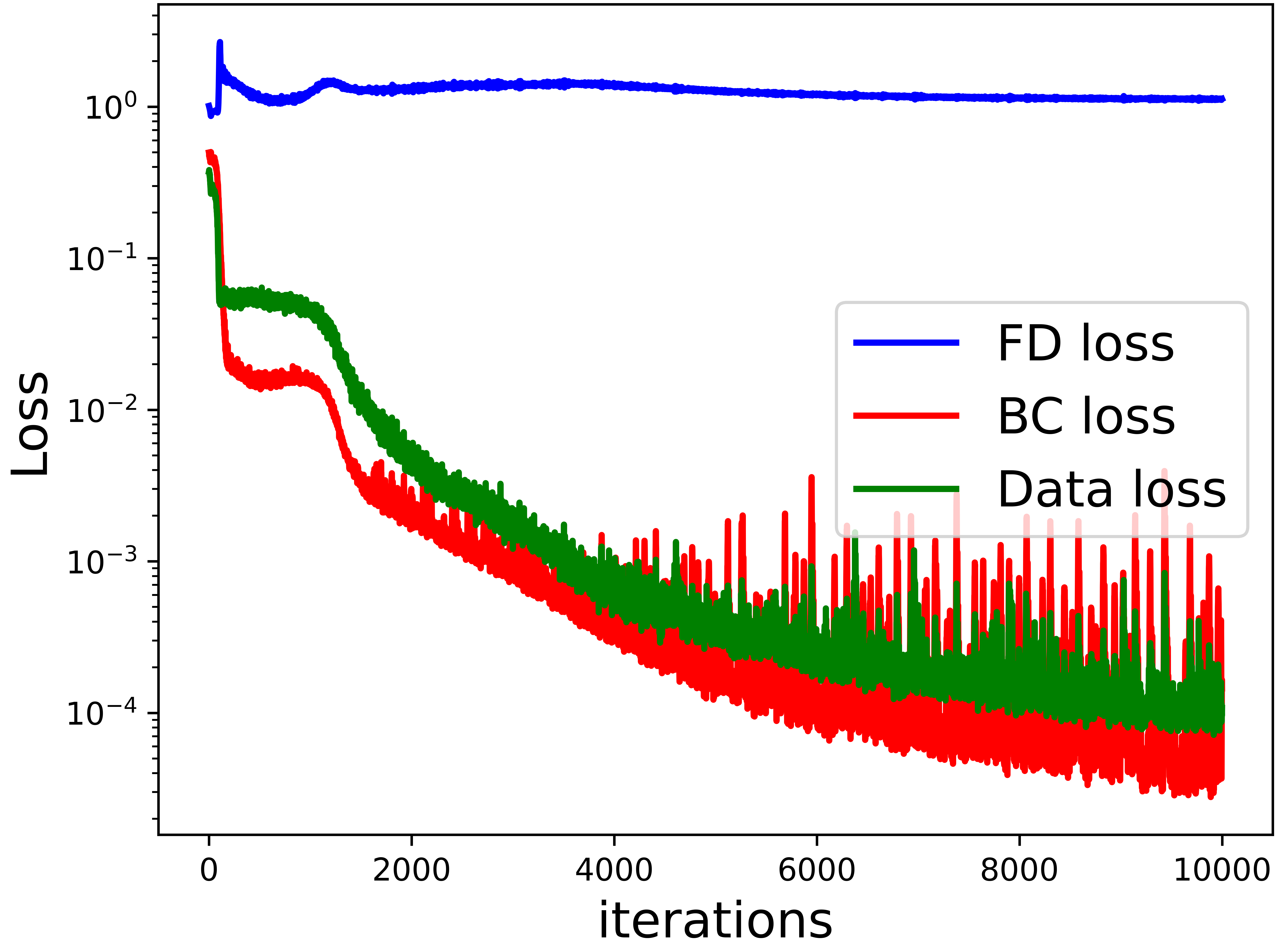}
	}%
	\subfigure[The test comparison]{
		\includegraphics[scale=0.35]{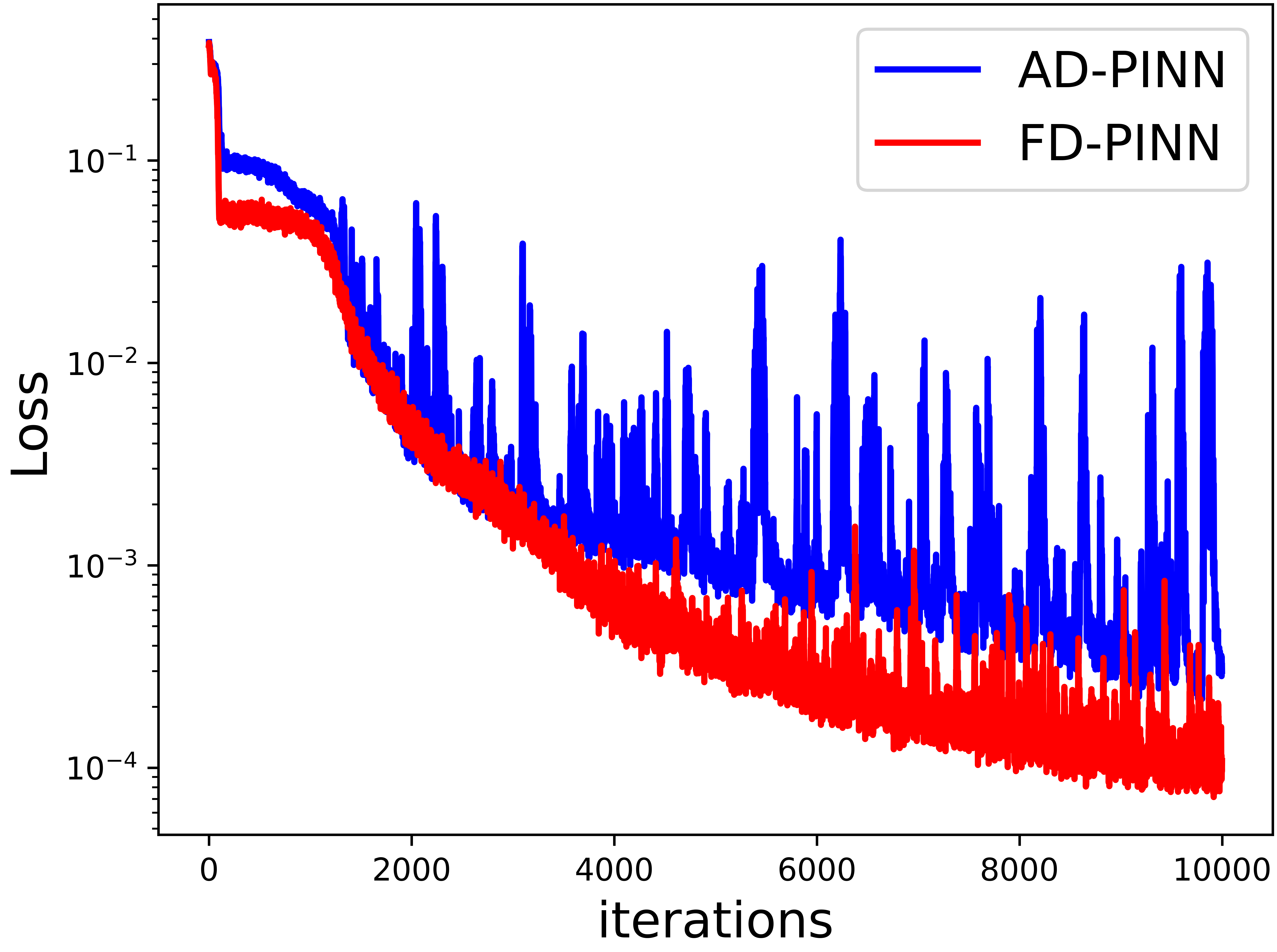}
	}%
	\centering
	\caption{Viscous Burgers equation: Evolution of the loss function along with the training of the AD-PINN (a) and FD-PINN (b). Besides, (c) The test comparison of the AD-PINN and FD-PINN.}
	\label{BurlossFD}
\end{figure}

The reference solution for this problem is available in \cite{raissi2019physics}. Fig. \ref{BurFDPINNpred} shows the prediction results of FD-PINN with 400 collocation points. In Fig. \ref{BurFDPINNtest}, according to three temporal snapshots $t=0.25,0.50,0.75$, we compare the predicted and exact solutions with FD-PINN in detail. It can be seen from Fig. \ref{BurADPINNerror} that the relative error of the FD-PINN prediction is less than that of AD-PINN, which indicates higher accuracy. Additionally, the decreasing trends of different losses in the 10k epochs of AD-PINN and FD-PINN are compared in Fig. \ref{BurlossFD}. Changes in the calculation of PDE residuals can allow neural networks to find optimization directions faster and more effectively, leading to higher prediction accuracy and robustness.

To further study the impact of the finite difference scheme on the learning performance of the newly proposed model, we chose three difference schemes for comparison in Table \ref{tab2}. Here the CDM \cite{2017Compact} of Eq. \eqref{Bureq} with $O(h^2+\tau^4)$ is defined as:

\begin{equation}
\begin{aligned}
u_{j}^{n+1}+\frac{\tau}{4}\left(u_{j}^{n+1}+u_{j}^{n}\right)\delta_{x} u_{j}^{n+1}-\frac{v \tau}{2}\delta_{x} u_{j}^{n+1}&= 
u_{j}^{n}-\frac{\tau}{4}\left(u_{j}^{n+1}+u_{j}^{n}\right)\delta_{x} u_{j}^{n}+\frac{v \tau}{2}\delta_{x} u_{j}^{n}, \\
\delta_{x} u_{j-1}^{n}+\delta_{x} u_{j+1}^{n}+4\delta_{x} u_{j}^{n}&=\frac{3}{h}(u_{j+1}^{n}-u_{j-1}^{n}),\\
\delta_{x}^{2} u_{j-1}^{n}+\delta_{x}^{2} u_{j+1}^{n}+10\delta_{x}^{2} u_{j}^{n}&=\frac{12}{h^2}(u_{j+1}^{n}-2u_{j}^{n}+u_{j-1}^{n}).
\end{aligned}
\end{equation}
The general form of the CNM \cite{2017Compact} with $O(h^2+\tau^2)$ as follows:
\begin{equation}
\begin{aligned}
&\left(-\frac{\tau}{8 h}\left(u_{j}^{n+1}+u_{j}^{n}\right)-\frac{v \tau}{2 h^{2}}\right) u_{j-1}^{n+1}+\left(1+\frac{v \tau}{h^{2}}\right) u_{j}^{n+1}+ \\
&\left(\frac{\tau}{8 h}\left(u_{j}^{n+1}+u_{j}^{n}\right)-\frac{v \tau}{2 h^{2}}\right) u_{j+1}^{n+1}= \\
&\left(\frac{\tau}{8 h}\left(u_{j}^{n+1}+u_{j}^{n}\right)+\frac{v \tau}{2 h^{2}}\right) u_{j-1}^{n}+\left(1-\frac{v \tau}{h^{2}}\right) u_{j}^{n}+ \\
&\left(-\frac{\tau}{8 h}\left(u_{j}^{n+1}+u_{j}^{n}\right)+\frac{v \tau}{2 h^{2}}\right) u_{j+1}^{n}.
\end{aligned}
\end{equation}

\begin{figure}[htbp]
	\centering
	\subfigure[Data Loss]{
		\includegraphics[scale=0.35]{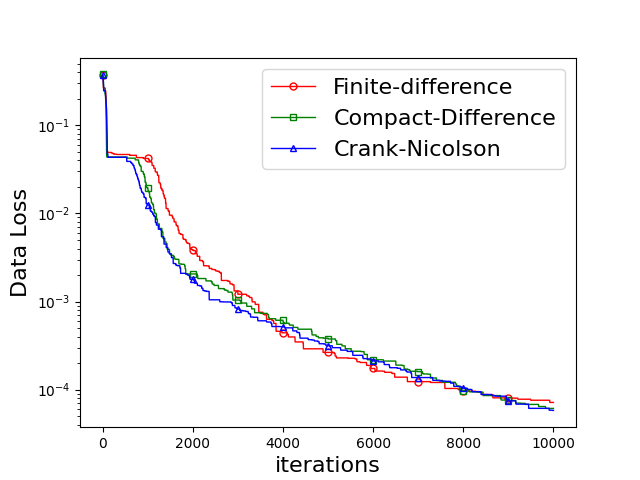}
	}%
	\subfigure[BC Loss]{
		\includegraphics[scale=0.35]{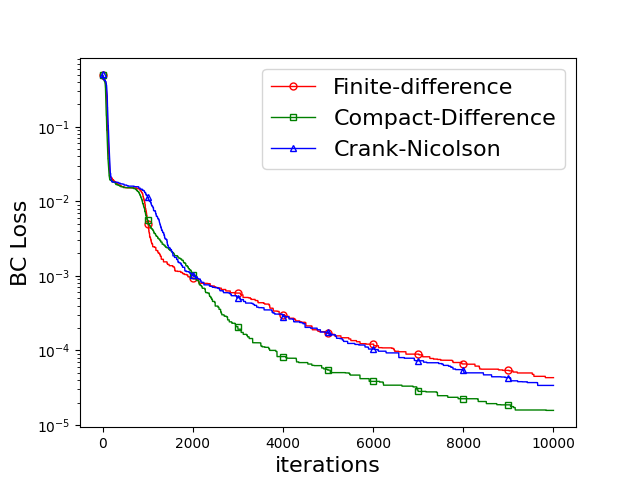}
	}%
	\centering
	\caption{Viscous Burgers equation: Evolution of Data loss (a) and BC loss (b) with three difference schemes along with the training of the HFD-PINN.}
	\label{BurFDPINNT}
\end{figure}

\begin{figure}[htbp]
	\centering
	\subfigure[FD-PINN]{
		\includegraphics[scale=0.35]{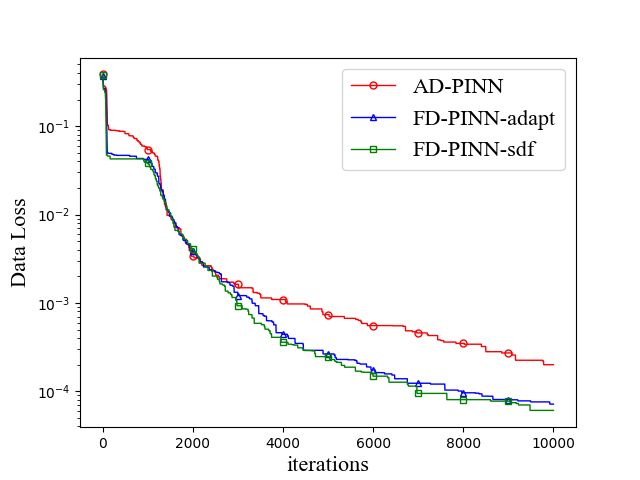}
	}%
	\subfigure[CD-PINN]{
		\includegraphics[scale=0.35]{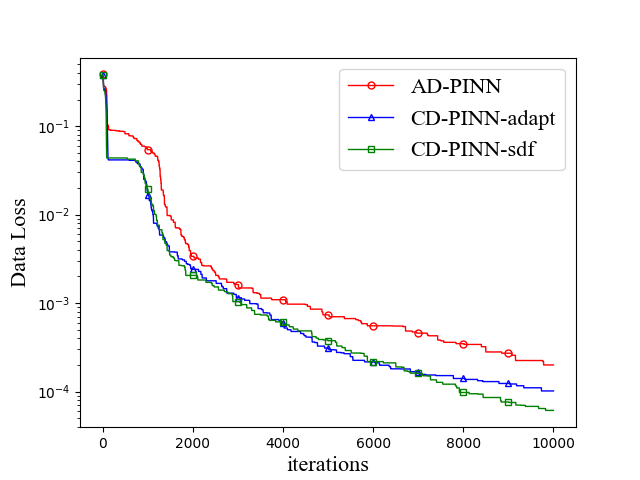}
	}%
	\subfigure[CN-PINN]{
		\includegraphics[scale=0.35]{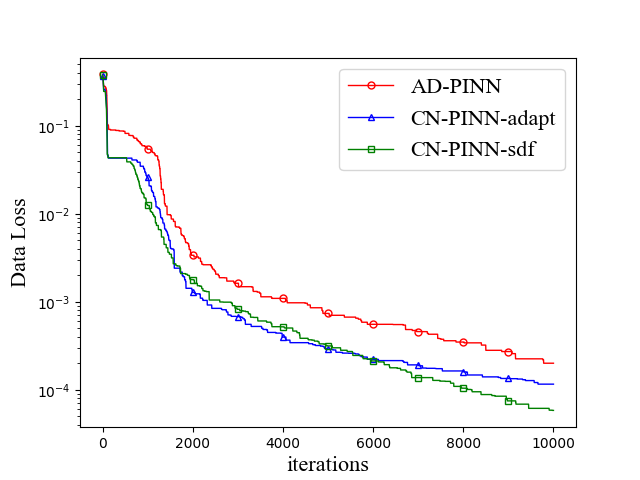}
	}%
	\centering
	\caption{Viscous Burgers equation: (a) The comparison of the AD-PINN, FD-PINN-adapt and FD-PINN-sdf. Similarly, the results of CD-PINN (b) and CN-PINN (c) are also shown.}
	\label{Burlossself}
\end{figure}

The results of FDM are unstable and easily affected by the difference scheme. However, the results shown in Fig. \ref{BurFDPINNT} indicate that HFD-PINN outperforms AD-PINN in terms of accuracy, no matter which difference scheme is used. Moreover, using Compact-Difference to fit the data for the BC could be more accurate. The three methods, the AD-PINN, FD-PINN-adapt, and FD-PINN-sdf, are comparable in this benchmark problem. The evolution is shown in Fig. \ref{Burlossself}(a) demonstrate that the proposed FD-PINN-sdf produces the most accurate approximation to the reference solution. Similarly, the self-adaptive finite difference method with sdf effectively improves the predictions of CD-PINN and CN-PINN.  
\begin{table}[tp]  
	
	\centering  
	\fontsize{8}{8}\selectfont  
	\begin{threeparttable}  
		\caption{Comparison of PINN using different Finite Difference Schemes to solve the Viscous Burgers equation.}  
		\label{tab2}  
		\begin{tabular}{ccccccc}  
			\toprule  
			\multirow{2}{*}{Loss}&  
			\multicolumn{2}{c}{Finite Difference Scheme}&\multicolumn{2}{c}{self-adaptive}&\multicolumn{2}{c}{self-adaptive with sdf}\cr  
			\cmidrule(lr){2-3} \cmidrule(lr){4-5} \cmidrule(lr){5-7}  
			&CD-PINN&CN-PINN&CD-PINN-adapt&CN-PINN-adapt&CD-PINN-sdf&CN-PINN-sdf\cr  
			\midrule  
			Data loss&1.3e-04&2.5e-04&1.4e-04&1.7e-04&1.2e-04&1.3e-04\cr  
			BC loss&2.1e-05&3.3e-05&2.2e-05&3.8e-05&1.6e-05&2.5e-05\cr    
			\bottomrule  
		\end{tabular}  
	\end{threeparttable}  
\end{table} 

\begin{figure}[htbp]
	\centering
	\subfigure{
		\includegraphics[scale=0.45]{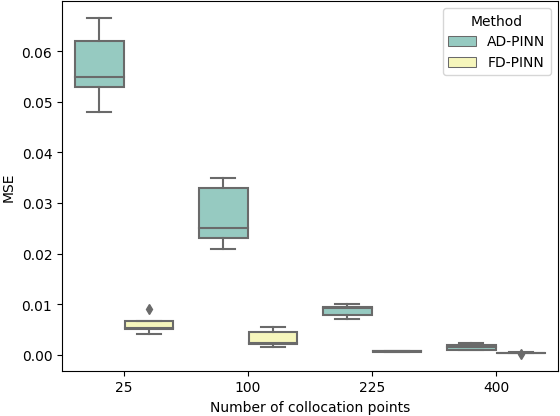}
	}
	\caption{One-dimensional Heat conduction problem: Boxplot of the relative L2 loss of AD-PINN and FD-PINN with different numbers of collocation points.}
	\label{TUFDPINNpa}
\end{figure} 
\subsection{ One-dimensional Heat conduction problem}
We chose the heat equation further to study the learning performance of the newly proposed model. The PDE in one-dimensional space is of the form:
\begin{equation}
\begin{array}{l}
u_{t}=a^{2} u_{xx}+f(x,t) \quad x \in[0,l], \quad t \in[0,T], \\
u(0, t)=u(l, t)=0, \\
u(x, 0)=0,
\end{array}
\label{1dPoisson Equation1}
\end{equation}
where $a=l=T=1$ and $f(x,t)=A \sin \frac{\pi x}{l}$. In particular, A is set as 40. The analytical solution reads as 
\begin{equation}
\begin{array}{l}
u(x, t)=\frac{A l^{2}}{\pi^{2} a^{2}}\left(1-e^{-n^{2} a^{2} l / t^{2}}\right) \sin \frac{\pi x}{l}.
\end{array}
\label{1dPoisson Equation2}
\end{equation}

\begin{figure}[htbp]
	\centering
	\subfigure[Exact]{
		\includegraphics[scale=0.45]{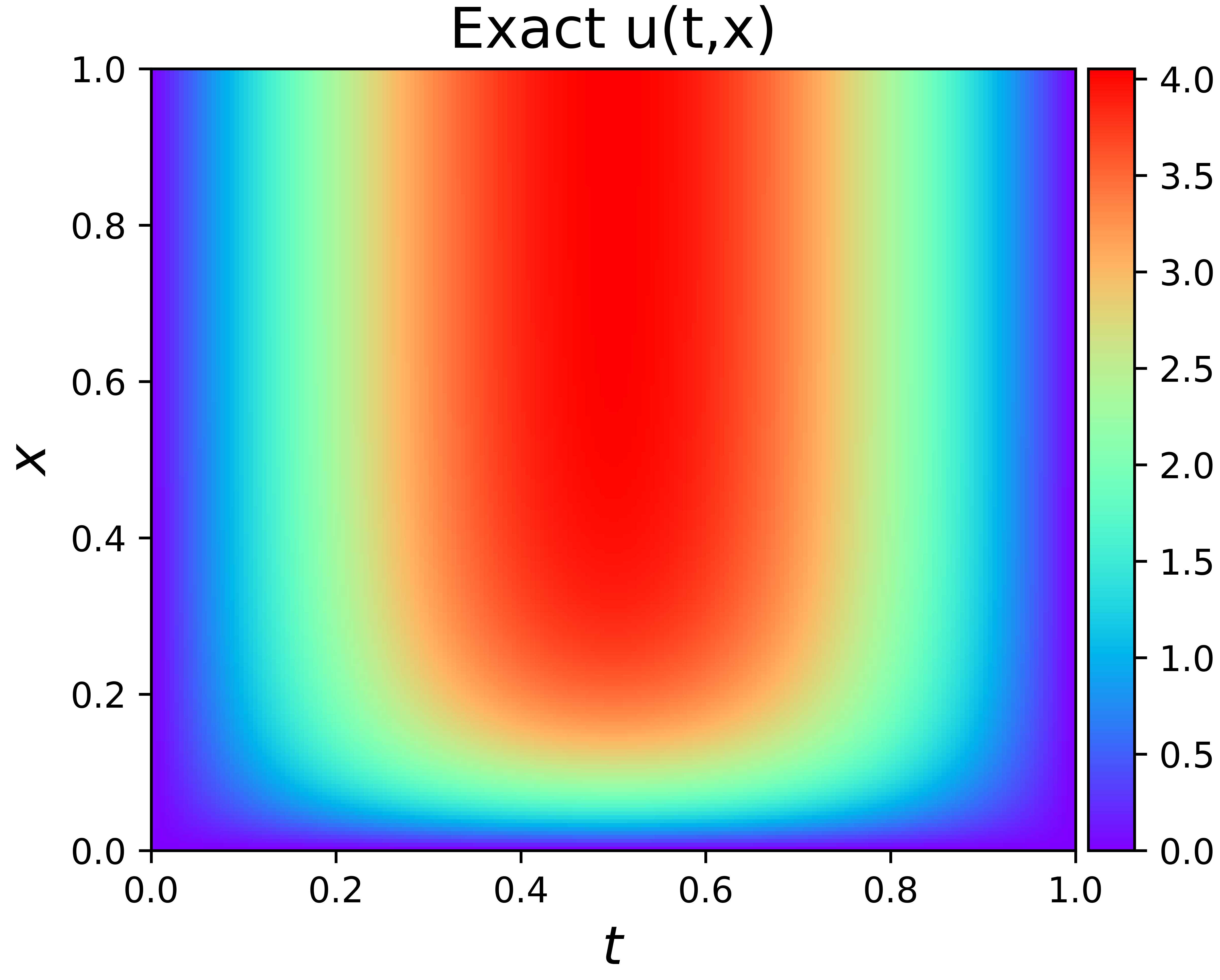}
	}%
	\subfigure[FD-PINN]{
		\includegraphics[scale=0.45]{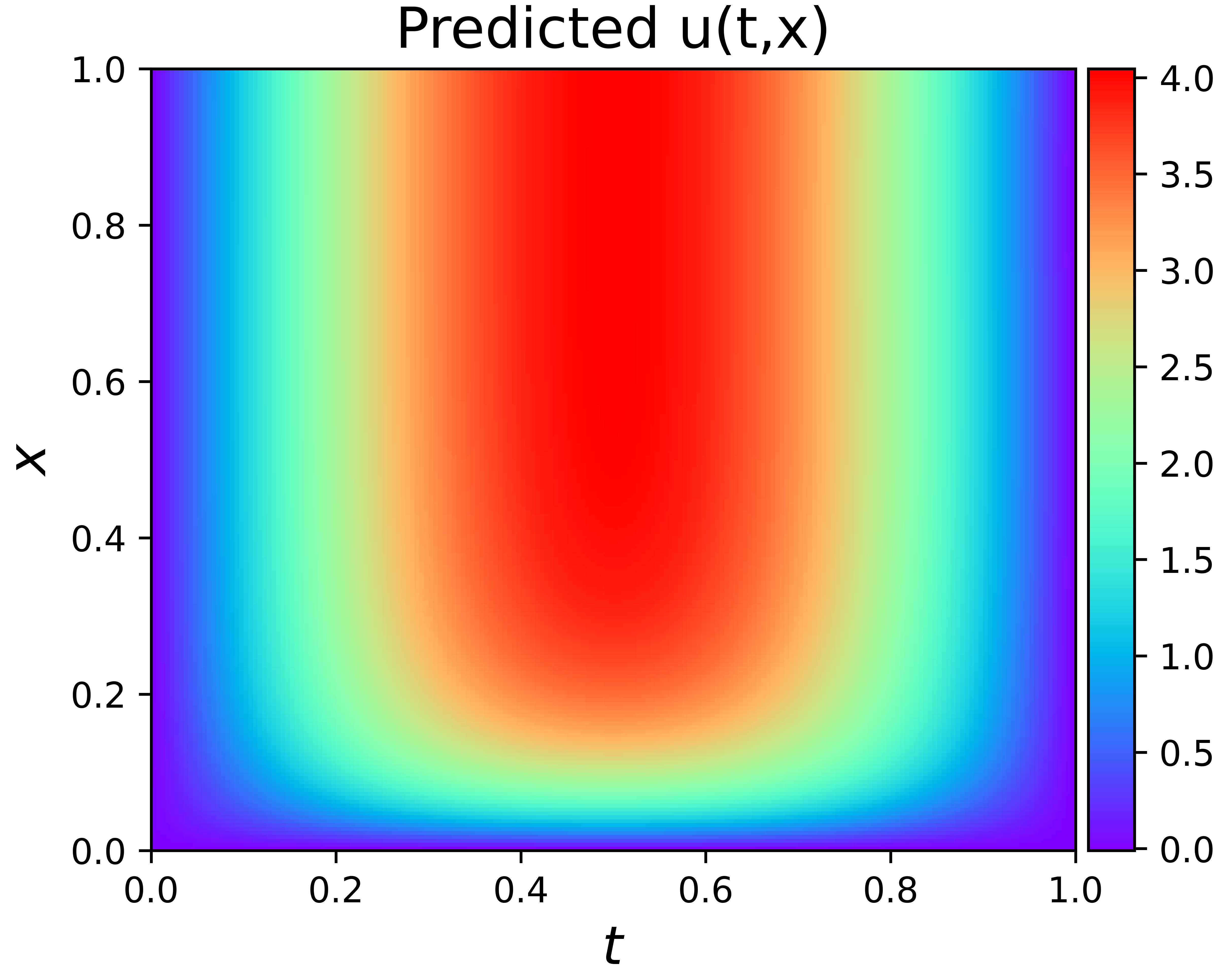}
	}%
	\centering
	\caption{One-dimensional Heat conduction problem: (a) The reference solution. (b) The prediction solution of FD-PINN.}
	\label{TUpred}
\end{figure}

\begin{figure}[htbp]
	\centering
	\subfigure[AD-PINN]{
		\includegraphics[scale=0.45]{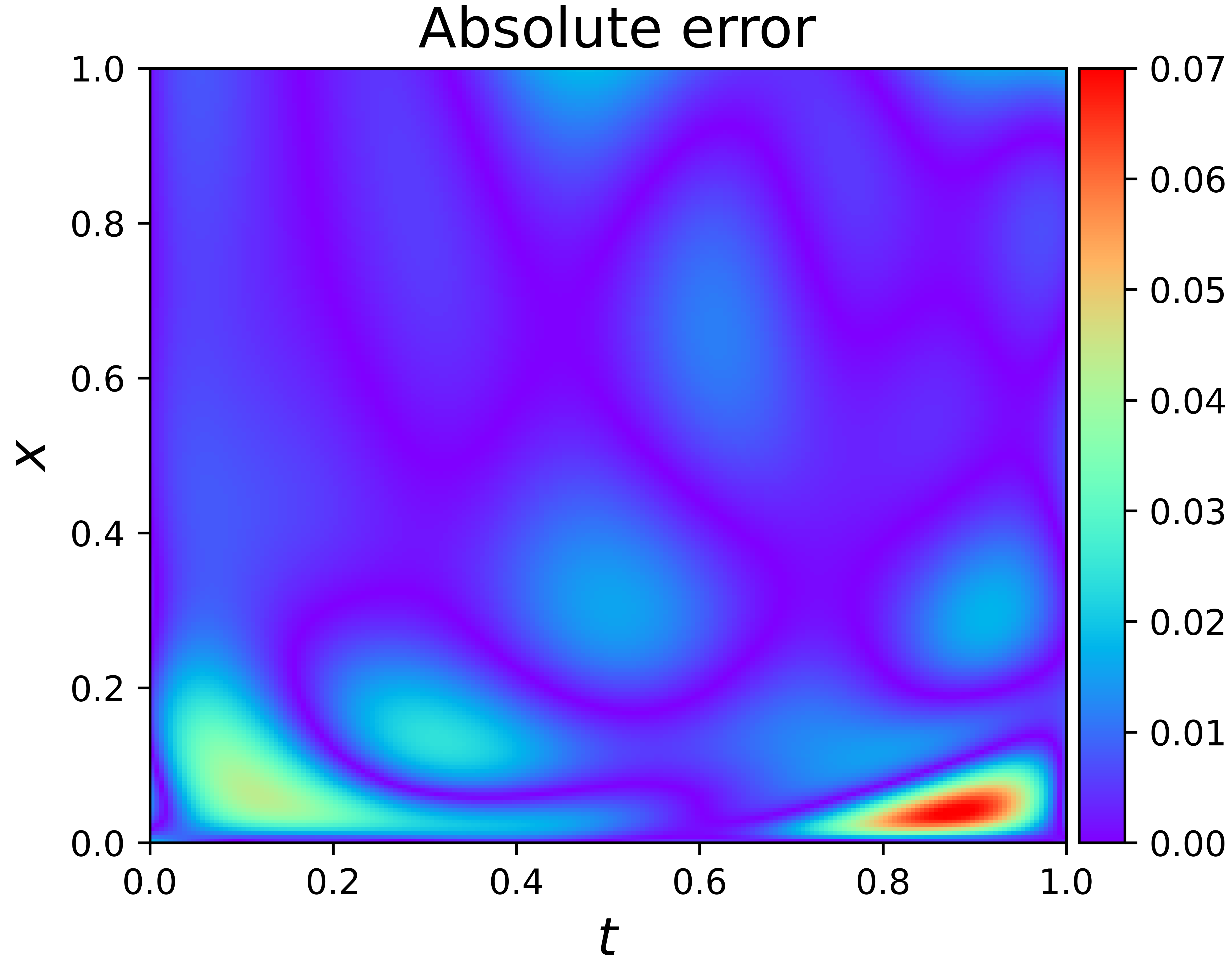}
	}%
	\subfigure[FD-PINN]{
		\includegraphics[scale=0.45]{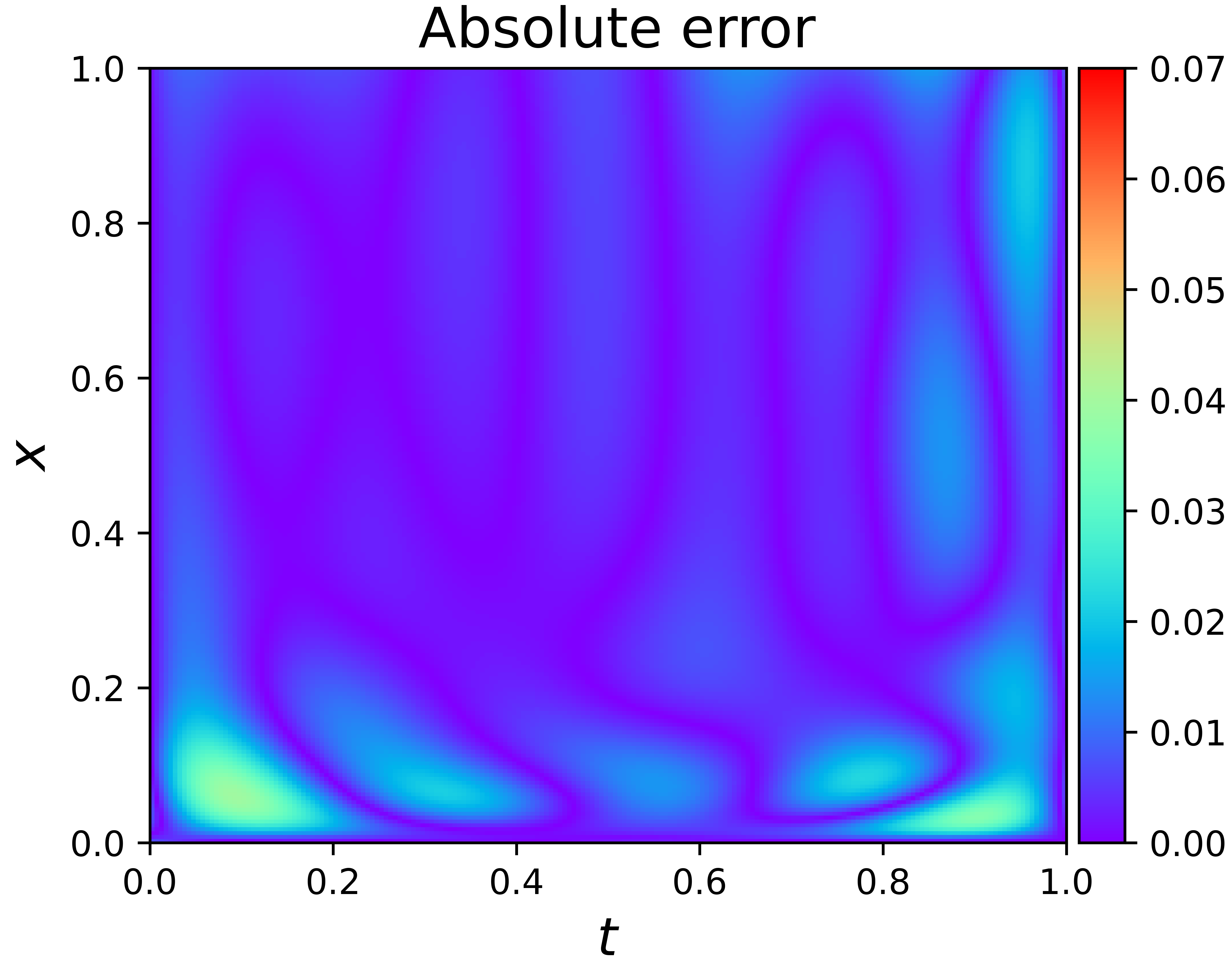}
	}%
	\centering
	\caption{One-dimensional Heat conduction problem: (a) The absolute error of AD-PINN. (b) The absolute error of FD-PINN}
	\label{TUerror}
\end{figure}

The network architecture, optimizer, and learning rate scheduler of training are kept the same as in section 3.1. We generate $N_f =N_b$ collocation points across the domain with two boundary conditions. The results of the AD-PINN and FD-PINN with the different number of collocation points are presented in Table \ref{tab3}. In addition, the impact of the collocation points becomes much more explicit by analyzing the error measures shown in Fig. \ref{TUFDPINNpa} for both the AD-PINN and FD-PINN. The exact solution and the result obtained with the FD-PINN are presented in Fig. \ref{TUpred}. In Fig. \ref{TUerror} we compare the evolution of the loss function during training and the validation MSE loss. It can be seen that FD-PINN performs significantly better than AD-PINN. 

\begin{table}[tp]  
	
	\centering  
	\fontsize{8}{8}\selectfont  
	\caption{Comparing the relative error with different number of collocation points when learning One-dimensional Heat conduction problem.}  
	\label{tab3}  
	\begin{tabular}{|c|c|c|c|c|c|c|}  
		\hline  
		\multirow{2}{*}{collocation points}&  
		\multicolumn{2}{c|}{AD-PINN}&\multicolumn{2}{c|}{FD-PINN}\cr\cline{2-5}  
		&BC loss&Data loss&BC loss&Data loss\cr  
		\hline  
		\hline  
		25&7.3e-02&{\bf 5.3e-02}&4.5e-02&{\bf 5.2e-03}\cr\hline  
		100&1.0e-02&{\bf 9.5e-03}&9.4e-03&{\bf 2.2e-03}\cr\hline  
		225&8.2e-03&{\bf 2.1e-03}&7.1e-03&{\bf 6.8e-04}\cr\hline  
		400&6.6e-03&{\bf 9.5e-04}&5.9e-03&{\bf 4.1e-04}\cr\hline  
	\end{tabular}  
\end{table} 

%
%

\begin{figure}[htbp]
	\centering
	\subfigure[AD-PINN]{
		\includegraphics[scale=0.35]{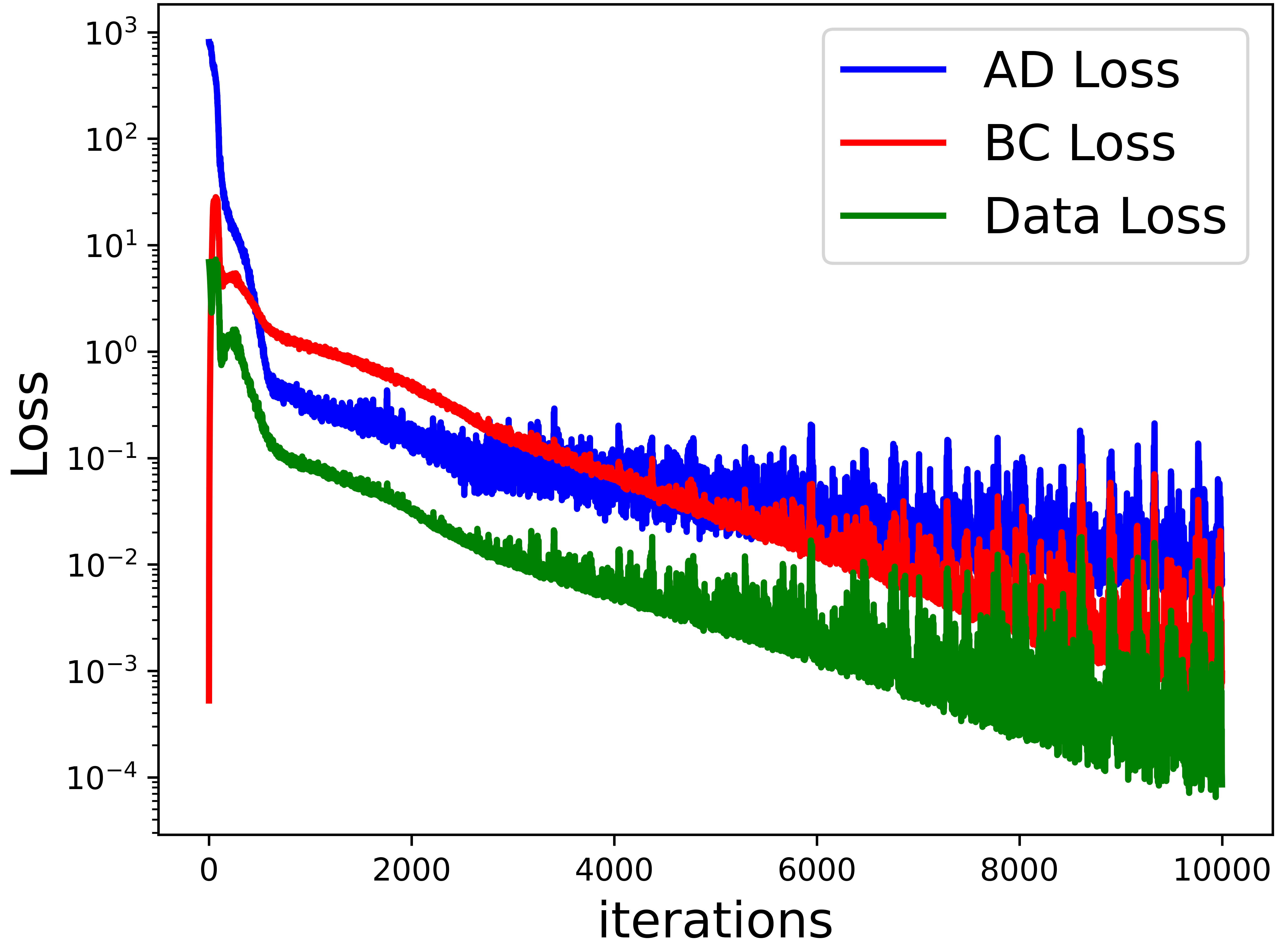}
	}%
	\subfigure[FD-PINN]{
		\includegraphics[scale=0.35]{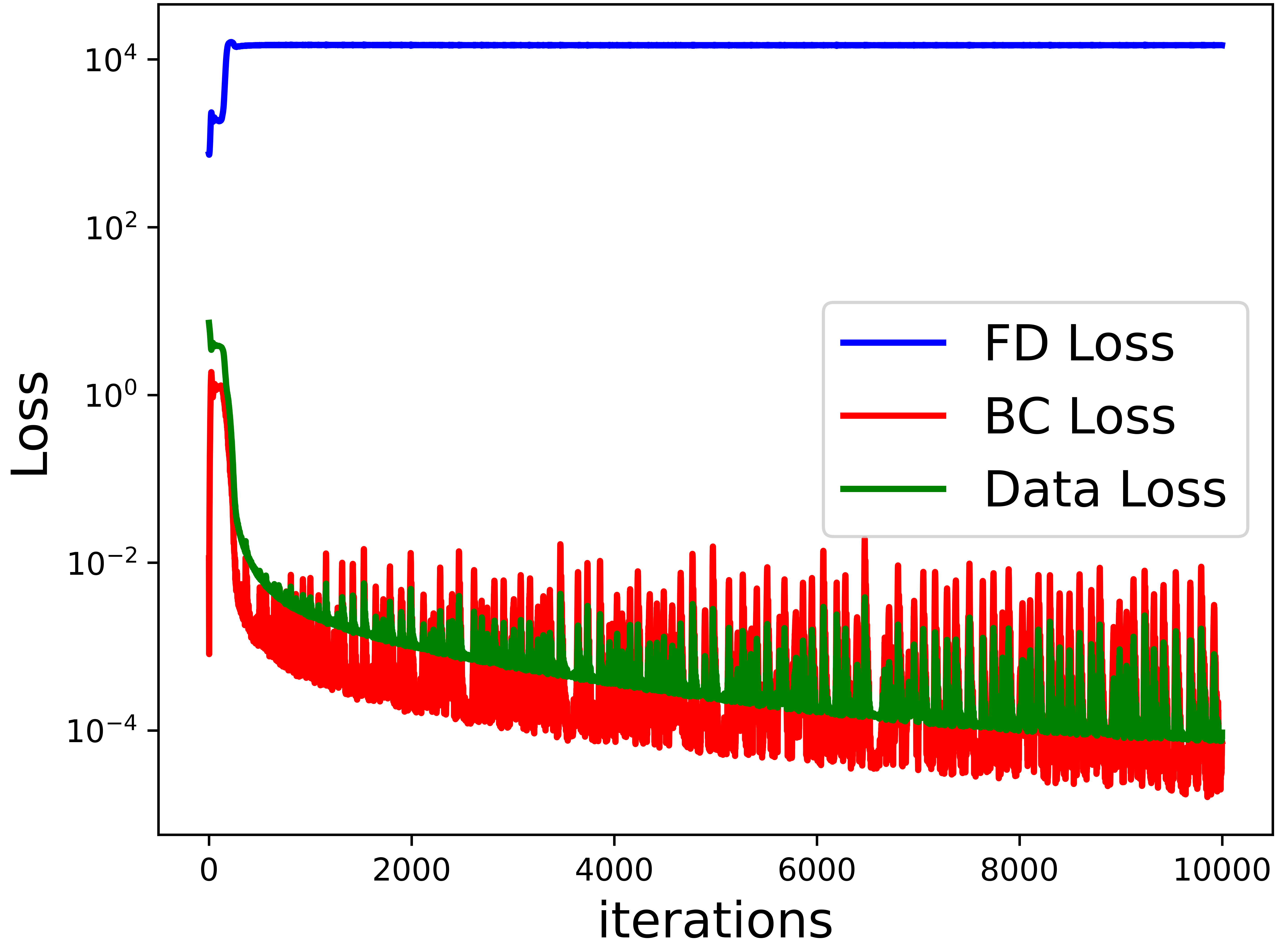}
	}%
	\subfigure[The test comparison]{
		\includegraphics[scale=0.35]{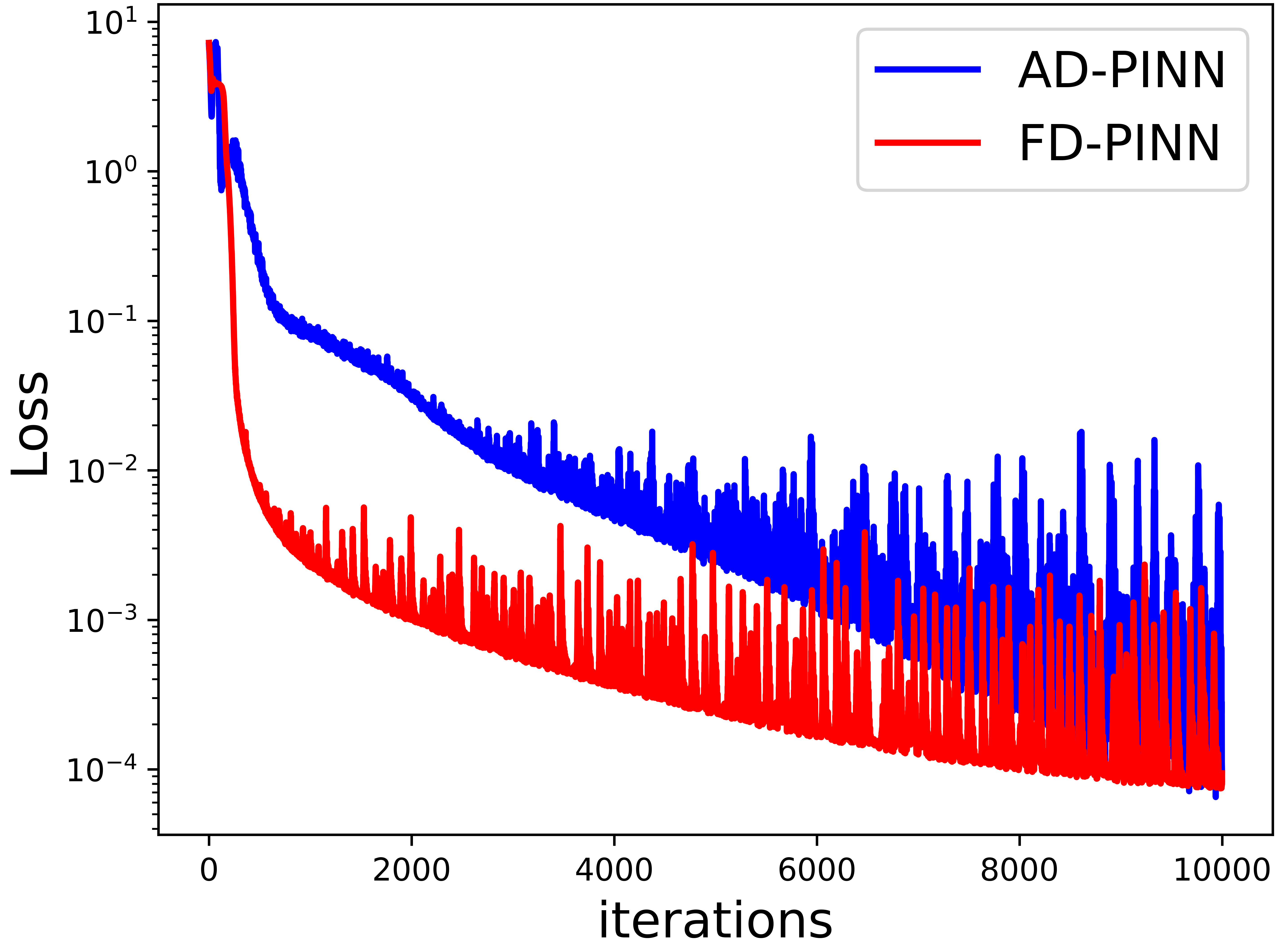}
	}%
	\centering
	\caption{One-dimensional Heat conduction problem: Evolution of the loss function along with the training of the AD-PINN (a) and FD-PINN (b). Besides, (c) The test comparison of the AD-PINN and FD-PINN.}
	\label{TUlossFD}
\end{figure}

In Table\ref{tab4}, We evaluate the performance of a fully connected deep neural network model that uses a finite difference scheme to infer the PDE. Here the CDM of Eq. \eqref{1dPoisson Equation1} is defined as:
\begin{equation}
\begin{aligned}
&\frac{1}{12}\left(\delta_{t} u_{i-1}^{k+\frac{1}{2}}+10 \delta_{t} u_{i}^{k+\frac{1}{2}}+\delta_{t} u_{i+1}^{k+\frac{1}{2}}\right)-a \delta_{t}^{2} u_{i}^{k+\frac{1}{2}} \\
&=\frac{1}{12}\left[f\left(x_{i-1}, t_{k+\frac{1}{2}}\right)+10 f\left(x_{i}, t_{k+\frac{1}{2}}\right)+f\left(x_{i+1}, t_{k+\frac{1}{2}}\right)\right] \\
&1 \leq i \leq m-1,0 \leq k \leq n-1.
\end{aligned}
\end{equation}
where we set $r=a \tau /h^2$. The general form of the CNM with $O(h^2+\tau^3)$ as follows:
\begin{equation}
\begin{aligned}
&\left(\frac{1}{12}-\frac{1}{2} r\right) u_{i-1}^{k+1}+\left(\frac{5}{6}+r\right) u_{i}^{k+1}+\left(\frac{1}{12}-\frac{1}{2} r\right) u_{i+1}^{k+1} \\
&=\left(\frac{1}{12}+\frac{1}{2} r\right) u_{i-1}^{k}+\left(\frac{5}{6}-r\right) u_{i}^{k}+\left(\frac{1}{12}+\frac{1}{2} r\right) u_{i+1}^{k} \\
&+\frac{\tau}{12}\left[f\left(x_{i-1}, t_{k+\frac{1}{2}}\right)+10 f\left(x_{i}, t_{k+\frac{1}{2}}\right)+f\left(x_{i+1}, t_{k+\frac{1}{2}}\right)\right] \\
&1 \leq i \leq m-1,0 \leq k \leq n-1.
\end{aligned}
\end{equation}

\begin{figure}[htbp]
	\centering
	\subfigure[Data Loss]{
		\includegraphics[scale=0.35]{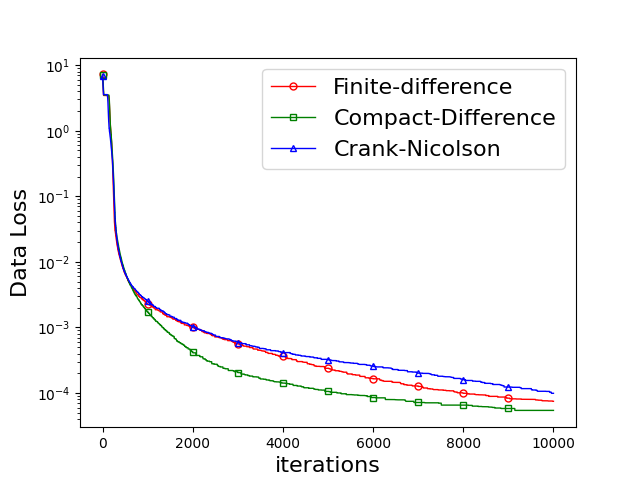}
	}%
	\subfigure[BC Loss]{
		\includegraphics[scale=0.35]{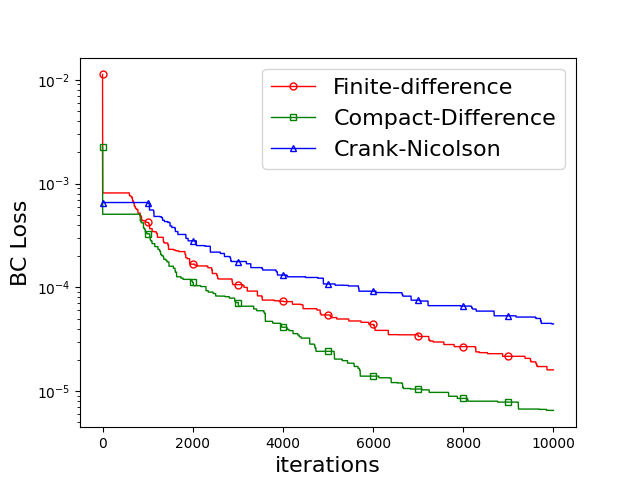}
	}%
	\centering
	\caption{One-dimensional Heat conduction problem: Evolution of Data loss (a) and BC loss (b) with three difference schemes along with the training of the HFD-PINN.}
	\label{TUFDPINNT}
\end{figure}

\begin{figure}[htbp]
	\centering
	\subfigure[CD-PINN]{
		\includegraphics[scale=0.35]{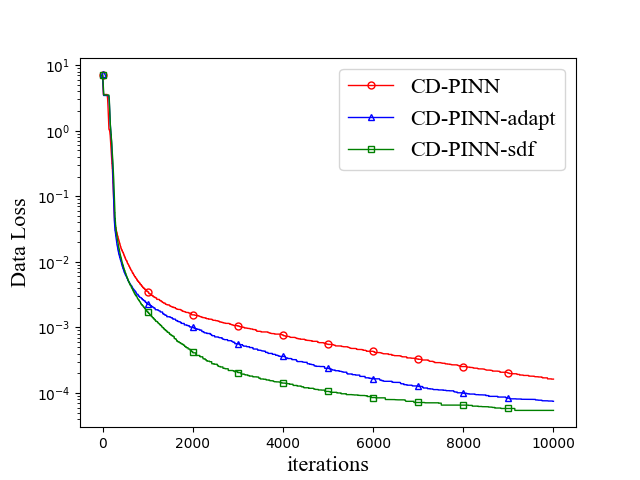}
	}%
	\subfigure[CN-PINN]{
		\includegraphics[scale=0.35]{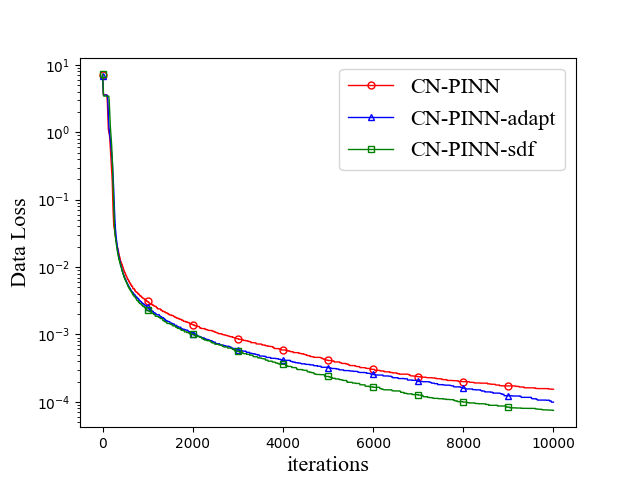}
	}%
	\centering
	\caption{One-dimensional Heat conduction problem: (a) The comparison of the CD-PINN, CD-PINN-adapt and CD-PINN-sdf. Similarly, the results of CN-PINN (b) is also shown.}
	\label{TUlossself}
\end{figure}

In Fig. \ref{TUFDPINNT}, the prediction accuracy of the model is slightly affected by the difference format, but the loss at the boundary with the Compact-Difference format is minimal. In particular, By applying the self-adaptive finite difference methods into CD-PINN and CN-PINN showed in Fig. \ref{TUlossself}, it is found that these methods work well, and Using sdf to define the difference interval is also more stable. So it can replace the essential finite difference that requires a background mesh in complex areas.  

\begin{table}[tp]  
	
	\centering  
	\fontsize{8}{8}\selectfont  
	\begin{threeparttable}  
		\caption{Comparison of PINN using different Finite Difference Schemes to solve the 1d Heat conduction problem.}  
		\label{tab4}  
		\begin{tabular}{ccccccc}  
			\toprule  
			\multirow{2}{*}{Loss}&  
			\multicolumn{2}{c}{Finite Difference Scheme}&\multicolumn{2}{c}{self-adaptive}&\multicolumn{2}{c}{self-adaptive with sdf}\cr  
			\cmidrule(lr){2-3} \cmidrule(lr){4-5} \cmidrule(lr){5-7}  
			&CD-PINN&CN-PINN&CD-PINN-adapt&CN-PINN-adapt&CD-PINN-sdf&CN-PINN-sdf\cr  
			\midrule  
			Data loss&3.2e-04&6.2e-04&6.5e-04&4.2e-04&2.2e-04&3.2e-04\cr  
			BC loss&1.4e-05&5.3e-04&4.2e-05&3.8e-04&2.4e-05&2.1e-04\cr    
			\bottomrule  
		\end{tabular}  
	\end{threeparttable}  
\end{table} 

\begin{figure}[htbp]
	\centering
	\subfigure[Boundary conditions]{
		\includegraphics[scale=0.22]{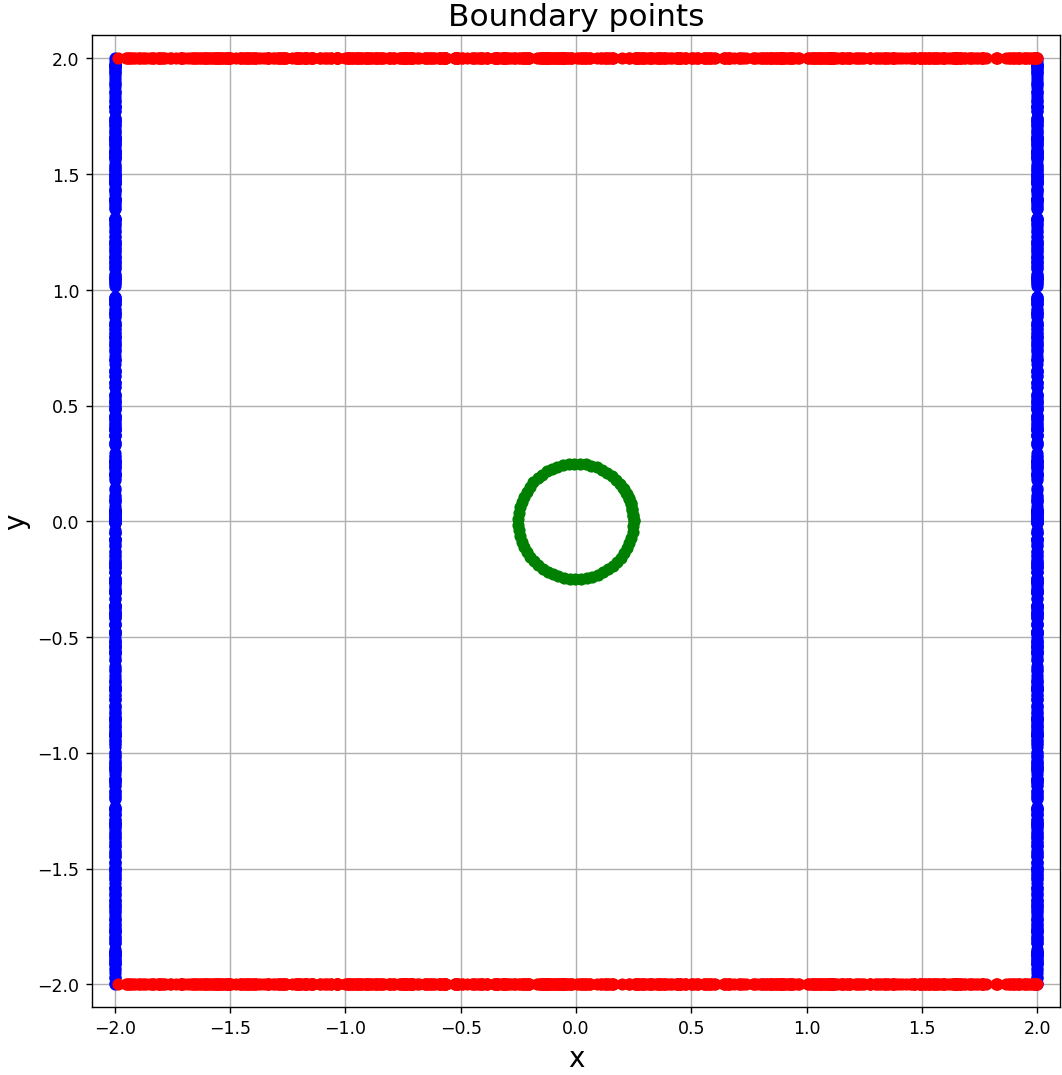}
	}%
	\subfigure[Training points]{
		\includegraphics[scale=0.22]{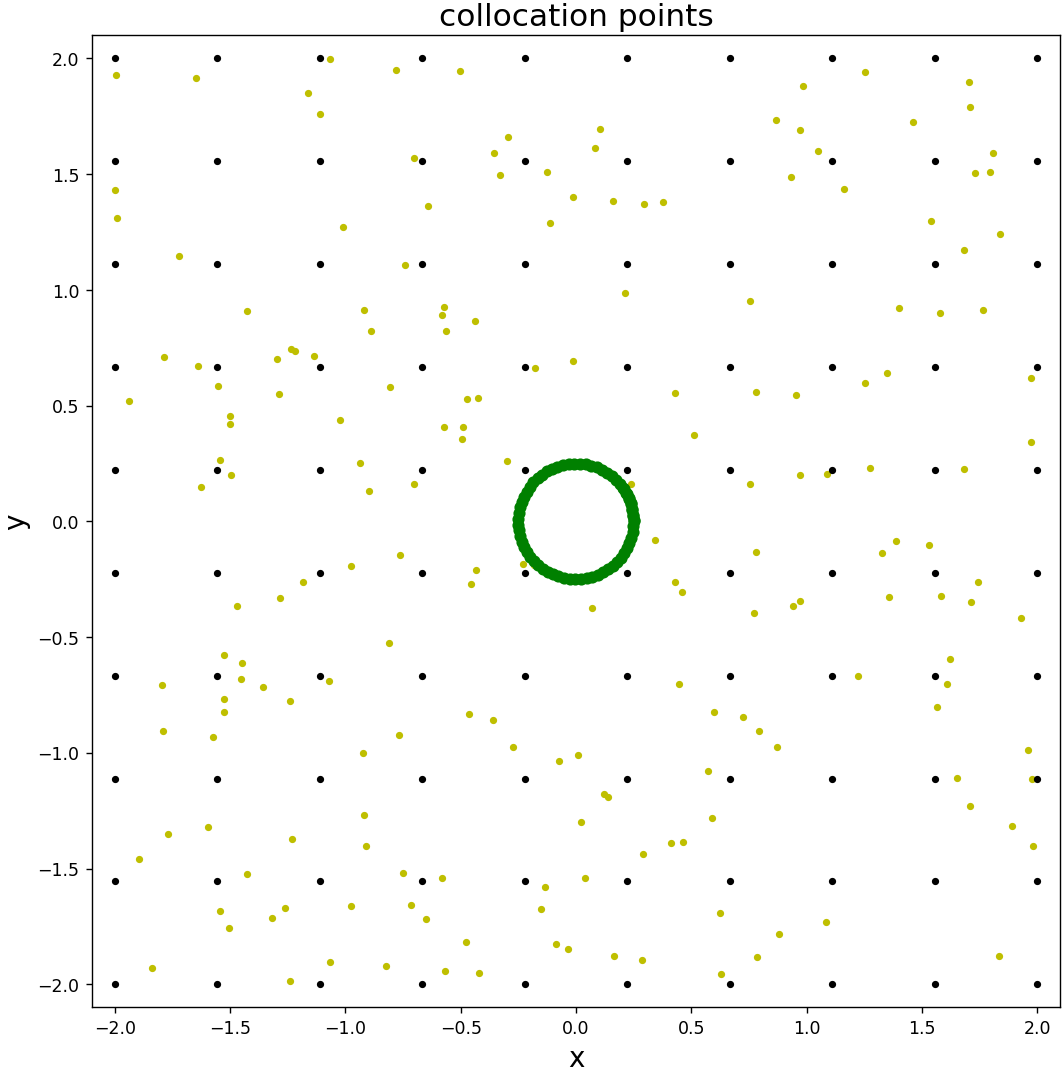}
	}%
	\centering
	\caption{Domain and input points for the heat transfer problem of square plate with a hole in the middle. (a) Blue and red markers are used for Dirichlet and Neumann boundary conditions enforcement. The hole boundary has 160 equidistant Dirichlet collocation points (green dot points). (b) Black markers are FDM mesh points. Yellow dot points show collocation points for imposing PDEs.}
	\label{2dhpoints}
\end{figure}

\begin{figure}[htbp]
	\centering
	\subfigure[AD-PINN]{
		\includegraphics[scale=0.35]{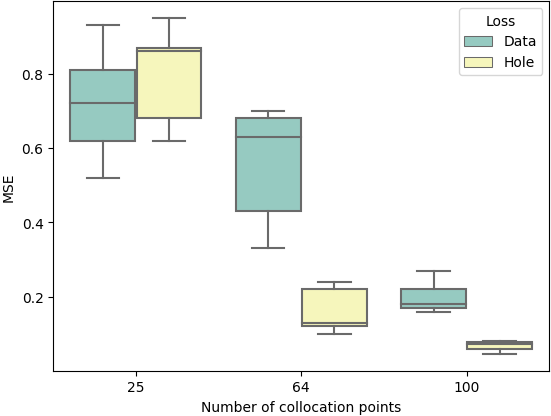}
	}%
	\subfigure[HFD-PINN]{
		\includegraphics[scale=0.35]{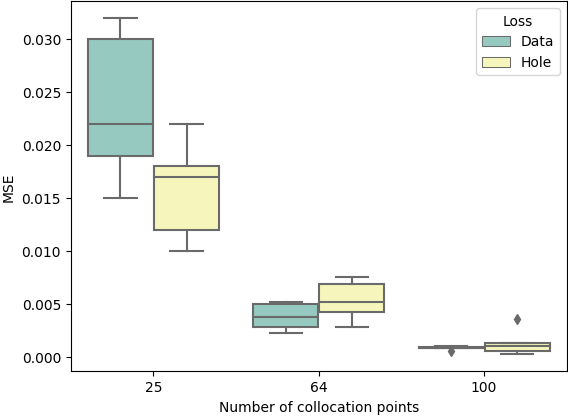}
	}%
	\caption{Heat transfer problem of square plate with a hole in the middle: Boxplot of the Hole and Data loss of AD-PINN (a) and FD-PINN (b) with different numbers of collocation points.}
	\label{2dhFDPINNpa}
\end{figure} 

\subsection{Heat transfer problem of square plate with a hole in the middle}
In this problem, we solve Heat transfer problem for the plate with a circular hole at its center. The boundary conditions of this problem are expressed in a mixed form, including Riemann and Dirichlet (see Fig. \ref{2dhpoints}(a)). The partial differential equation reads as:

\begin{equation}
\begin{array}{c}
u_{xx} +u_{yy} = u_{t}, \quad (x, y) \in \Omega,  t \in [0, 2],\\
\Omega = \Omega_{1}/\Omega_{2}, \Omega_{1}=[x_l, x_r] \times [y_l, y_r], \Omega_{2}=(x^{2} + y^{2} \leqslant r^{2}), \\
u(x_l, y) = u(x_r, y) = u(x_{h}, y_{h}) = T_{0}, (x_{h}, y_{h}) \in \partial \Omega, \\
u_{x}(x, y_l) = u_{x}(x, y_r) = 0. 
\end{array}
\label{2dPoisson Equation}
\end{equation}

We solve the problem where $\Omega$ is a square domain with $[x_l, x_r]=[-2,2]$. We take the hole radius $r=0.25$ and $T_{0}=5$ to obtain the solution. The loss function for the AD-PINN and HFD-PINN solver reads as:
\begin{equation}
\begin{aligned}
\mathcal{L}^{\mathrm{AD-PINN}}(\theta)=& \mathcal{L}_{\mathrm{AD}}(\theta)+\mathcal{L}_{\mathrm{hole}}(\theta)+\mathcal{L}_{\mathrm{BC}}(\theta), \\
\mathcal{L}^{\mathrm{HFD-PINN}}(\theta)=& \mathcal{L}_{\mathrm{FD}}(\theta)+\mathcal{L}_{\mathrm{hole}}(\theta)+\mathcal{L}_{\mathrm{BC}}(\theta). \\
\end{aligned}
\end{equation}

As for AD-PINN, all partial derivatives $\partial t$ and $u_{xx}$ are obtained via the automatic differentiation to define $\mathcal{L}_{\mathrm{AD}}$. HFD-PINN uses the regular background mesh to define $\mathcal{L}_{\mathrm{FD}}$. Fig. \ref{2dhpoints}(b) shows the collocation points of AD-PINN and HFD-PINN with black markers and yellow dot points. 

\begin{equation}
\begin{aligned}
\mathcal{L}_{\mathrm{AD}}(\theta)=& \frac{1}{N_{\mathrm{f}}} \sum_{i=1}^{N_{\mathrm{f}}}\left\|\hat{u}_{xx} +\hat{u}_{yy} - \hat{u}_{t}\right\|_{2}^{2} \\
\mathcal{L}_{\mathrm{FD}}(\theta)=& \frac{1}{N_{\mathrm{f}}} \sum_{i=1}^{N_{\mathrm{f}}}\left\|\frac{\hat{u}(x+\Delta x,y,t)-2\hat{u}(x,y,t)+\hat{u}(x-\Delta x,y,t)}{{\Delta x}^{2}}\right\|_{2}^{2} \\+&\frac{1}{N_{\mathrm{f}}} \sum_{i=1}^{N_{\mathrm{f}}}\left\| \frac{\hat{u}(x,y+\Delta y,t)-2\hat{u}(x,y,t)+\hat{u}(x,y-\Delta y,t) }{{\Delta y}^{2}}\right\|_{2}^{2} \\-& \frac{1}{N_{\mathrm{f}}} \sum_{i=1}^{N_{\mathrm{f}}}\left\|\frac{\hat{u}(x,y,t + \Delta t)-\hat{u}(x,y,t)}{\Delta t} \right\|_{2}^{2}. \\
\end{aligned}
\label{ADFDEquation}
\end{equation}

We also use the the boundary points of the hole $(x_{h}, y_{h}, t_{h})$ to define $\mathcal{L}_{\mathrm{Hole}}$. The hole loss includes the boundary error at the hole and the PDE residual error calculated by AD. The BC loss represents the mixed boundary error in the square domain:
\begin{equation}
\begin{aligned}
\mathcal{L}_{\mathrm{Hole}}(\theta)=&\frac{1}{N_{\mathrm{h}}} \sum_{i=1}^{N_{\mathrm{h}}}\left\|\hat{u}_{x_{h}x_{h}} +\hat{u}_{y_{h}y_{h}}- \hat{u}_{t_{h}}\right\|_{2}^{2} + \sum_{i=1}^{N_{\mathrm{h}}}\left\|\hat{u}(x_{h},y_{h},t_{h} )- T_{0}\right\|_{2}^{2},  \\
\mathcal{L}_{\mathrm{BC}}(\theta)=& \frac{1}{N_{\mathrm{b}}} \sum_{i=1}^{N_{\mathrm{b}}}\left\|\hat{u}(-2,y) + \hat{u}(2,y) + \hat{u}_{y}(x,-2) + \hat{u}_{y}(x,2) - 2T_{0} \right\|_{2}^{2}. \\
\end{aligned}
\end{equation}

The neural network has four hidden layers with the tanh activation function, and each layer has 50 hidden units. The training is performed by the Adam optimizer with a learning rate of 0.001. we compare the performance of AD-PINN and HFD-PINN at different numbers of observations. These observations can be further quantified in Table \ref{tab6}. Fig. \ref{2dhFDPINNpa} represents the mean and standard deviation of the Data and Hole loss for the AD-PINN and HFD-PINN with different numbers of collocation points. The prediction accuracy and robustness of the HFD-PINN are always higher than AD-PINN, irrespective of the number of collocation points.

\begin{figure}[htbp]
	\centering
	\subfigure[Exact]{
		\includegraphics[scale=0.35]{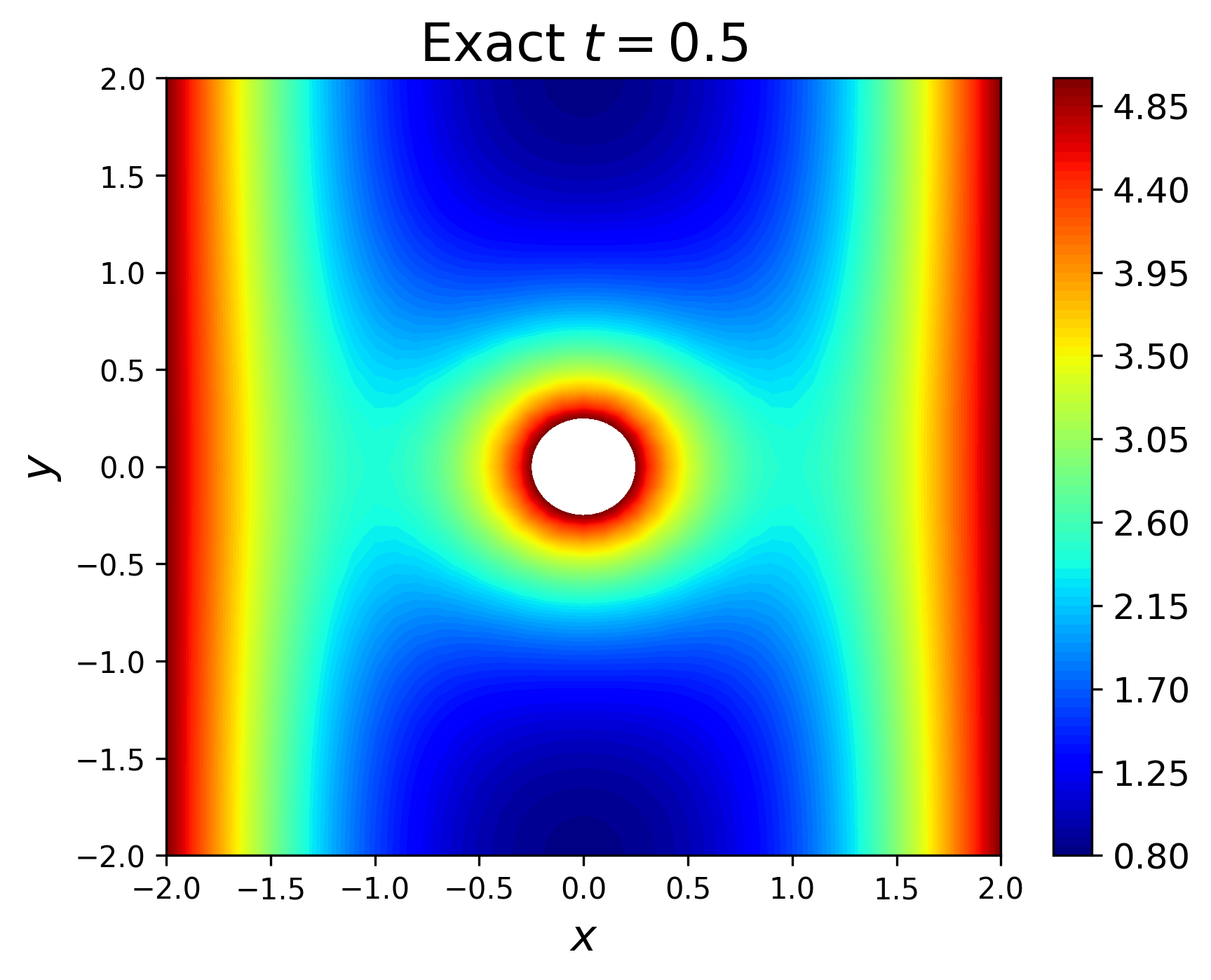}
	}%
	\subfigure[HFD-PINN]{
		\includegraphics[scale=0.35]{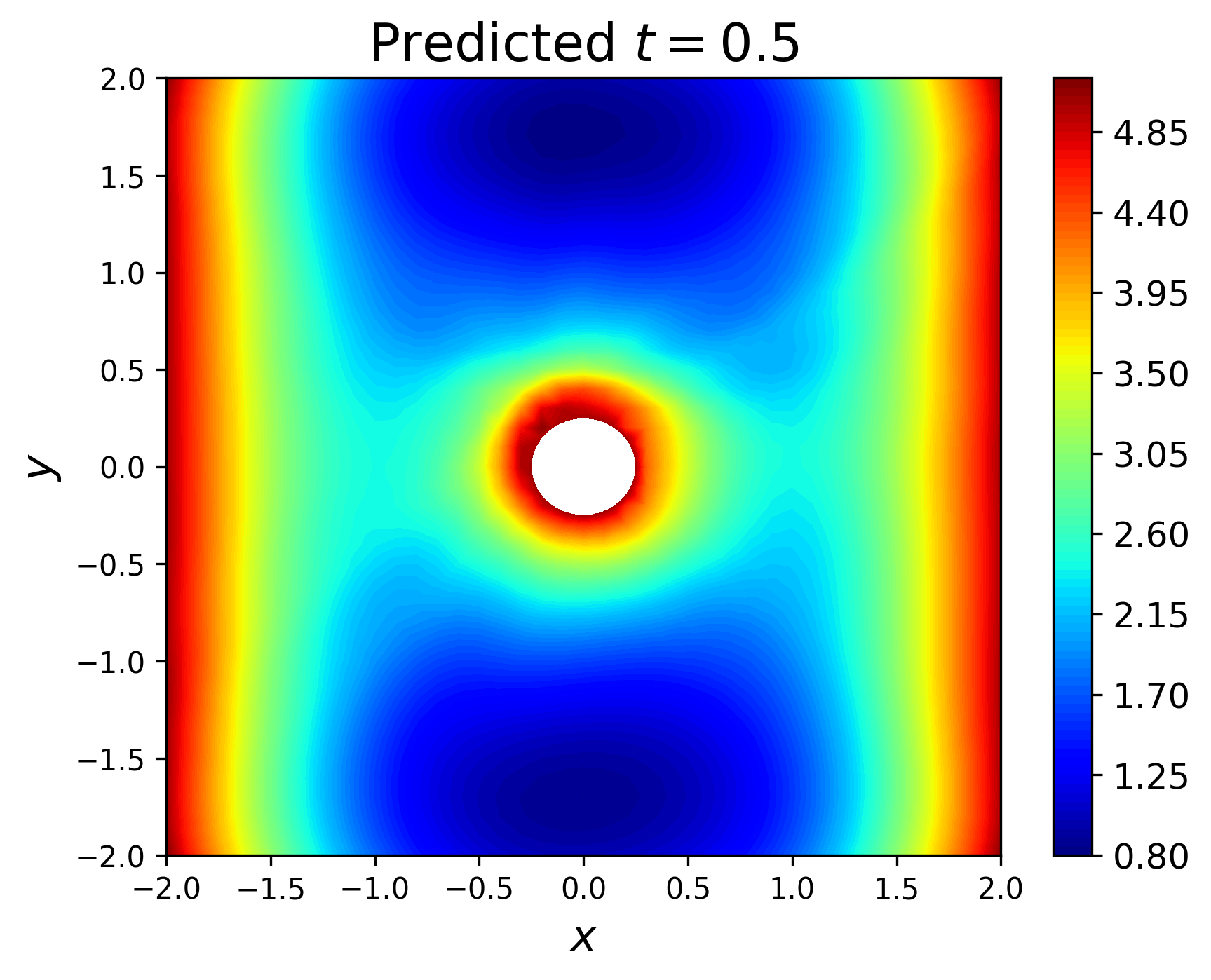}
	}%
	\centering
	\caption{Heat transfer problem of square plate with a hole in the middle. (a) The reference solution at $t=0.5$. (b) The prediction solution of HFD-PINN at $t=0.5$.}
	\label{FDPINNpred}
\end{figure}

\begin{figure}[htbp]
	\centering
	\subfigure[AD-PINN]{
		\includegraphics[scale=0.35]{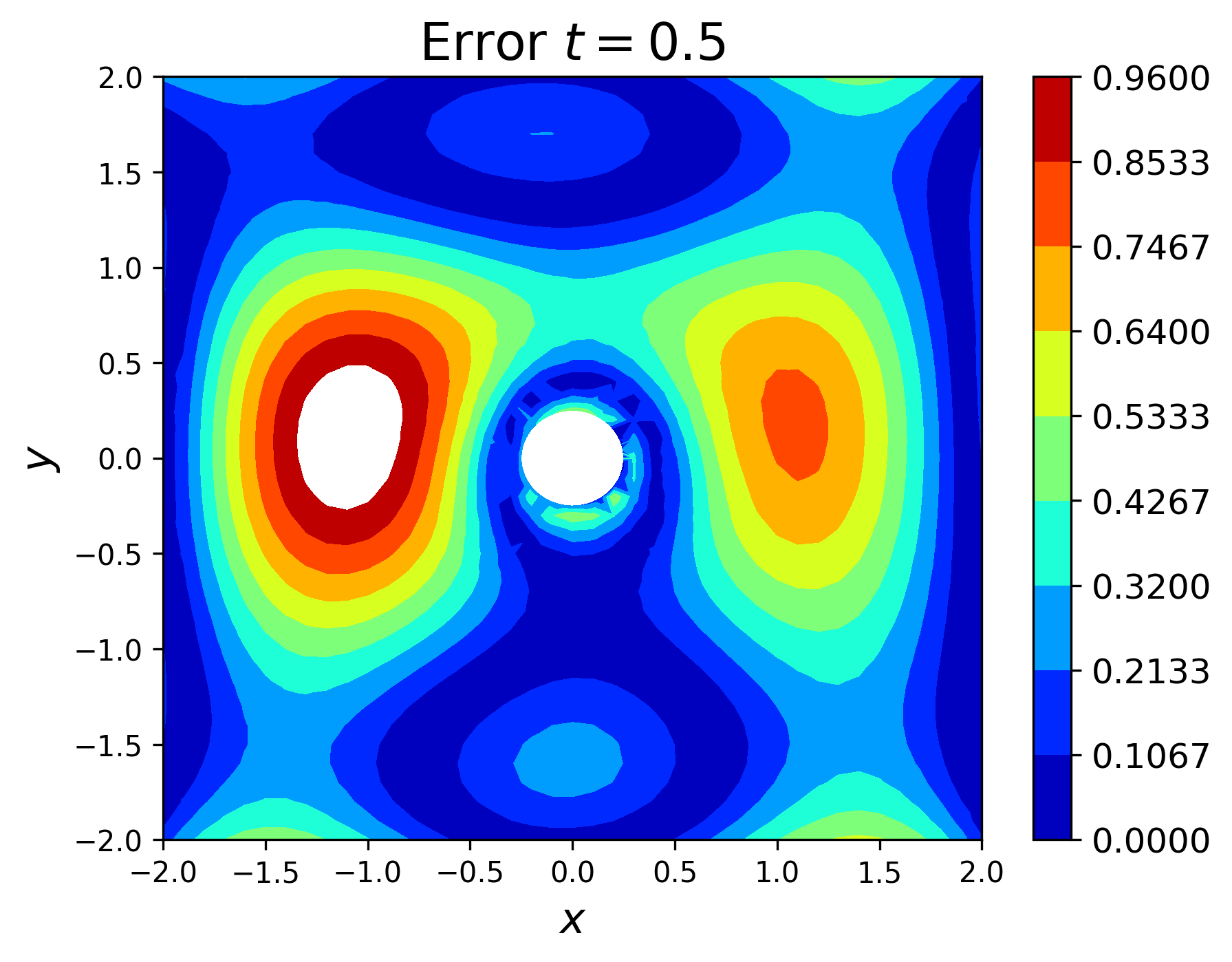}
	}%
	\subfigure[HFD-PINN]{
		\includegraphics[scale=0.35]{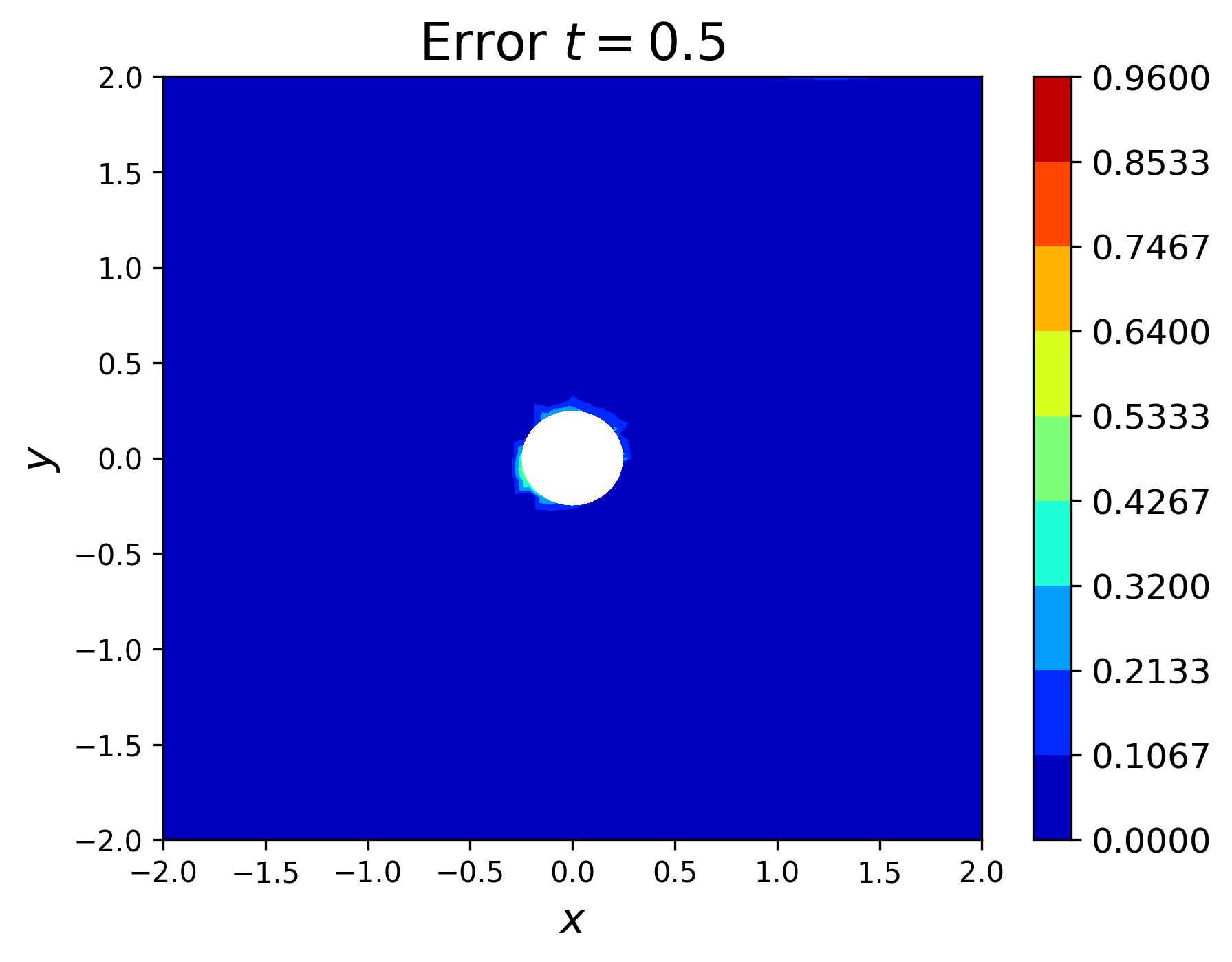}
	}%
	\centering
	\caption{Heat transfer problem of square plate with a hole in the middle. (a) The absolute error of AD-PINN. (b) The absolute error of HFD-PINN.}
	\label{FDPINNerror}
\end{figure}

\begin{figure}[htbp]
	\centering
	\subfigure[AD-PINN]{
		\includegraphics[scale=0.35]{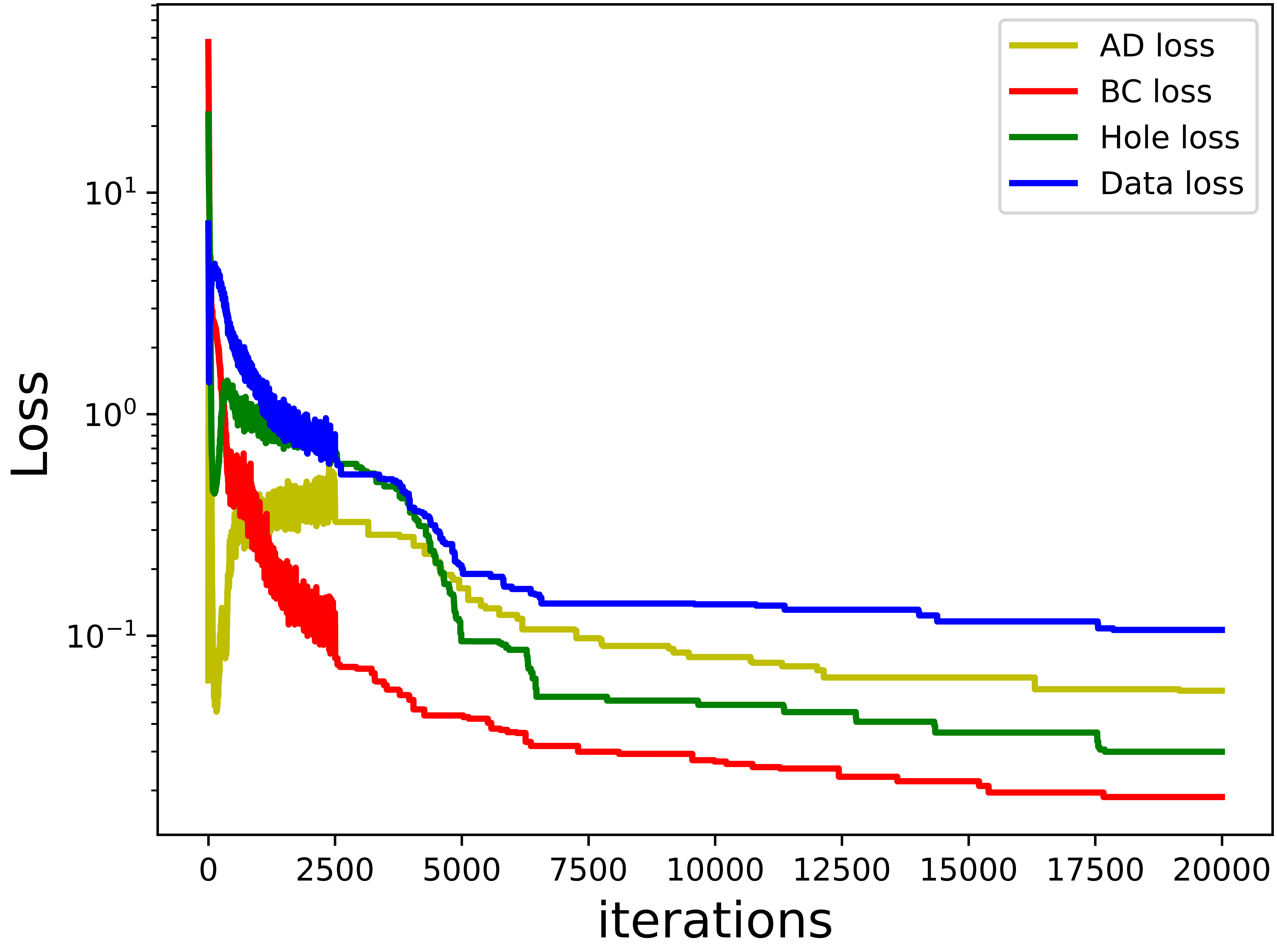}
	}%
	\subfigure[HFD-PINN]{
		\includegraphics[scale=0.35]{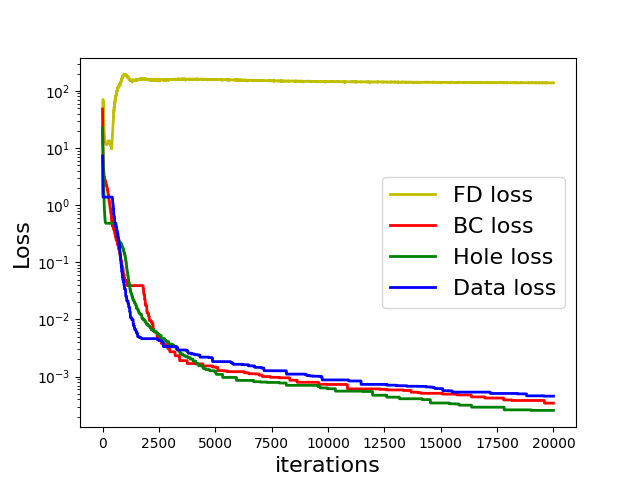}
	}%
	\subfigure[The test comparison]{
		\includegraphics[scale=0.35]{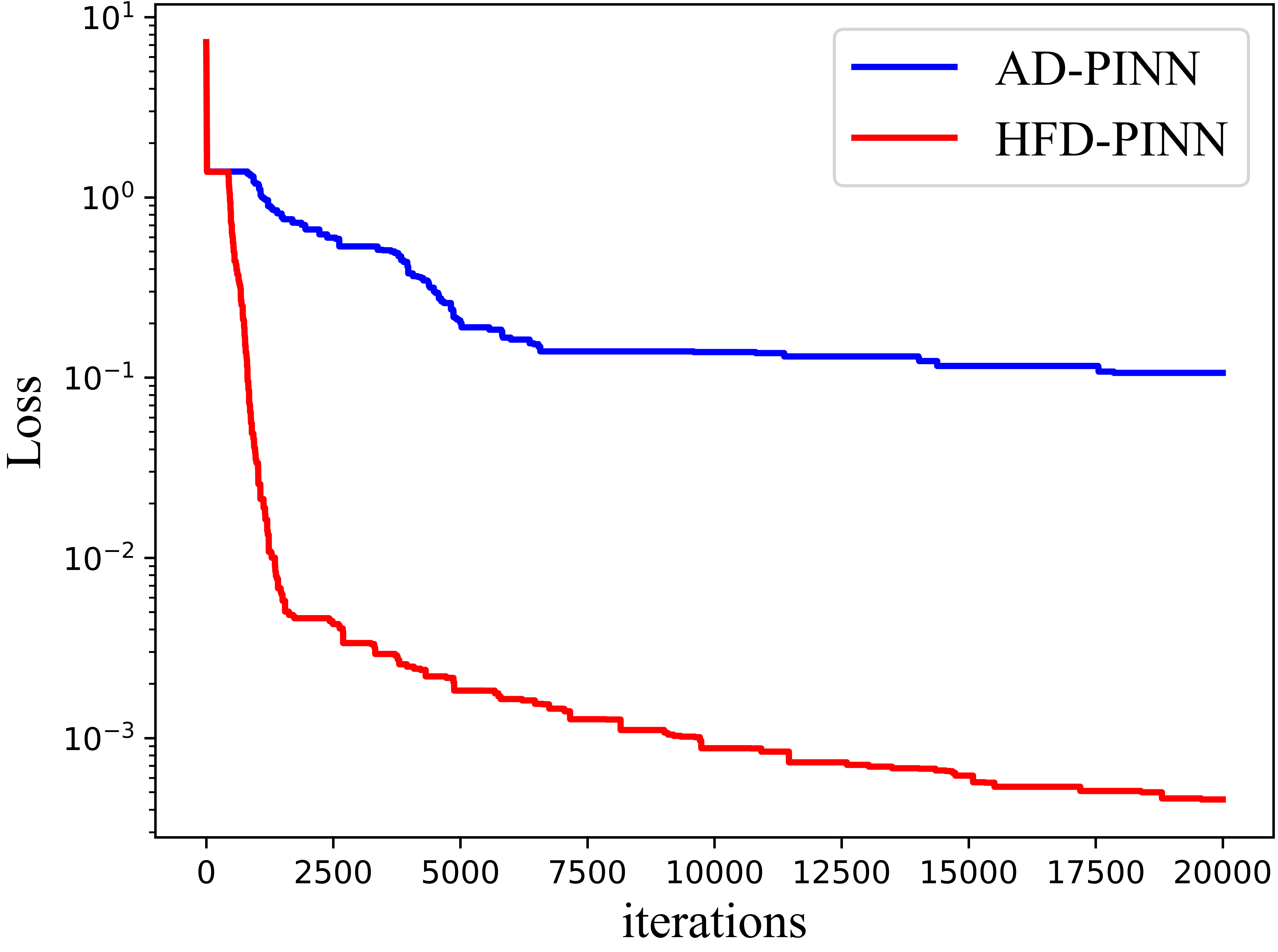}
	}%
	\centering
	\caption{Heat transfer problem of square plate with a hole in the middle. Evolution of the loss function along with the training of the AD-PINN (a) and HFD-PINN (b). Besides, (c) The test comparison of the AD-PINN and HFD-PINN.}
	\label{FDPINNloss}
\end{figure}

Here, we show the solution at $t=0.5$ obtained by FD simulation and HFD-PINN prediction in Fig. \ref{FDPINNpred}. Fig. \ref{FDPINNerror} compare the absolute error between AD-PINN and HFD-PINN. It can be seen that the HFD-PINN can achieve an accurate approximation. To better understand the differences between the two methods, the values of the loss functions during training are plotted in Fig. \ref{FDPINNloss}. As a result indicates, the HFD-PINN behave better during the training. The HFD-PINN can effectively fuse information obtained from data and domain for model inference.
\begin{table}[tp]  
	
	\centering  
	\fontsize{8}{8}\selectfont  
	\caption{Comparing the relative error with the different number of collocation points when learning heat transfer problem of square plate with a hole in the middle.}  
	\label{tab6}  
	\begin{tabular}{|c|c|c|c|c|c|c|c|c|}  
		\hline  
		\multirow{2}{*}{collocation points}&  
		\multicolumn{4}{c|}{AD-PINN}&\multicolumn{4}{c|}{ HFD-PINN}\cr\cline{2-9}  
		&AD loss&BC loss&Hole loss&Data loss&FD loss&BC loss&Hole loss&Data loss\cr  
		\hline  
		\hline  
		25&1.4e+00&4.7e-01&{\bf 7.0e-01}&{\bf 8.3e-01}&2.3e+01&2.5e-02&{\bf 1.7e-02}&{\bf 2.3e-02}\cr\hline  
		64&4.1e-01&9.5e-02&{\bf 6.1e-01}&{\bf 1.4e-01}&3.7e+01&7.5e-03&{\bf 5.1e-03}&{\bf 4.6e-03}\cr\hline  
		100&6.2e-02&3.3e-02&{\bf 1.9e-01}&{\bf 6.0e-02}&1.9e+02&9.0e-04&{\bf 8.8e-04}&{\bf 8.1e-04}\cr\hline   
	\end{tabular}  
\end{table}


\begin{figure}[htbp]
	\centering
	\subfigure[Boundary conditions]{
		\includegraphics[scale=0.22]{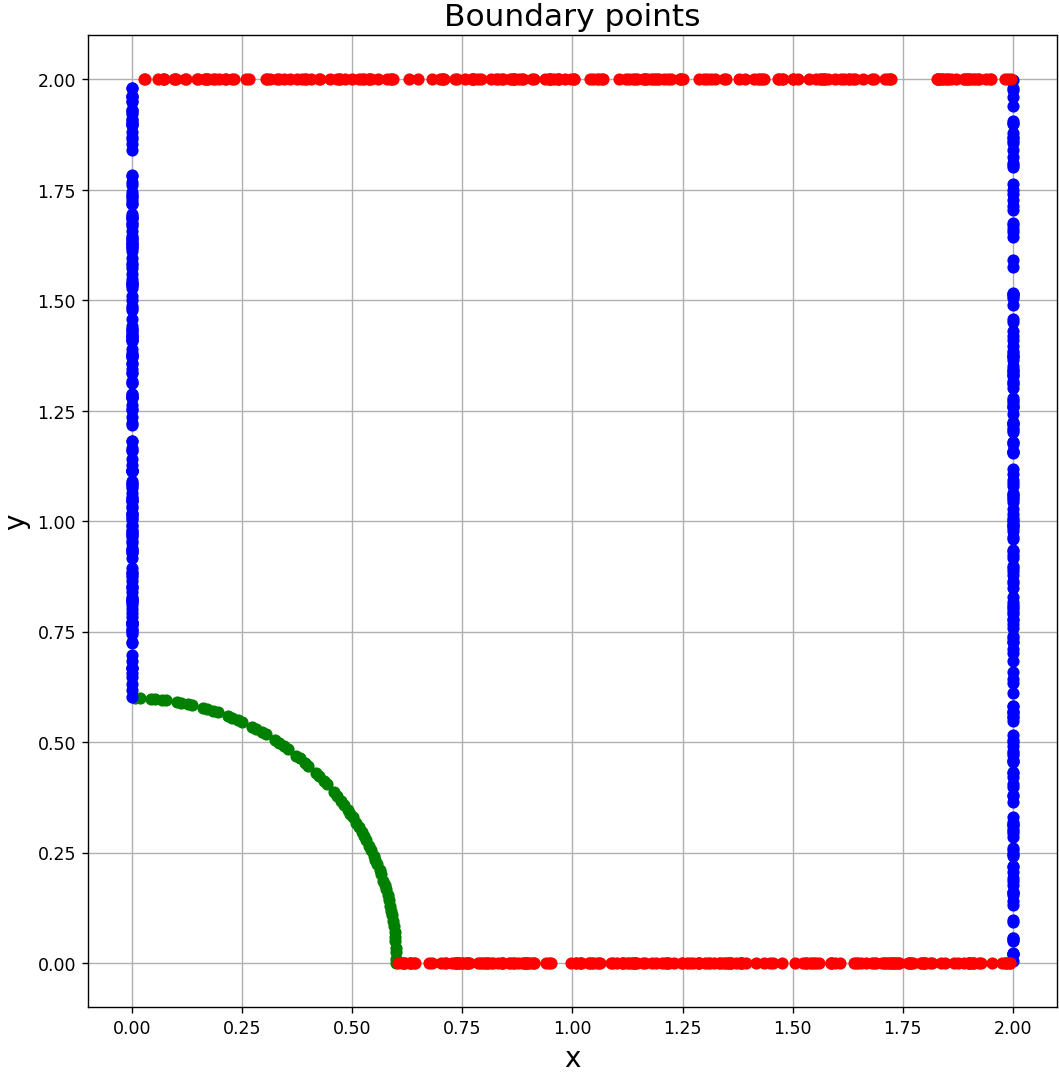}
	}%
	\subfigure[Training points]{
		\includegraphics[scale=0.225]{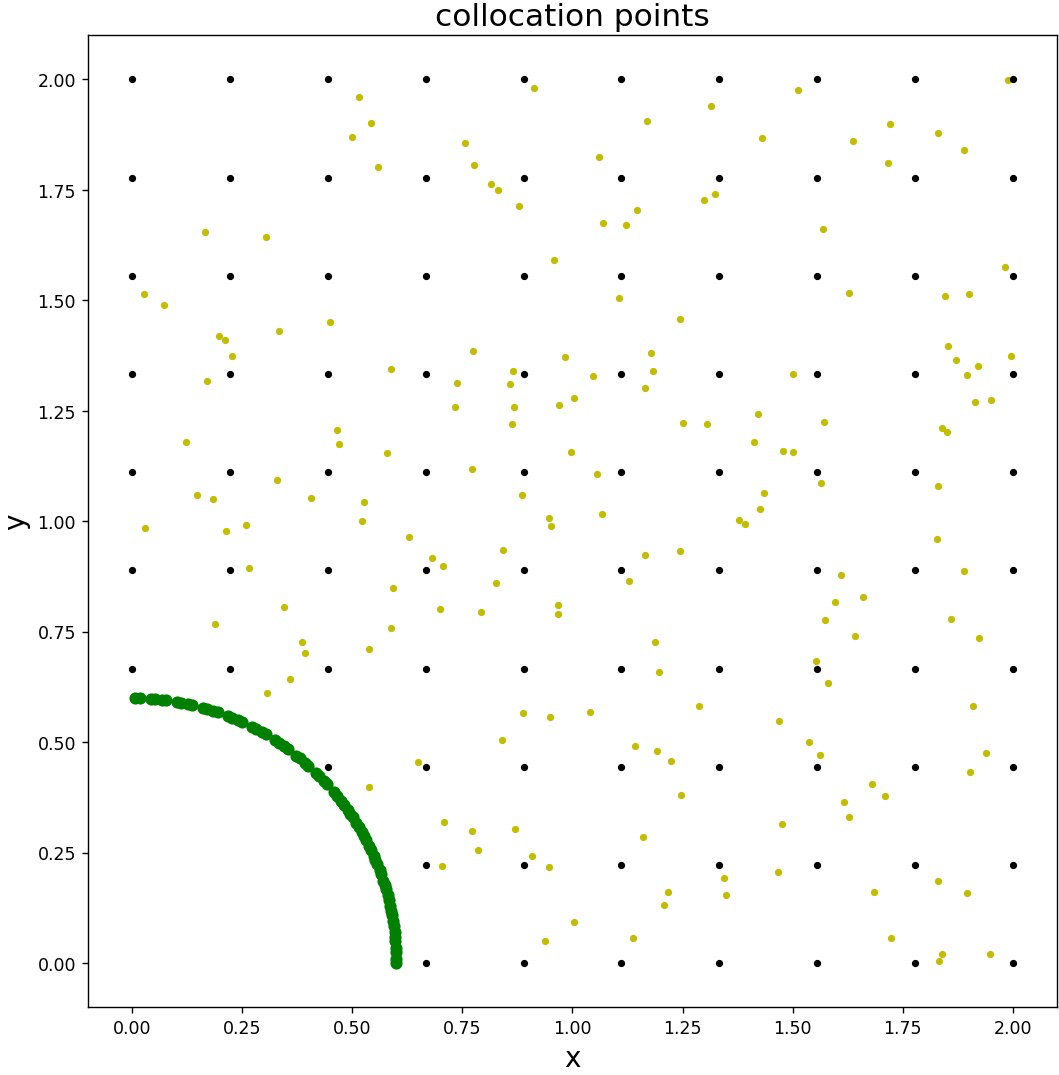}
	}%
	\centering
	\caption{Heat transfer problem of square plate with a hole in the corner. (a) Blue and red markers are used for Dirichlet and Neumann boundary conditions enforcement. The arc boundary has 160 equidistant Dirichlet collocation points (green dot points). (b) Black markers are FDM mesh points. Collocation points for imposing PDE are shown by yellow dot points.}
	\label{2dpoisson4k}
\end{figure}
\subsection{Heat transfer problem of square plate with a hole in the corner}
Let us consider the Heat transfer problem with a 1/4 circular hole in the lower-left corner. Express the boundary conditions of the problem in a mixed form, including Riemann and Dirichlet (see Fig. \ref{2dpoisson4k}(a)). The boundary value problem corresponding to the heat equation can also be defined by Eq. \eqref{2dPoisson Equation}. We consider problems on the square domain $\Omega$ with $[x_l, x_r]=[0,2]$ and the hole radius $r=0.6$ and $T_{0}=5$. 

\begin{figure}[htbp]
	\centering
	\subfigure[AD-PINN]{
		\includegraphics[scale=0.35]{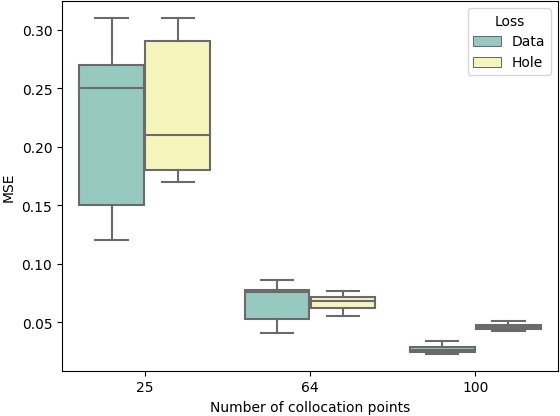}
	}%
	\subfigure[HFD-PINN]{
		\includegraphics[scale=0.35]{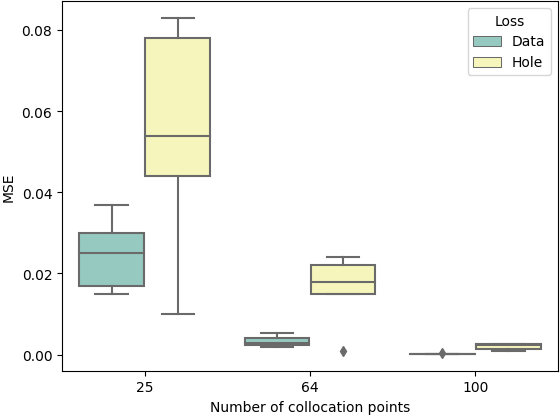}
	}%
	\caption{Heat transfer problem of square plate with a hole in the corner: Boxplot of the Hole and Data loss of AD-PINN (a) and HFD-PINN (b) with different numbers of collocation points.}
	\label{2d4kFDPINNpa}
\end{figure}

We use the collocation points shown in Fig. \ref{2dpoisson4k}(b) to define the $\mathcal{L}_{\mathrm{AD}}$ and $\mathcal{L}_{\mathrm{FD}}$ in Eq. \eqref{ADFDEquation}. We fix the network architecture with hidden layers $ = 4$ and layer sizes $ = [3, 50, 50, 50, 50, 1]$ and the activation function is tanh. We train the AD-PINN and HFD-PINN for 2k epochs with Adam gradient descent (learning rate 0.001). The numerical simulation results contain a total of 20 time steps of the heat field. The data of the $t=1$ is used as observations in the training data. Moreover, the data of $t=0.5$ are used as the test data. We attempt to predict the development of the heat field through domain knowledge and a few observations. The total number of training points is maintained as $N_b = 300, N_{data} = 100$. In addition, we explore the influence of different numbers of collocation points $N_{f}$ demonstrated in Table \ref{tab8}. A boxplot is drawn to reflect the Hole and data error distribution of the AD-PINN and HFD-PINN in Fig. \ref{2d4kFDPINNpa}.

\begin{figure}[htbp]
	\centering
	\subfigure[Exact]{
		\includegraphics[scale=0.085]{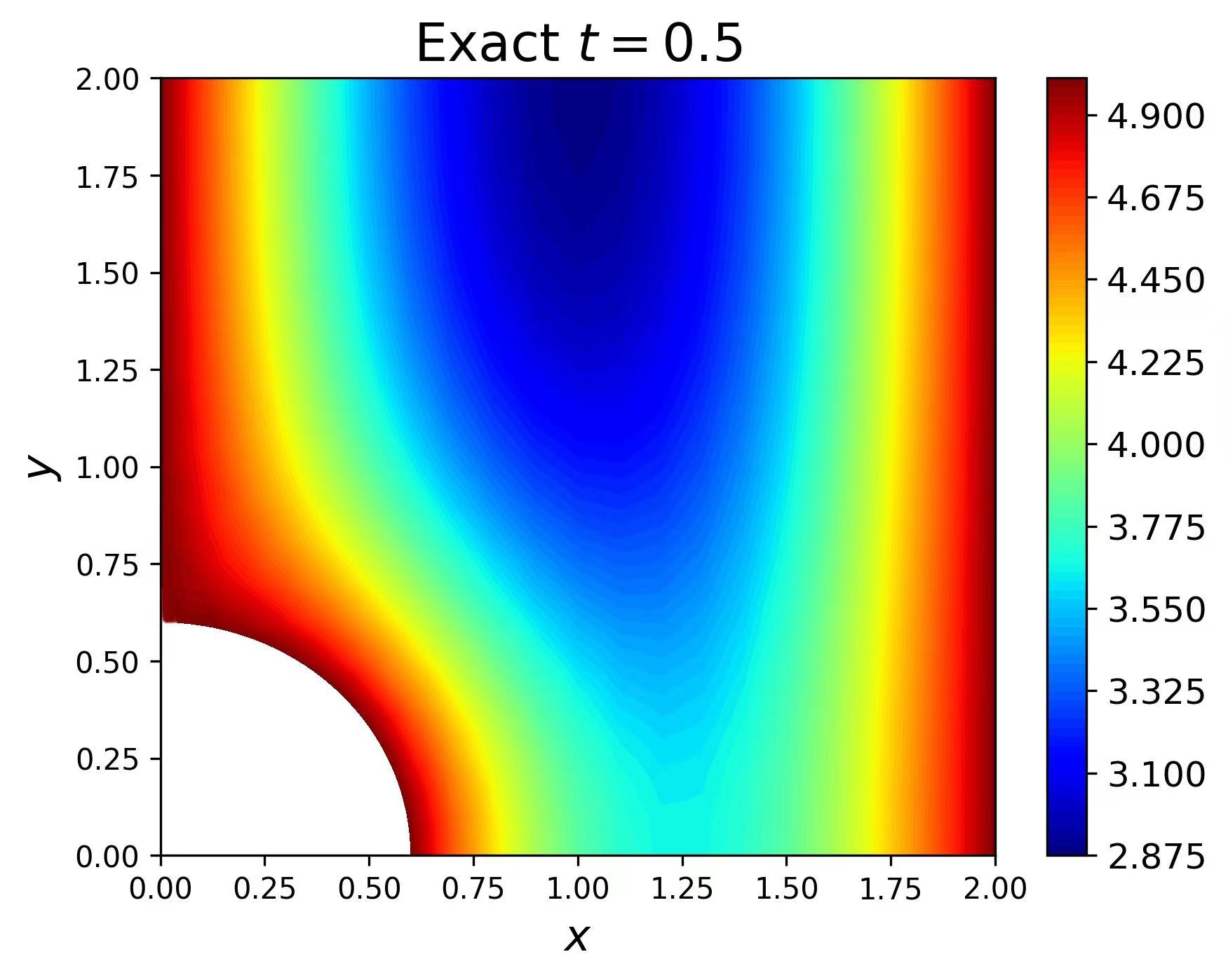}
	}%
	\subfigure[HFD-PINN]{
		\includegraphics[scale=0.35]{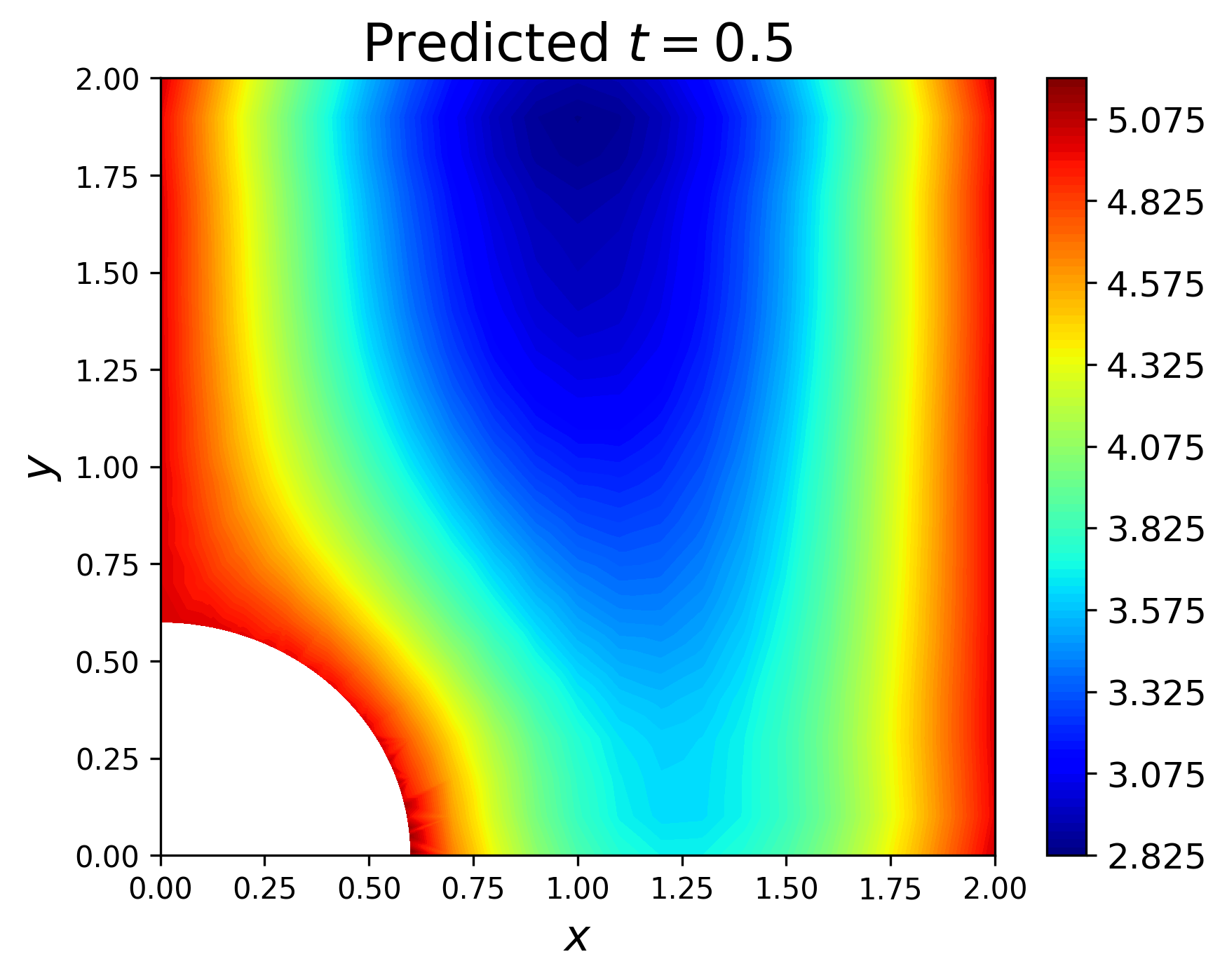}
	}%
	\centering
	\caption{Heat transfer problem of square plate with a hole in the corner. (a) The reference solution at $t=0.5$. (b) The prediction solution of HFD-PINN at $t=0.5$.}
	\label{FDPINNpred4k}
\end{figure}

\begin{figure}[htbp]
	\centering
	\subfigure[AD-PINN]{
		\includegraphics[scale=0.35]{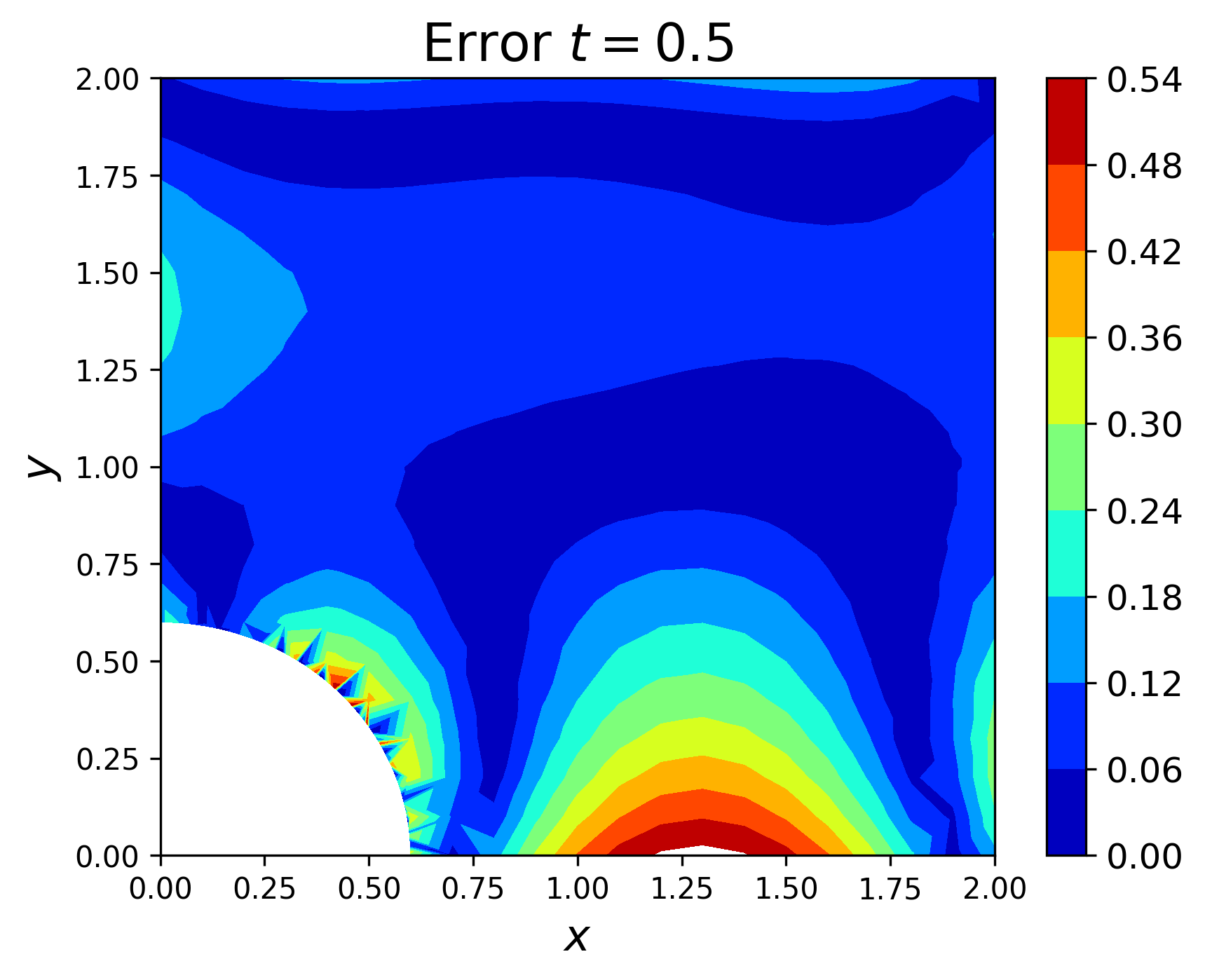}
	}%
	\subfigure[HFD-PINN]{
		\includegraphics[scale=0.35]{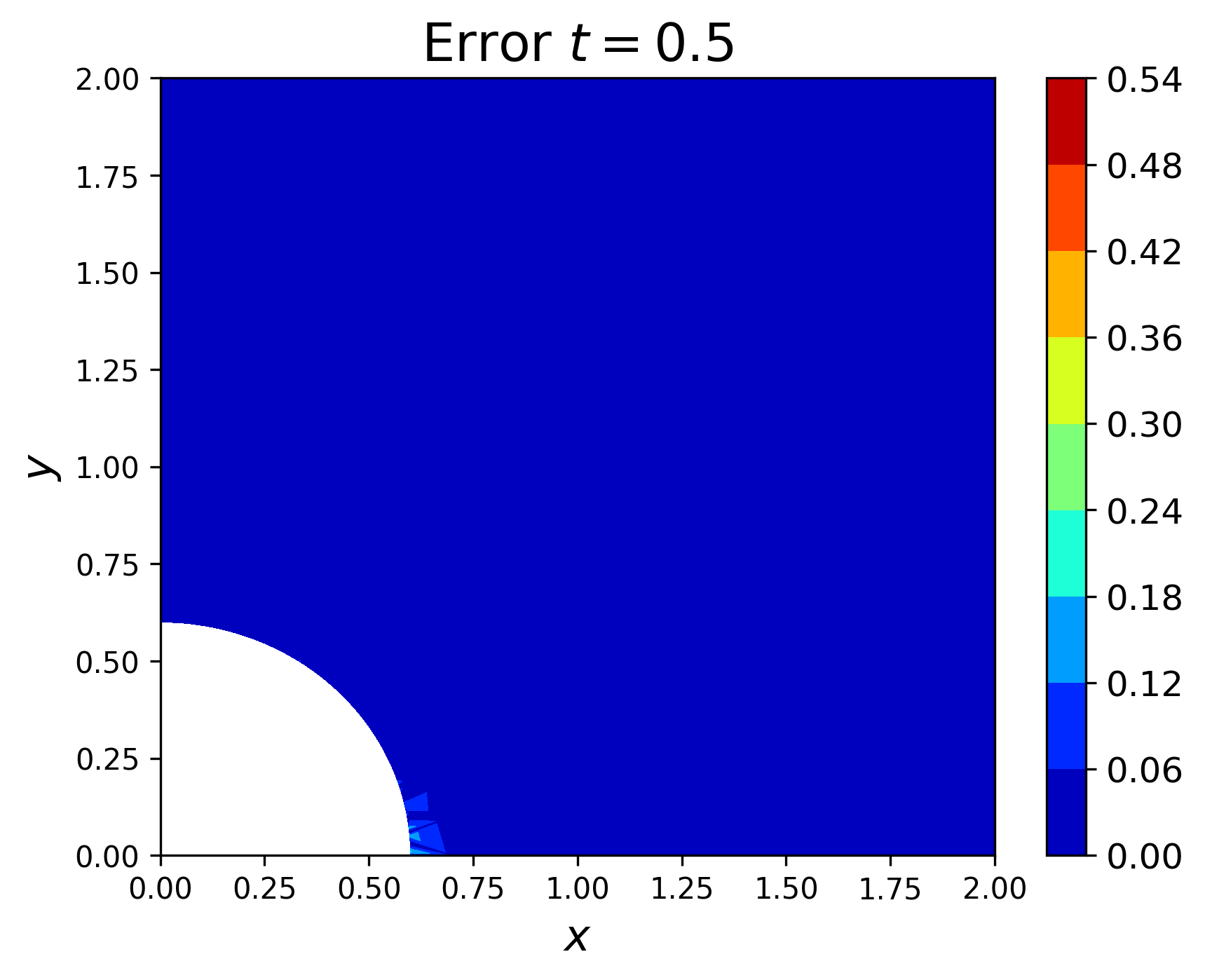}
	}%
	\centering
	\caption{Heat transfer problem of square plate with a hole in the corner. (a) The absolute error of AD-PINN. (b) The absolute error of HFD-PINN.}
	\label{FDPINNerror4k}
\end{figure}

\begin{figure}[htbp]
	\centering
	\subfigure[AD-PINN]{
		\includegraphics[scale=0.35]{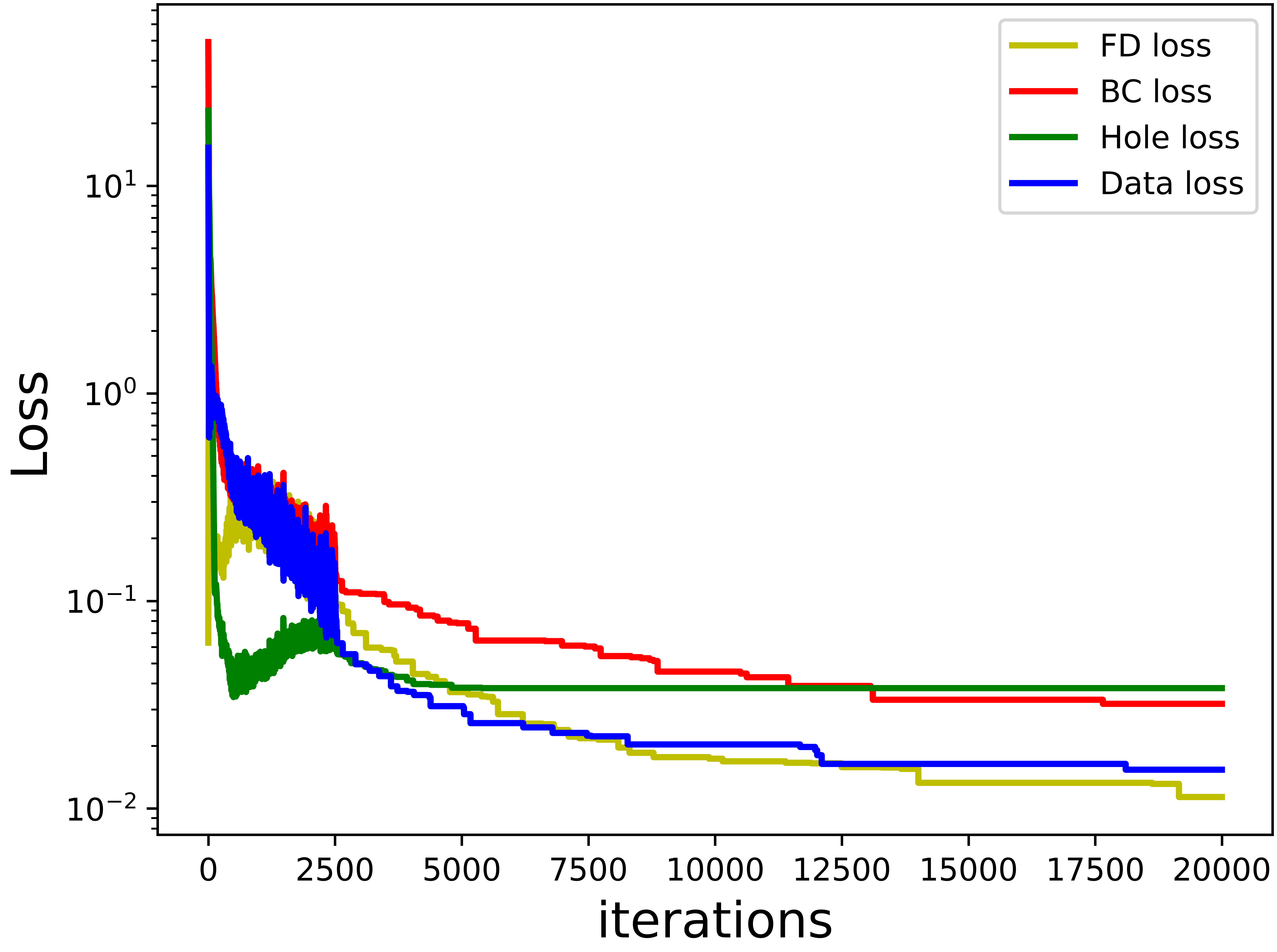}
	}%
	\subfigure[HFD-PINN]{
		\includegraphics[scale=0.35]{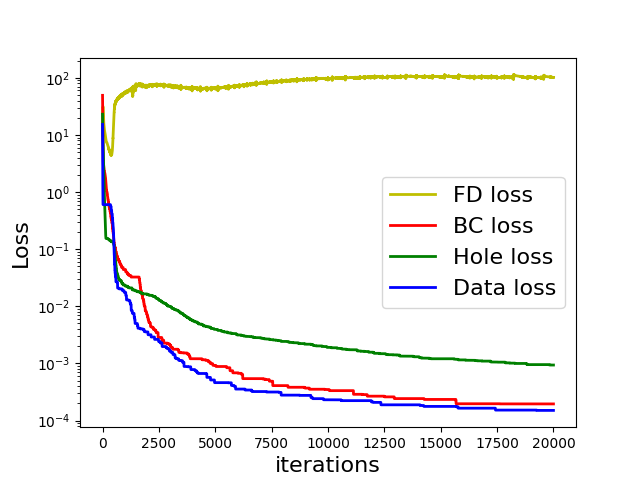}
	}%
	\subfigure[The test comparison]{
		\includegraphics[scale=0.35]{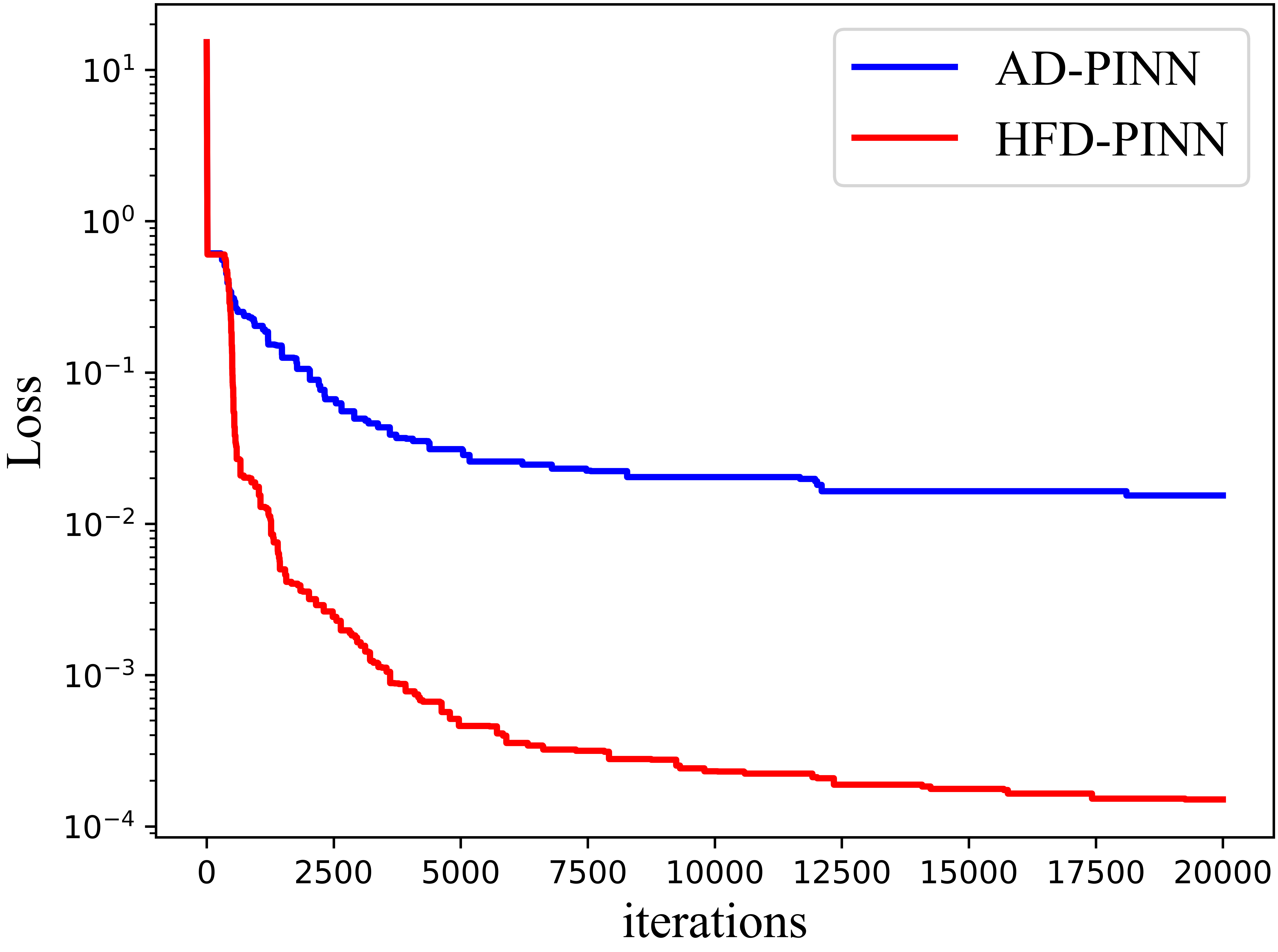}
	}%
	\centering
	\caption{Heat transfer problem of square plate with a hole in the corner. Evolution of the loss function along with the training of the AD-PINN (a) and HFD-PINN (b). Besides, (c) The test comparison of the AD-PINN and HFD-PINN.}
	\label{loss4k}
\end{figure}

In Fig. \ref{FDPINNpred4k}, the solution at $t=0.5$ obtained by FD simulation and the prediction of the HFD-PINN is presented. For a better comparison, the difference between the exact solution and AD-PINN or HFD-PINN are reported in Fig. \ref{FDPINNerror4k}. Training performance of the AD-PINN and FD-PINN is shown in Fig. \ref{loss4k}. The prediction error of both AD-PINN and HFD-PINN is significantly degraded as iterations increases. As the Fig. \ref{loss4k}(c) indicates, the HFD-PINN shows a much more accurate prediction. The predictive error computed with L2 error over 20k epochs could attain 8.8e-04 $\pm$ 8.1e-04. It indicates that the HFD-PINN is capable of producing a better approximation than the AD-PINN.
%

\begin{table}[tp]  
	
	\centering  
	\fontsize{8}{8}\selectfont  
	\caption{Comparing the relative error with the different number of collocation points when learning heat transfer problem of square plate with a hole in the corner.}  
	\label{tab8}  
	\begin{tabular}{|c|c|c|c|c|c|c|c|c|}  
		\hline  
		\multirow{2}{*}{collocation points}&  
		\multicolumn{4}{c|}{AD-PINN}&\multicolumn{4}{c|}{ HFD-PINN}\cr\cline{2-9}  
		&AD loss&BC loss&Hole loss&Data loss&FD loss&BC loss&Hole loss&Data loss\cr  
		\hline  
		\hline  
		25&1.4e+00&4.7e-01&{\bf 7.0e-01}&{\bf 8.3e-01}&2.3e+01&2.5e-02&{\bf 1.7e-02}&{\bf 2.3e-02}\cr\hline  
		64&4.1e-01&9.5e-02&{\bf 6.1e-01}&{\bf 1.4e-01}&3.7e+01&7.5e-03&{\bf 5.1e-03}&{\bf 4.6e-03}\cr\hline  
		100&6.2e-02&3.3e-02&{\bf 1.9e-01}&{\bf 6.0e-02}&1.9e+02&9.0e-04&{\bf 8.8e-04}&{\bf 8.1e-04}\cr\hline   
	\end{tabular}  
\end{table}


\subsection{Two-dimensional Poisson Equation}
This section covers the Poisson equation, which describes the heat equation with a source
function. As mentioned, this PDE is well understood and therefore serves as the starting point of the experimental analysis. We consider the following Two-dimensional Poisson Equation:

\begin{equation}
\begin{array}{c}
u_{xx} +u_{yy} = f, \quad x \in [0, 1], y \in[0, 1], \\
f = \sum_{0\leq i\leq 2,0\leq j\leq 2}\zeta_{ij}\cdot\sin(ix\pi)\cdot\sin(jy\pi)\cdot(\pi^{2})\cdot(i^{2} + j^{2}),\\
u(0, y) = u(1, y) = u(x, 0) = u(x, 1). \\
\end{array}
\label{2dPoisson Equation1}
\end{equation}

We take $ \zeta_{ij}=[\zeta_{0j}, \zeta_{1j}, \zeta_{2j}],0\leq j\leq 2,\zeta_{0j}= [-0.16, 0.81, -0.62], \zeta_{1j}=[0.16, -0.36, 0.71], \zeta_{2j}=[-0.98, -0.01, -0.4]$ to obtain the analytical solution. In particular, the true solution is $u=\sum_{0\leq i\leq 2,0\leq j\leq 2}\zeta_{ij}\cdot\sin(ix\pi)\cdot\sin(jy\pi)$. For AD-PINN, we select random points $N_f$ to compute the AD loss via the automatic differentiation. For HFD-PINN, to solve the problem, the domain is discretized with a $N_f=m\times m$ mesh. And these grid points are then used to define the FD loss via the finite differentiation. For both AD-PINN and HFD-PINN, we use additional points $N_b = 700, N_{data} = 100$ to define the BC and Data loss. we use a 4-layer network with the layer sizes $ = [2, 50, 50, 50, 25, 1]$ activated by Tanh. We train AD-PINN and HFD-PINN for 10k using Adam with a learning rate of 0.001.

\begin{figure}[htbp]
	\centering
	\subfigure{
		\includegraphics[scale=0.45]{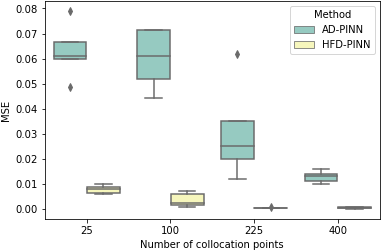}
	}
	\caption{Two-dimensional Poisson Equation: Boxplot of the relative L2 loss of AD-PINN and HFD-PINN with different numbers of collocation points.}
	\label{2dFDPINNpa}
\end{figure}

\begin{figure}[htbp]
	\centering
	\subfigure[Exact]{
		\includegraphics[scale=0.35]{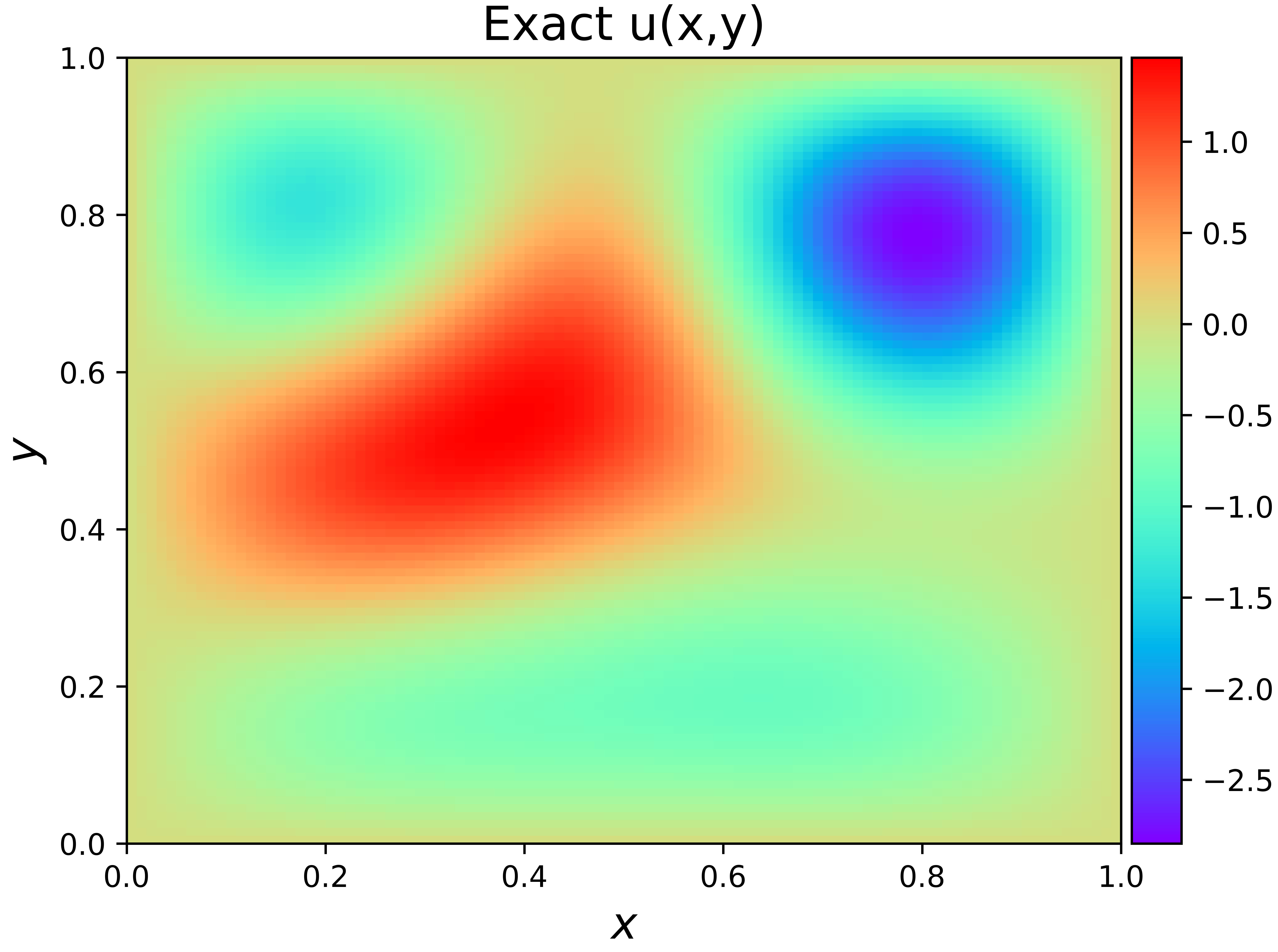}
	}%
	\subfigure[HFD-PINN]{
		\includegraphics[scale=0.35]{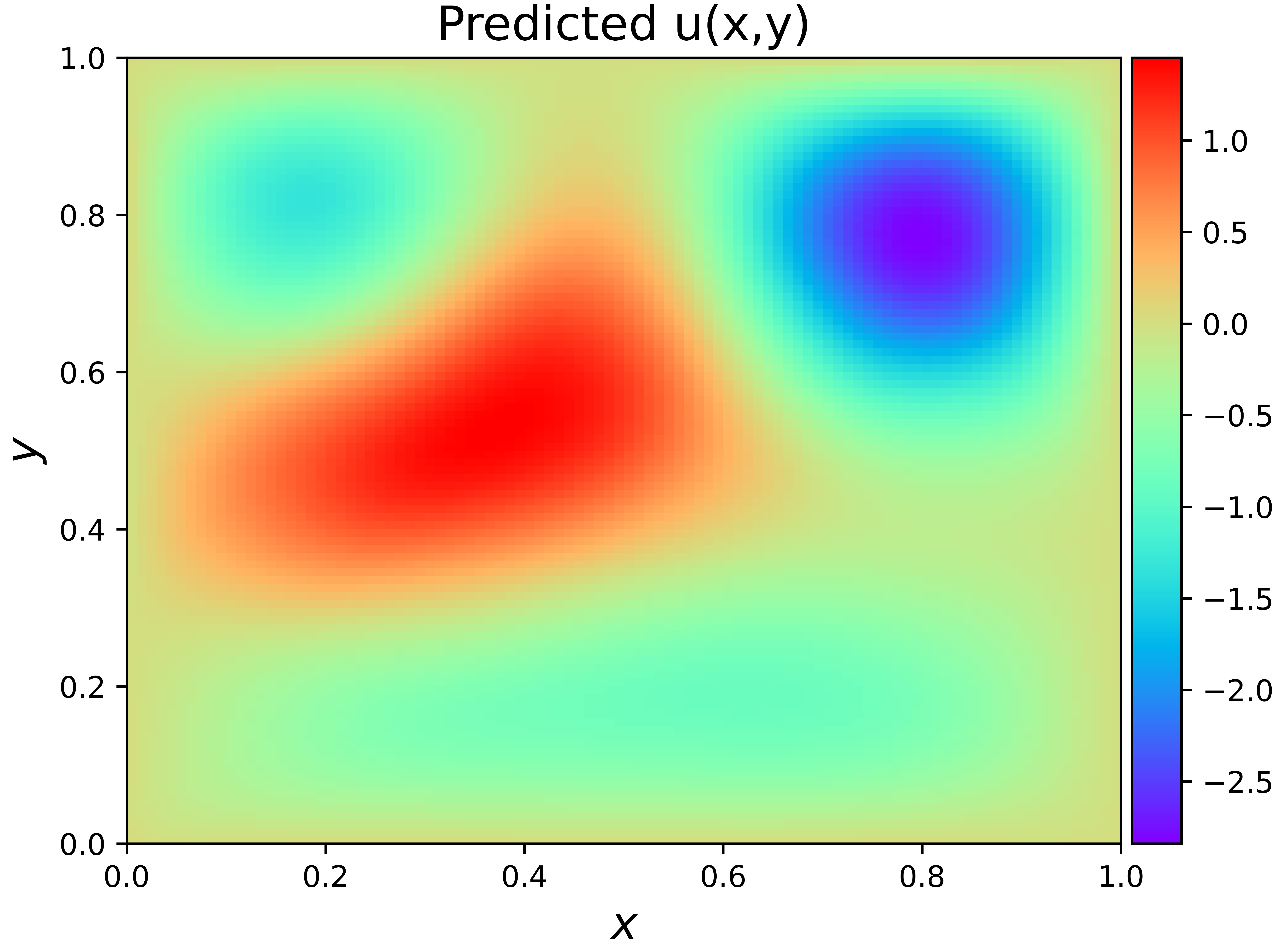}
	}%
	\centering
	\caption{Two-dimensional Poisson Equation: (a) The reference solution. (b) The prediction solution of HFD-PINN.}
	\label{2dFDPINNpred}
\end{figure}

\begin{figure}[htbp]
	\centering
	\subfigure[AD-PINN]{
		\includegraphics[scale=0.35]{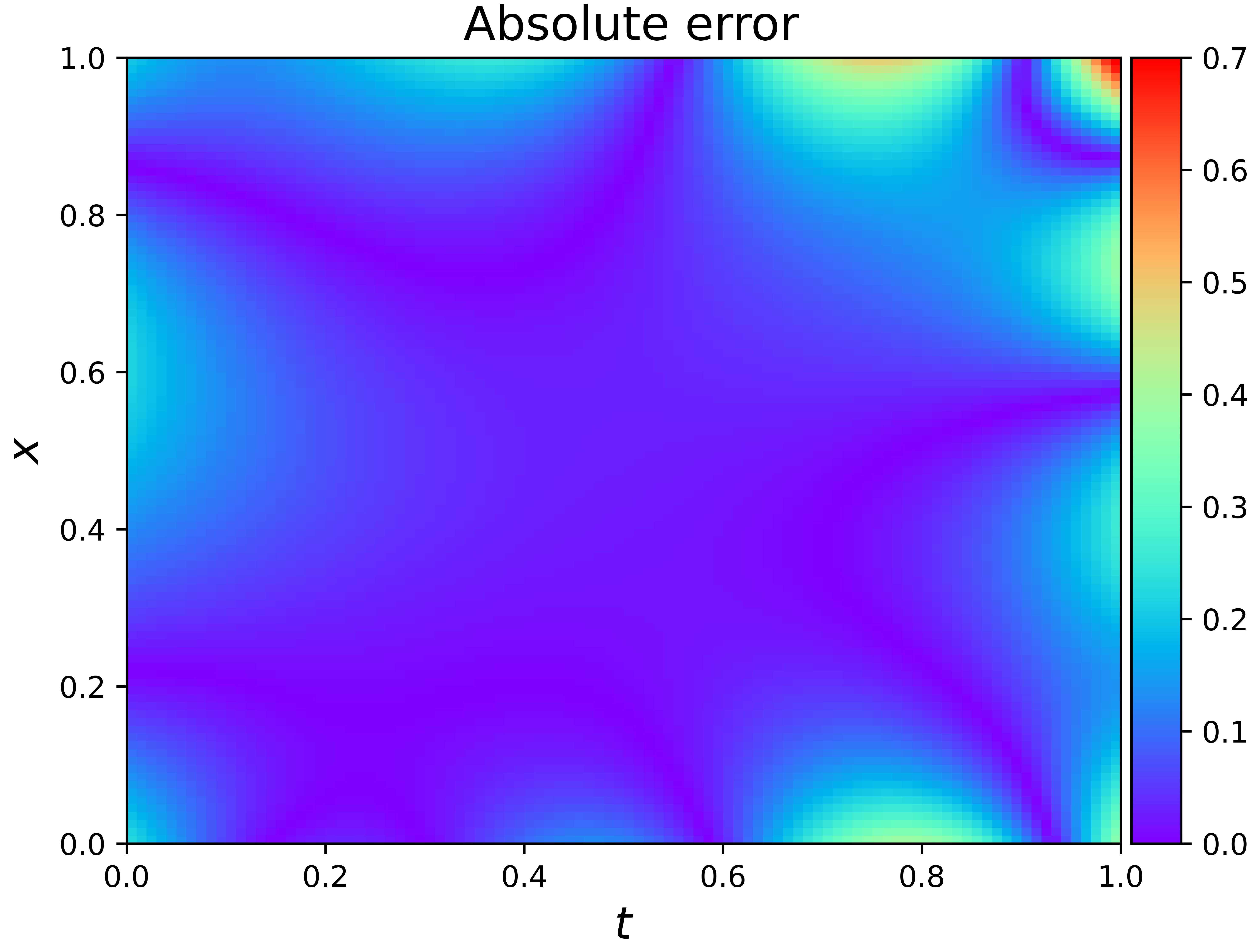}
	}%
	\subfigure[HFD-PINN]{
		\includegraphics[scale=0.35]{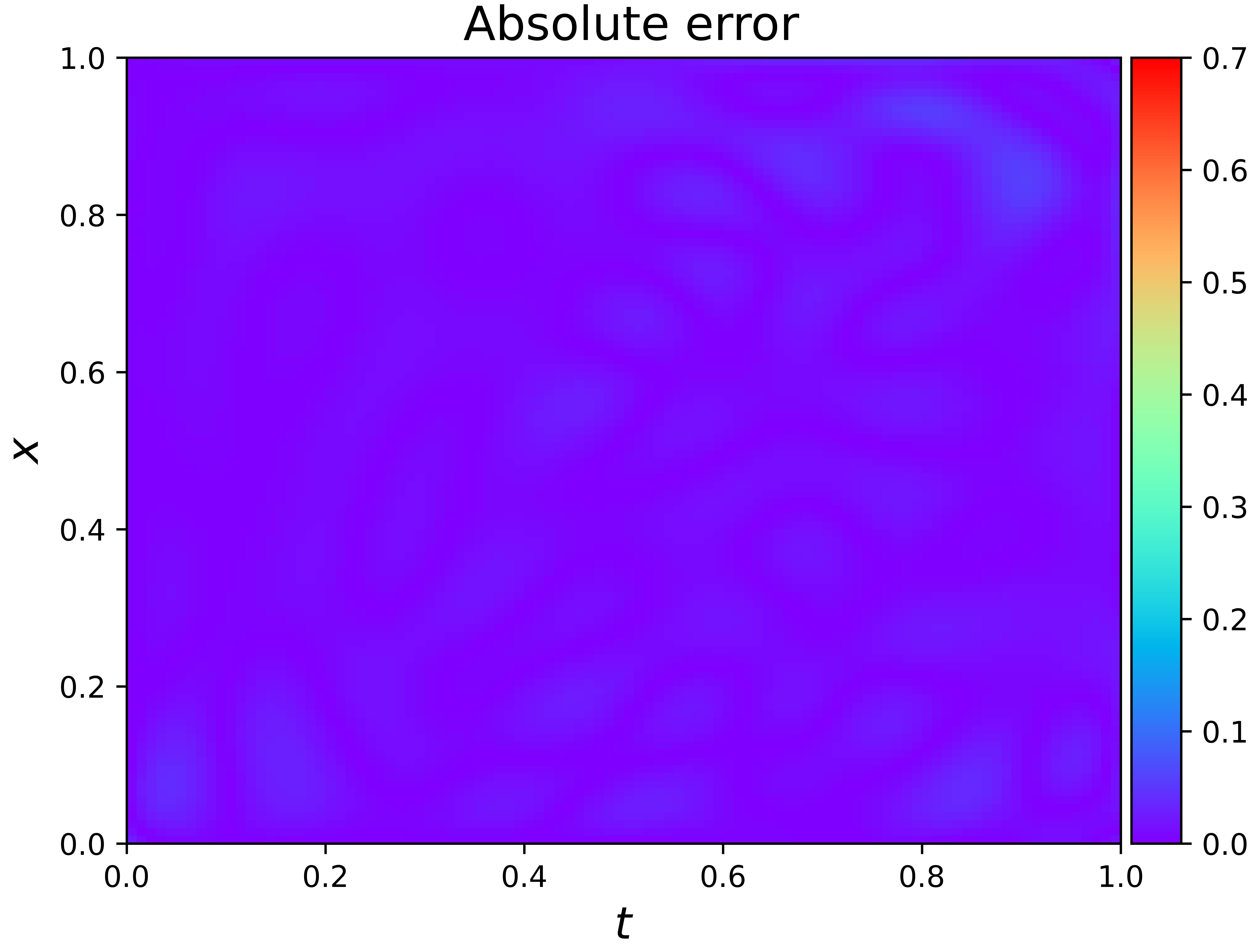}
	}%
	\centering
	\caption{Two-dimensional Poisson Equation. (a) The absolute error of AD-PINN. (b) The absolute error of HFD-PINN}
	\label{2dADPINNerror}
\end{figure}

\begin{figure}[htbp]
	\centering
	\subfigure[AD-PINN]{
		\includegraphics[scale=0.35]{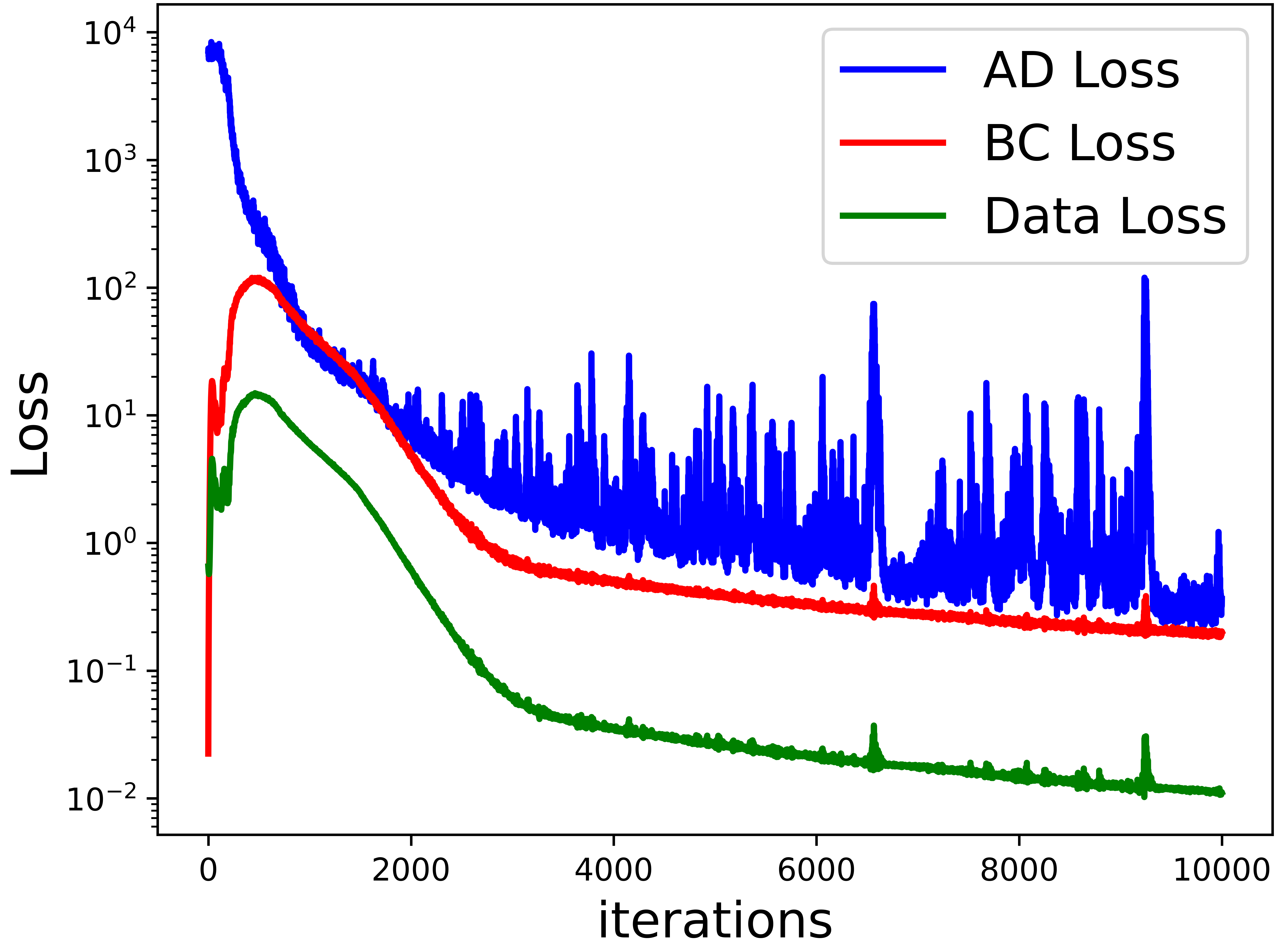}
	}%
	\subfigure[HFD-PINN]{
		\includegraphics[scale=0.35]{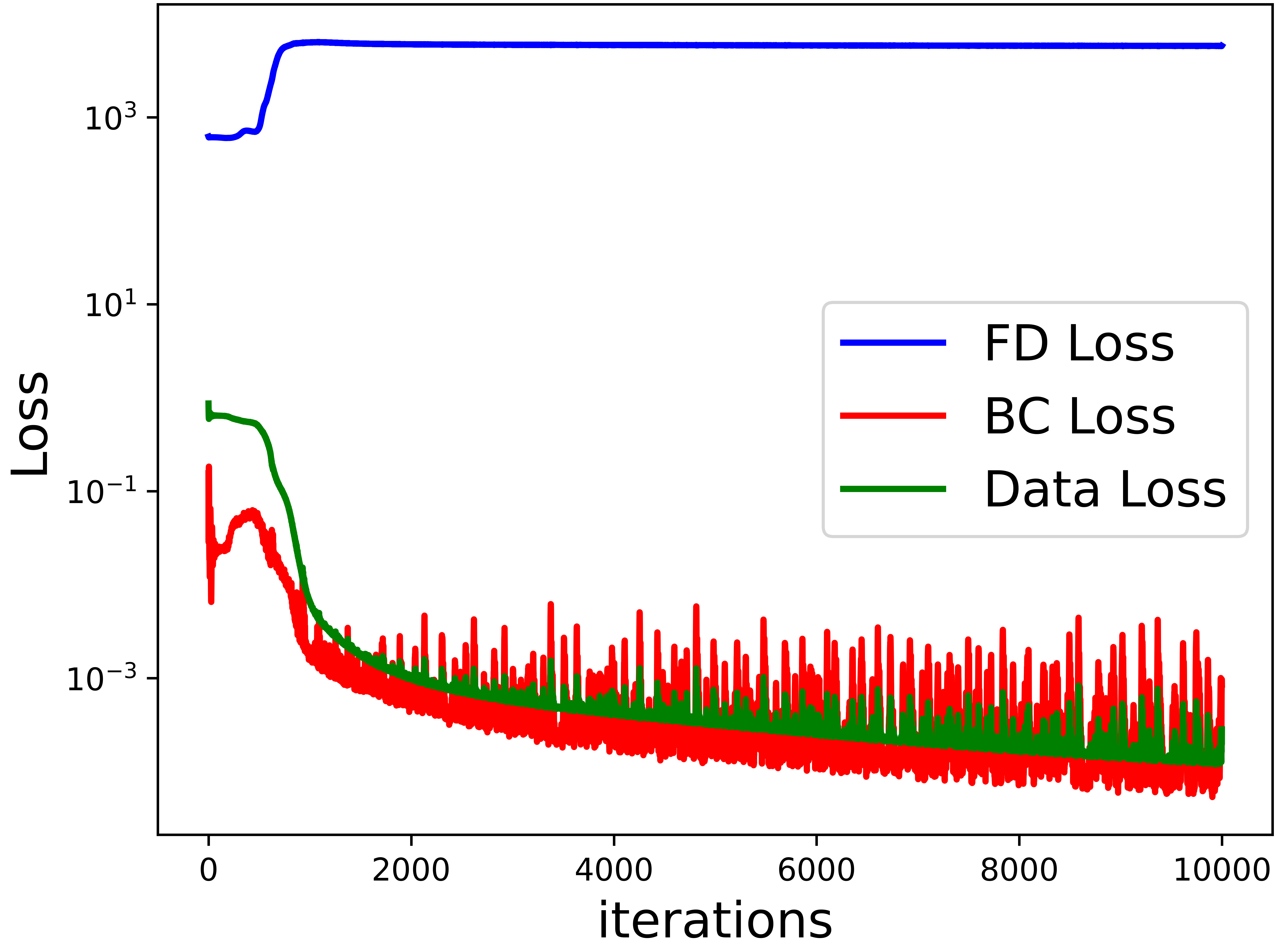}
	}%
	\subfigure[The test comparison]{
		\includegraphics[scale=0.35]{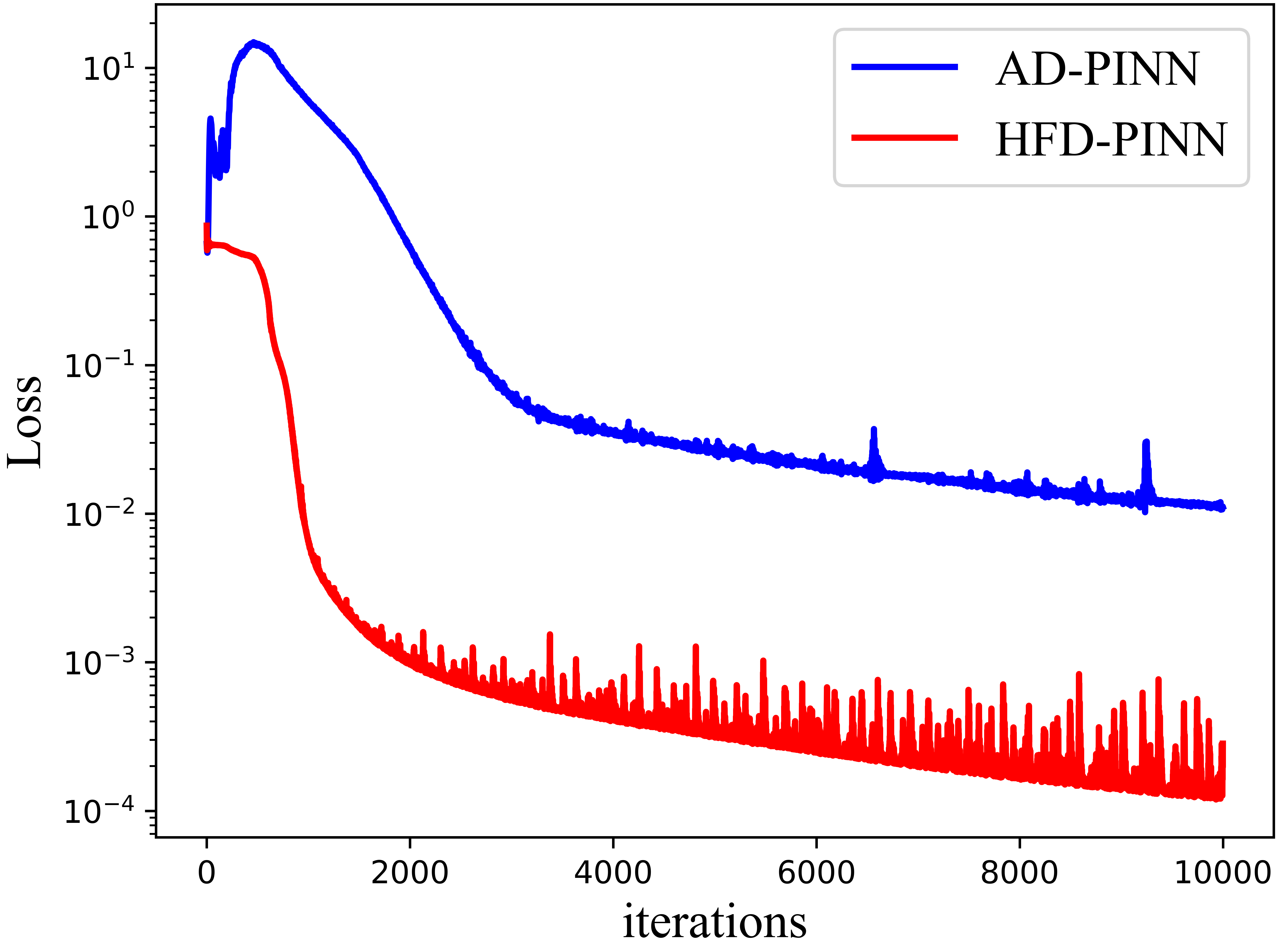}
	}%
	\centering
	\caption{Two-dimensional Poisson Equation: Evolution of the loss function along with the training of the AD-PINN (a) and HFD-PINN (b). Besides, (c) The test comparison of the AD-PINN and HFD-PINN.}
	\label{2dlossFD}
\end{figure}

All results of AD-PINN and HFD-PINN with the different number of collocation points are shown in Table  \ref{tab5}. The mean and standard deviation of the error for the AD-PINN and HFD-PINN are displayed in Fig. \ref{2dFDPINNpa}. We first present the exact solution and prediction using HFD-PINN in Fig. \ref{2dFDPINNpred}. The absolute error of AD-PINN and HFD-PINN is shown in Fig. \ref{2dADPINNerror}. In Fig. \ref{2dlossFD}, the evolution of $\mathcal{L}^{\mathrm{AD-PINN}}$ and $\mathcal{L}^{\mathrm{HFD-PINN}}$ is displayed. The focus is on comparing the test error of AD-PINN and HFD-PINN. The test error of AD-PINN is 5.120e-02 $\pm$ 2.203e-02, while HFD-PINN is 2.833e-04 $\pm$ 1.463e-04. It indicates that the AD-PINN would be at a disadvantage against the HFD-PINN.  

%

\begin{table}[tp]  
	
	\centering  
	\fontsize{8}{8}\selectfont  
	\caption{Comparing the relative error with the different number of collocation points when learning Two-dimensional Poisson Equation.}  
	\label{tab5}  
	\begin{tabular}{|c|c|c|c|c|c|c|}  
		\hline  
		\multirow{2}{*}{collocation points}&  
		\multicolumn{3}{c|}{AD-PINN}&\multicolumn{3}{c|}{ HFD-PINN}\cr\cline{2-7}  
		&AD loss&BC loss&Data loss&FD loss&BC loss&Data loss\cr  
		\hline  
		\hline  
		25&4.0e+01&4.5e+01&{\bf 6.0e-01}&6.3e+03&1.6e-03&{\bf 6.6e-03}\cr\hline  
		100&7.4e+00&4.9e+00&{\bf 6.1e-01}&6.0e+03&6.0e-04&{\bf 1.0e-03}\cr\hline  
		225&1.0e+00&2.18e-01&{\bf 1.2e-02}&5.8e+03&8.5e-04&{\bf 2.8e-04}\cr\hline  
		400&3.6e-01&2.0e-01&{\bf 1.1e-02}&5.8e+03&9.6e-05&{\bf 1.4e-04}\cr\hline  
	\end{tabular}  
\end{table} 


\begin{figure}[htbp]
	\centering
	\subfigure{
		\includegraphics[scale=0.22]{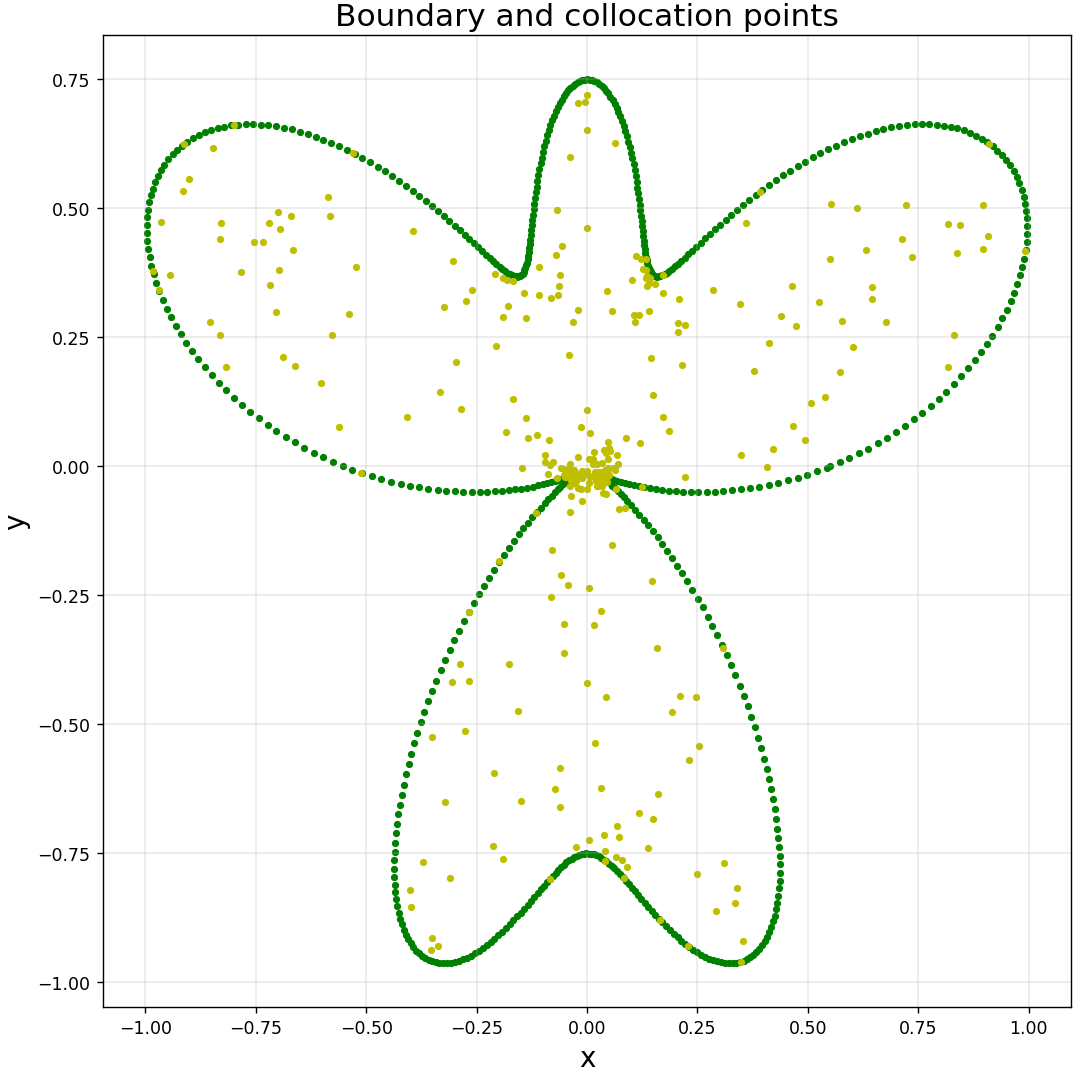}
	}
	\caption{Heat conduction problem on complex domain: Boundary (green dot points) and collocation points (yellow dot points).}
	\label{Buterpoints}
\end{figure} 

\subsection{Two-dimensional Poisson Equation on complex domain}
To illustrate this aspect of the proposed method, we solve the following two-dimensional steady-state heat equation on an irregular domain:
\begin{equation}
u_{xx} +u_{yy}=-37 \pi^{2} \cos (\pi x) \cos (6 \pi y), \quad(x, y) \in \Omega,
\end{equation}
where $\Omega=\{(x, y) \mid x=0.55 \rho(\theta) \cos (\theta), y=0.75 \rho(\theta) \sin (\theta)\} \text { and } \rho(\theta)=1+\cos (\theta) \sin (4 \theta),0\leqslant \theta\leqslant 2\pi$. In this case we consider using all points along the boundary as seen in Fig. \ref{Buterpoints}. The boundary conditions should be imposed as:
\begin{equation}
u=\cos (\pi x) \cos (6 \pi y), \quad(x, y) \in \partial \Omega.
\end{equation}

\begin{figure}[htbp]
	\centering
	\subfigure[Exact]{
		\includegraphics[scale=0.35]{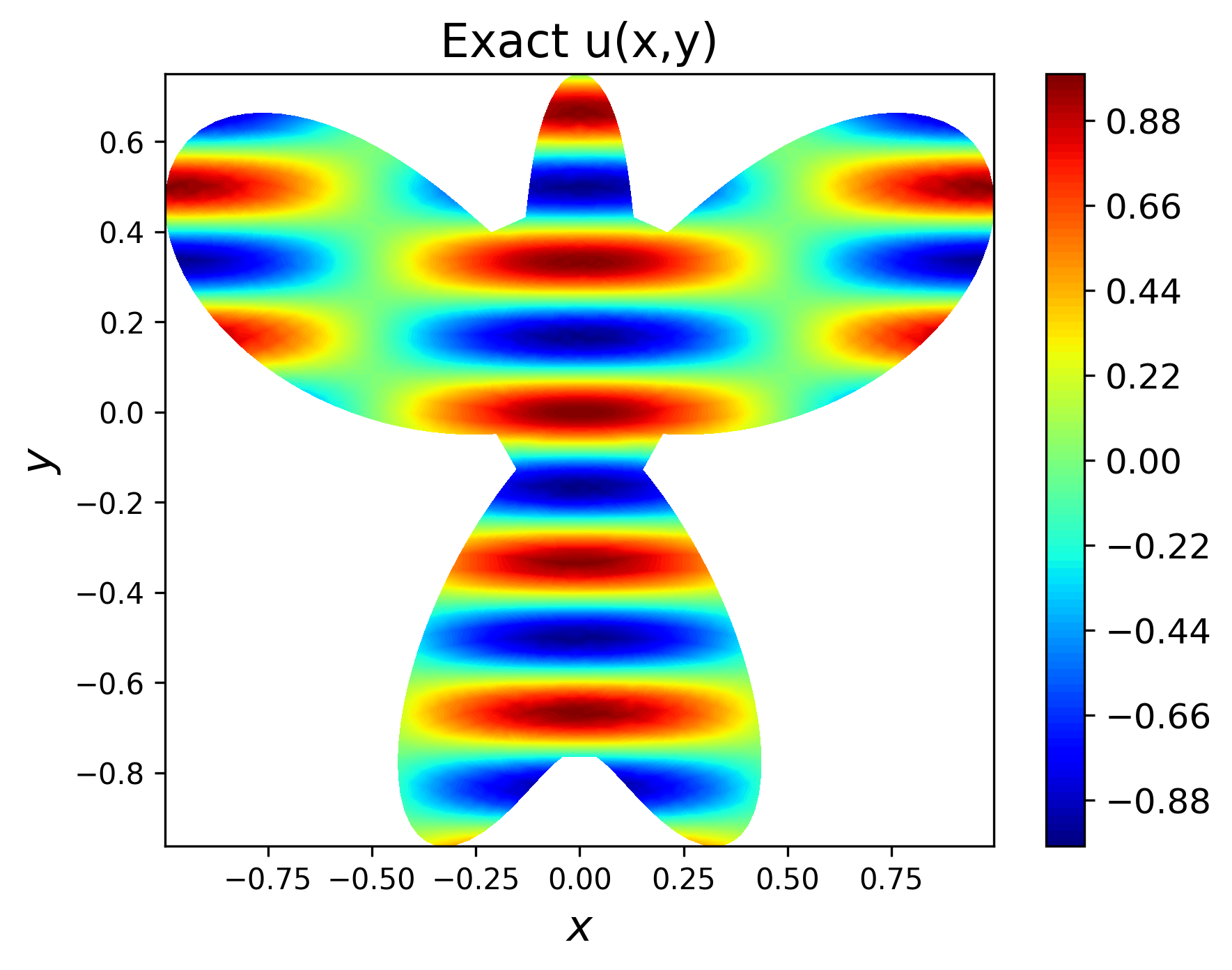}
	}%
	\subfigure[HFD-PINN-sdf]{
		\includegraphics[scale=0.35]{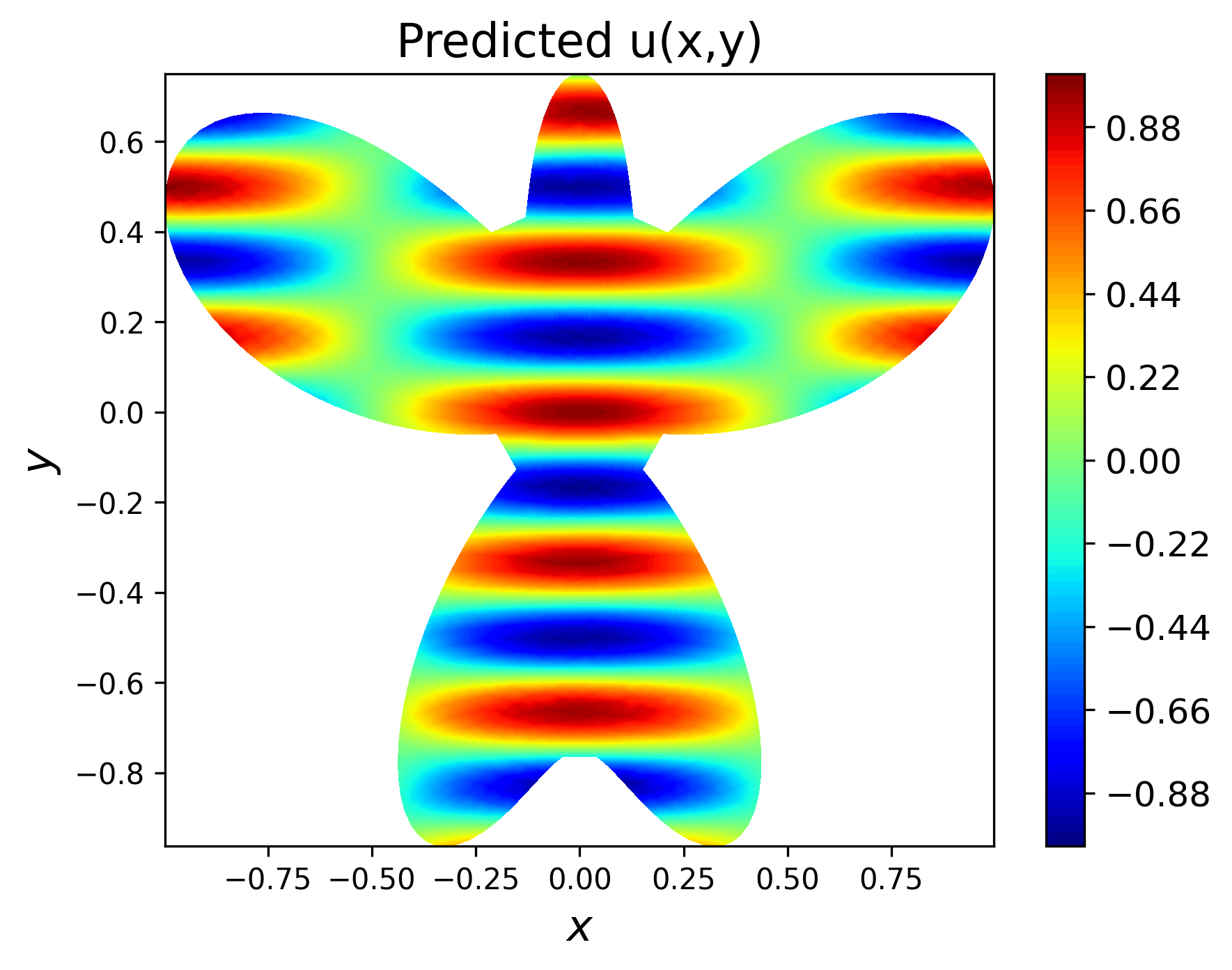}
	}%
	\subfigure[Error]{
		\includegraphics[scale=0.35]{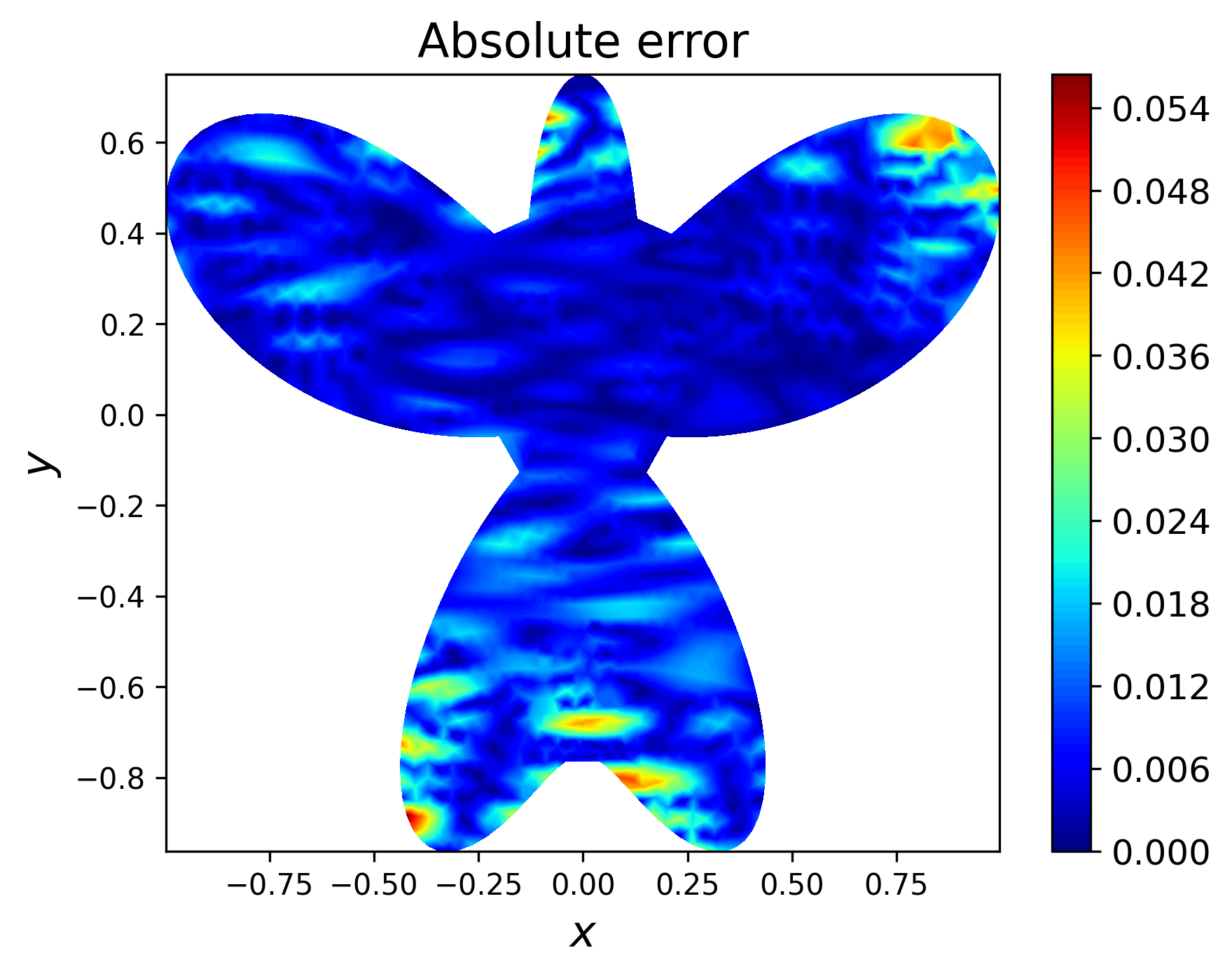}
	}%
	\centering
	\caption{Two-dimensional Poisson Equation on complex domain: Solution and error for the HFD-PINN-sdf.}
	\label{ButerFDPINNpred}
\end{figure}

\begin{figure}[htbp]
	\centering
	\subfigure[AD-PINN]{
		\includegraphics[scale=0.35]{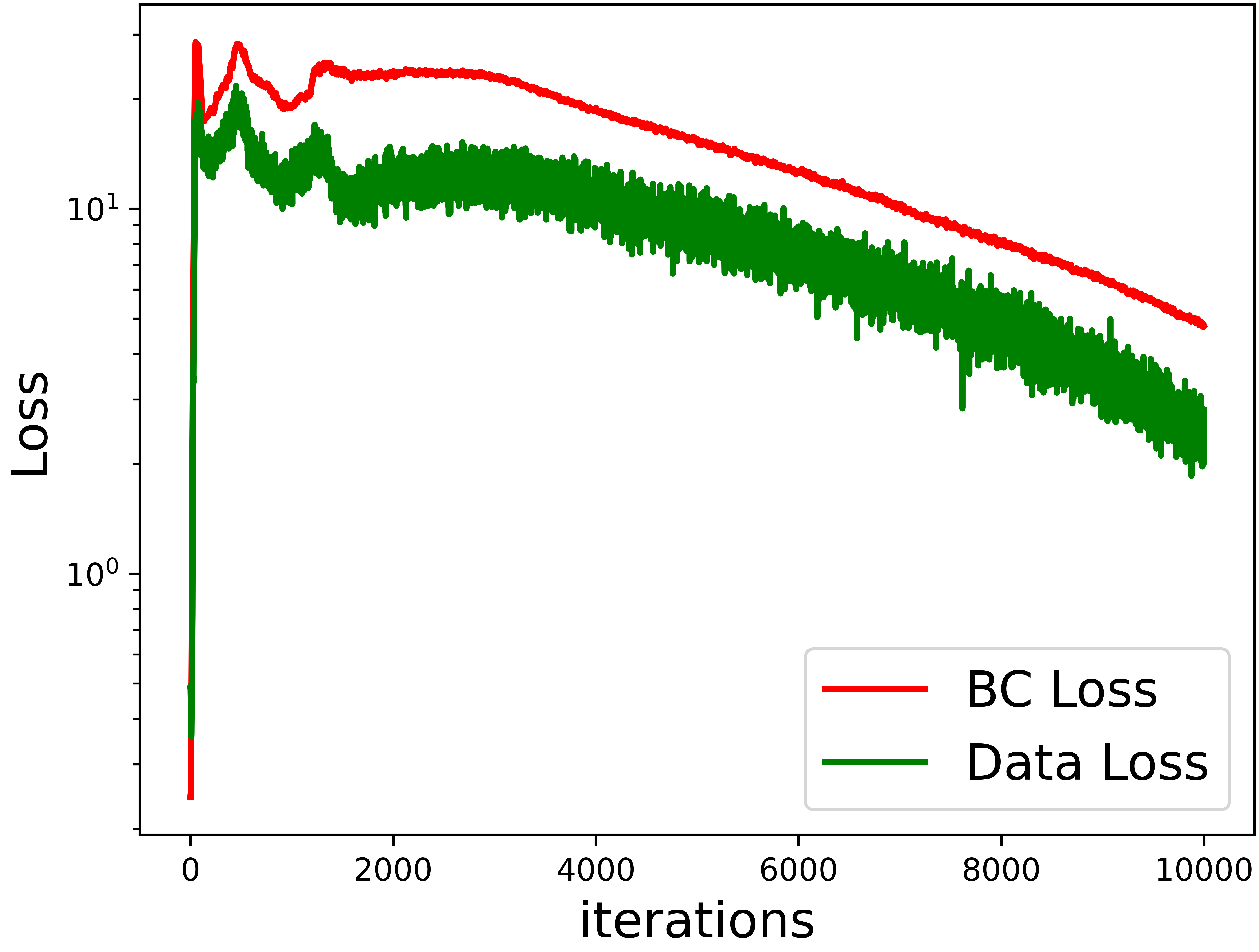}
	}%
	\subfigure[HFD-PINN-sdf]{
		\includegraphics[scale=0.35]{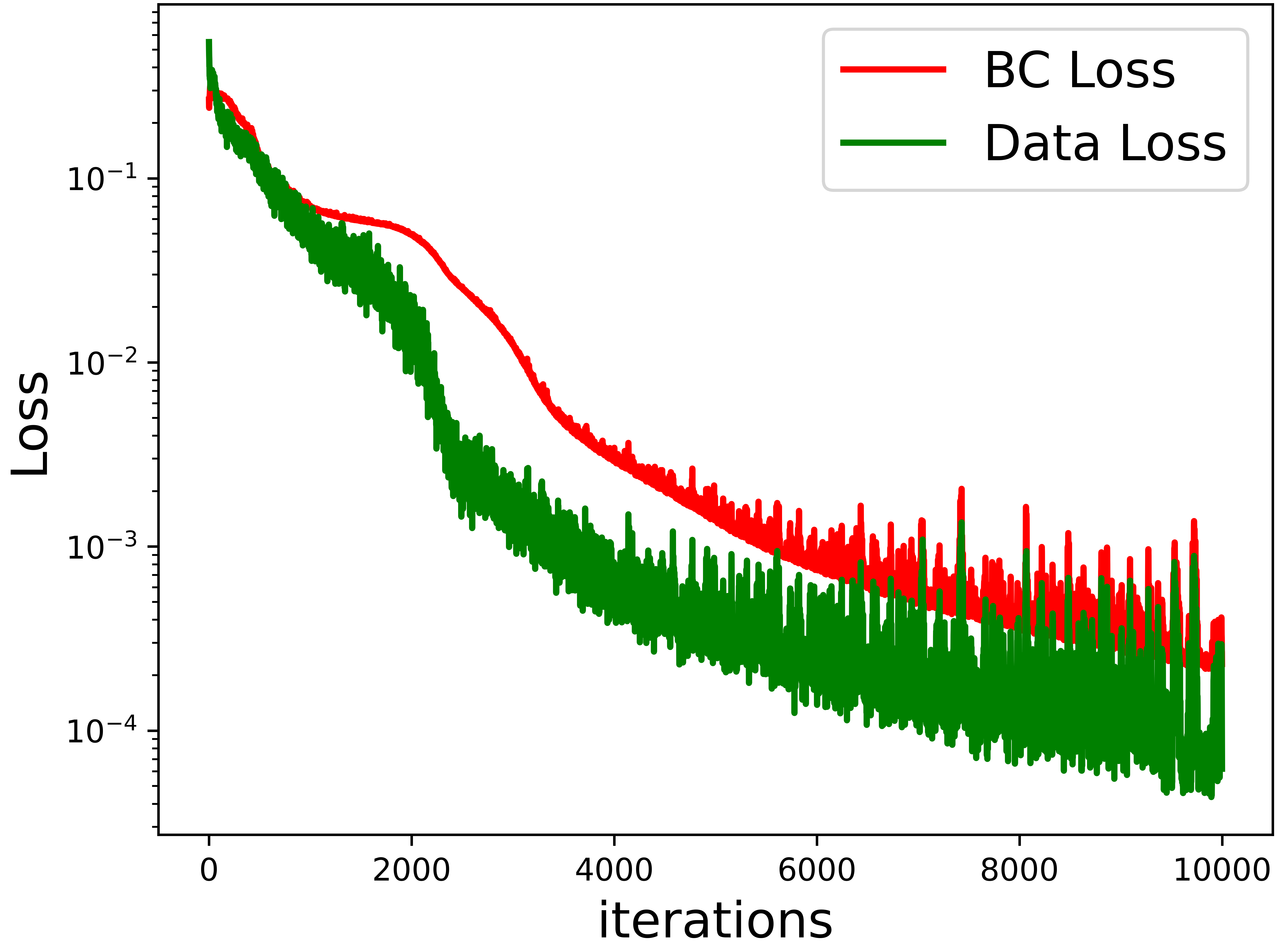}
	}%
	\subfigure[The test comparison]{
		\includegraphics[scale=0.35]{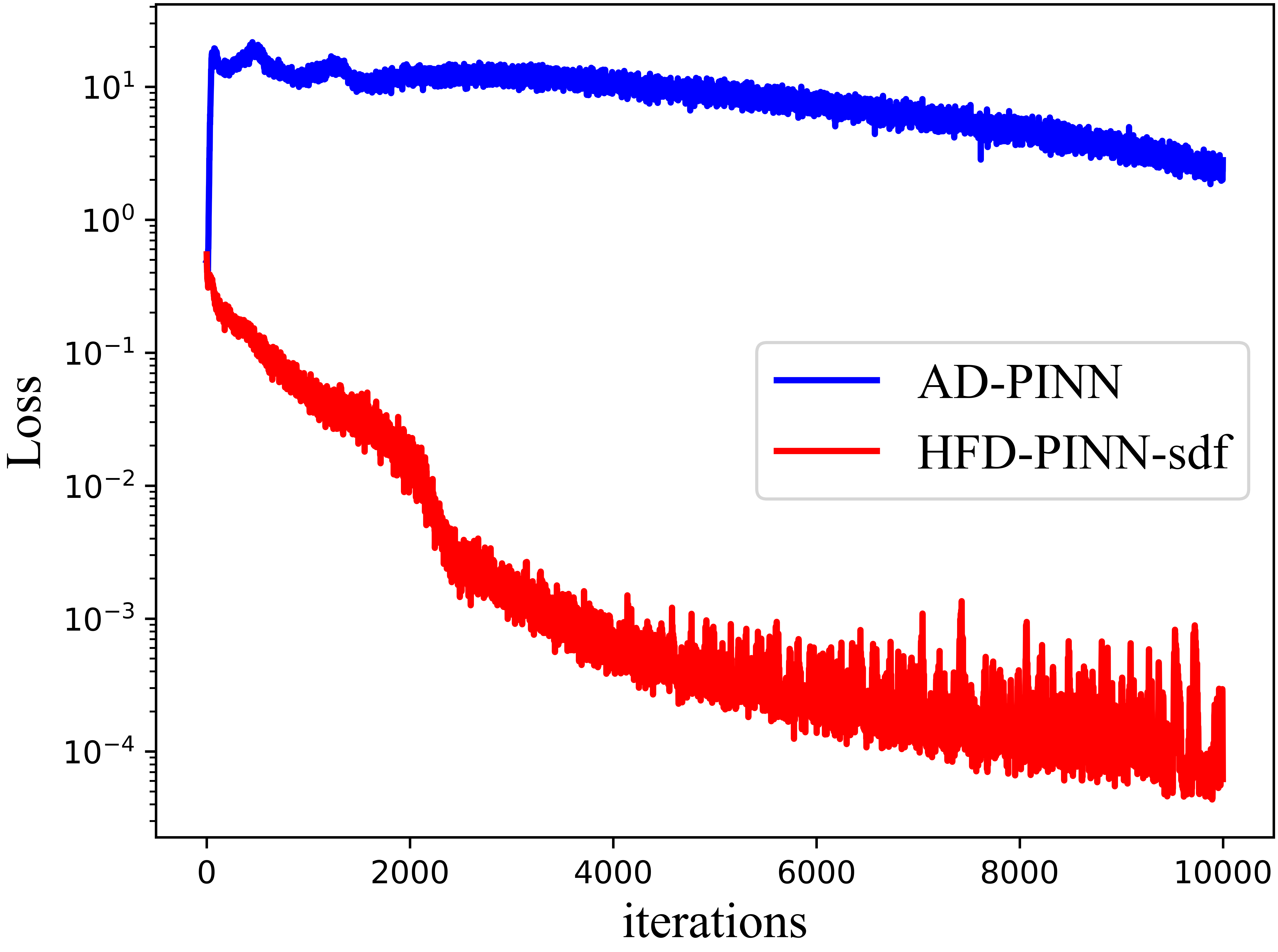}
	}%
	\centering
	\caption{Two-dimensional Poisson Equation on complex domain: Evolution of the loss function along with the training of the AD-PINN (a) and HFD-PINN-sdf (b). Besides, (c) The test comparison of the AD-PINN and HFD-PINN-sdf.}
	\label{ButerFDPINNloss}
\end{figure}

The solution is computed using an MLP with four hidden layers with 50 neurons each. We run 10000 iterations with stochastic gradient descent with a learning rate of 0.001. Since it is difficult to generate a background mesh, the FD loss is defined by a five-point difference scheme at each random point. Fig. \ref{ButerFDPINNpred} shows a detailed visual assessment of the predictability of the HFD-PINN-sdf. Fig. \ref{ButerFDPINNpred}(c) shows that the ground truth and the predicted solution are almost the same, especially at the boundary, and the maximum is 0.054, whose prediction accuracy is about 30 times better than AD-PINN. Fig. \ref{ButerFDPINNloss} show the evolution of the loss function during the continuous iteration via AD-PINN and HFD-PINN-sdf. The AD-PINN cannot get an accurately predicted solution, which its absolute error is magnified at the boundary as shown in Fig. \ref{ButerFDPINNloss}(a). From Fig. \ref{ButerFDPINNloss}(c), it can be seen that the Data loss term becomes smoother and smaller, and all these are finally reflected in the more accurate prediction solution of HFD-PINN-sdf.

\subsection{Two-dimensional Poisson Equation on irregular domain}
The example shows how to use the proposed method on nontrivial geometries. Let us now consider the following two-dimensional equation:
\begin{equation}
\begin{array}{ll}
u_{xx} +u_{yy}=f, & (x, y) \in \Omega, \\ u=g, & (x, y) \in \partial \Omega.
\end{array}
\end{equation}

To get an analytic solution we take, for example,
\begin{equation}
u=exp(-(2x^2+4y^2))+\frac{1}{2}.
\end{equation}

\begin{figure}[htbp]
	\centering
	\subfigure{
		\includegraphics[scale=0.22]{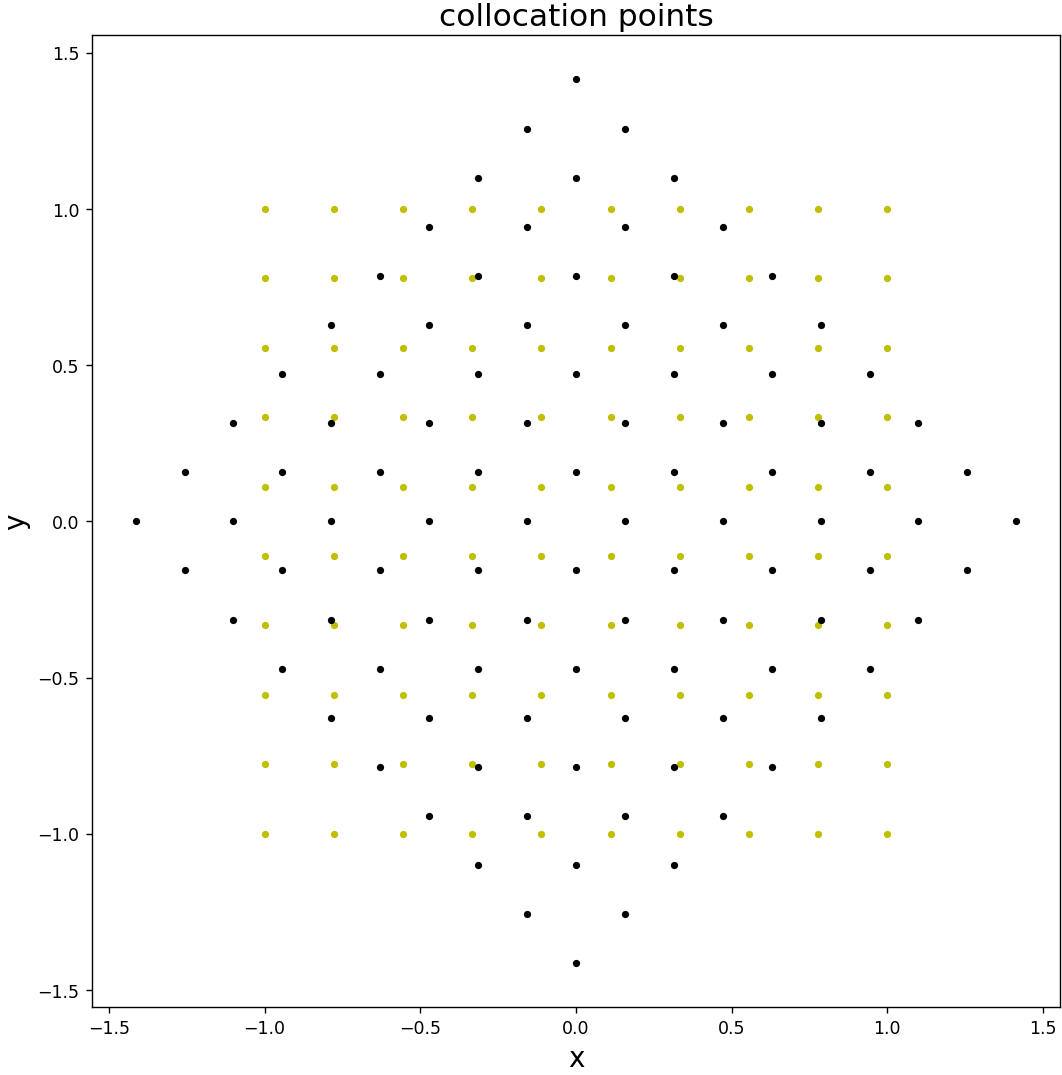}
	}
	\caption{Two-dimensional Poisson Equation on irregular domain: Collocation points.}
	\label{Ocpoints}
\end{figure}

We take the domain $\Omega$ to be a star shape and generate the collocation points with black and yellow dots, see Fig. \ref{Ocpoints}. The solution is defined as a deep fully connected neural network including four hidden layers and 50 neurons in each hidden layer, and the nonlinear activation function is designated as a hyperbolic tangent function. The FD loss is still defined using the self-adaptive finite difference method. Then we use the Adam optimizer to minimize the loss function with 10000 iterations of SGD. 

\begin{figure}[htbp]
	\centering
	\subfigure[Exact]{
		\includegraphics[scale=0.35]{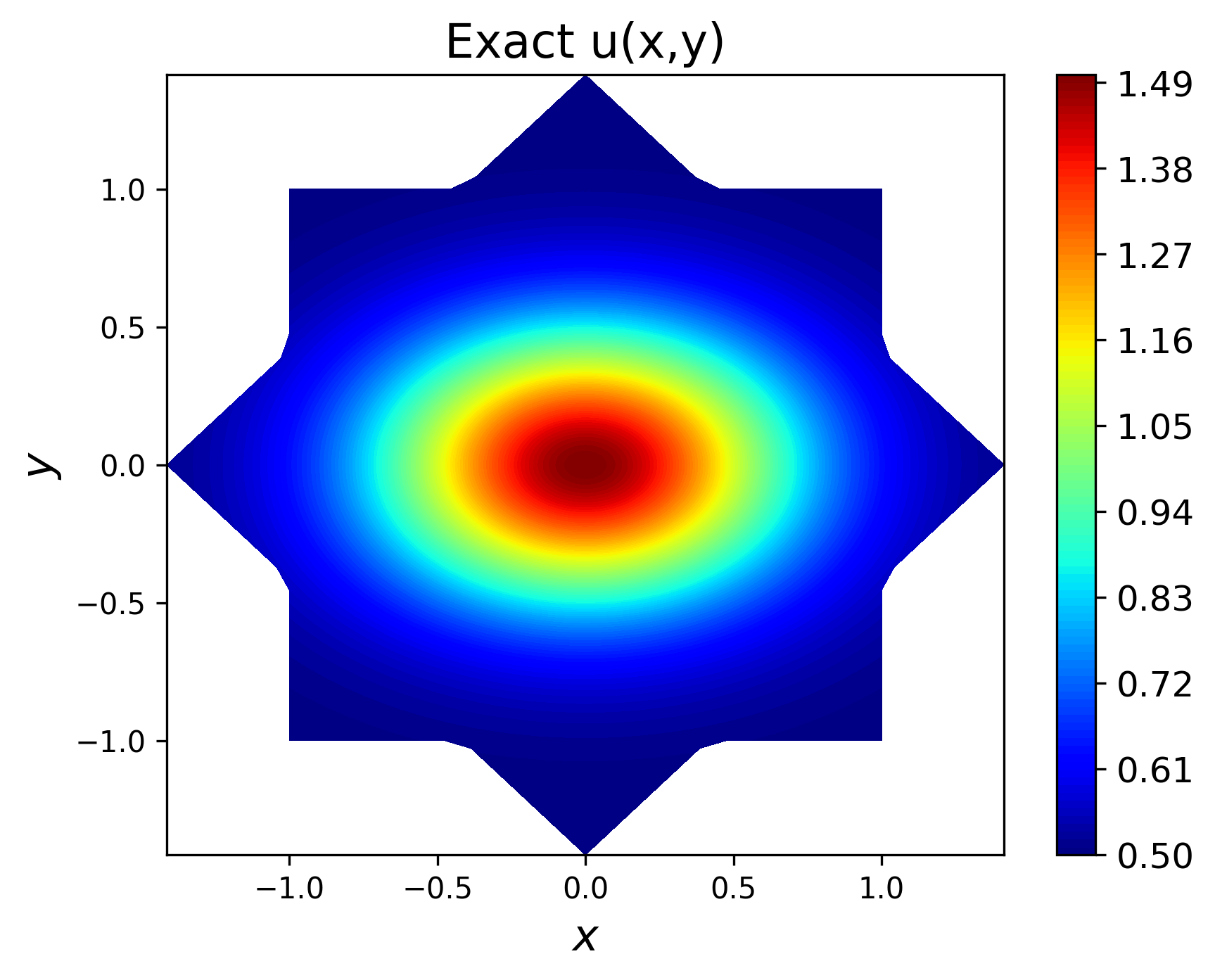}
	}
	\subfigure[AD-PINN]{
		\includegraphics[scale=0.35]{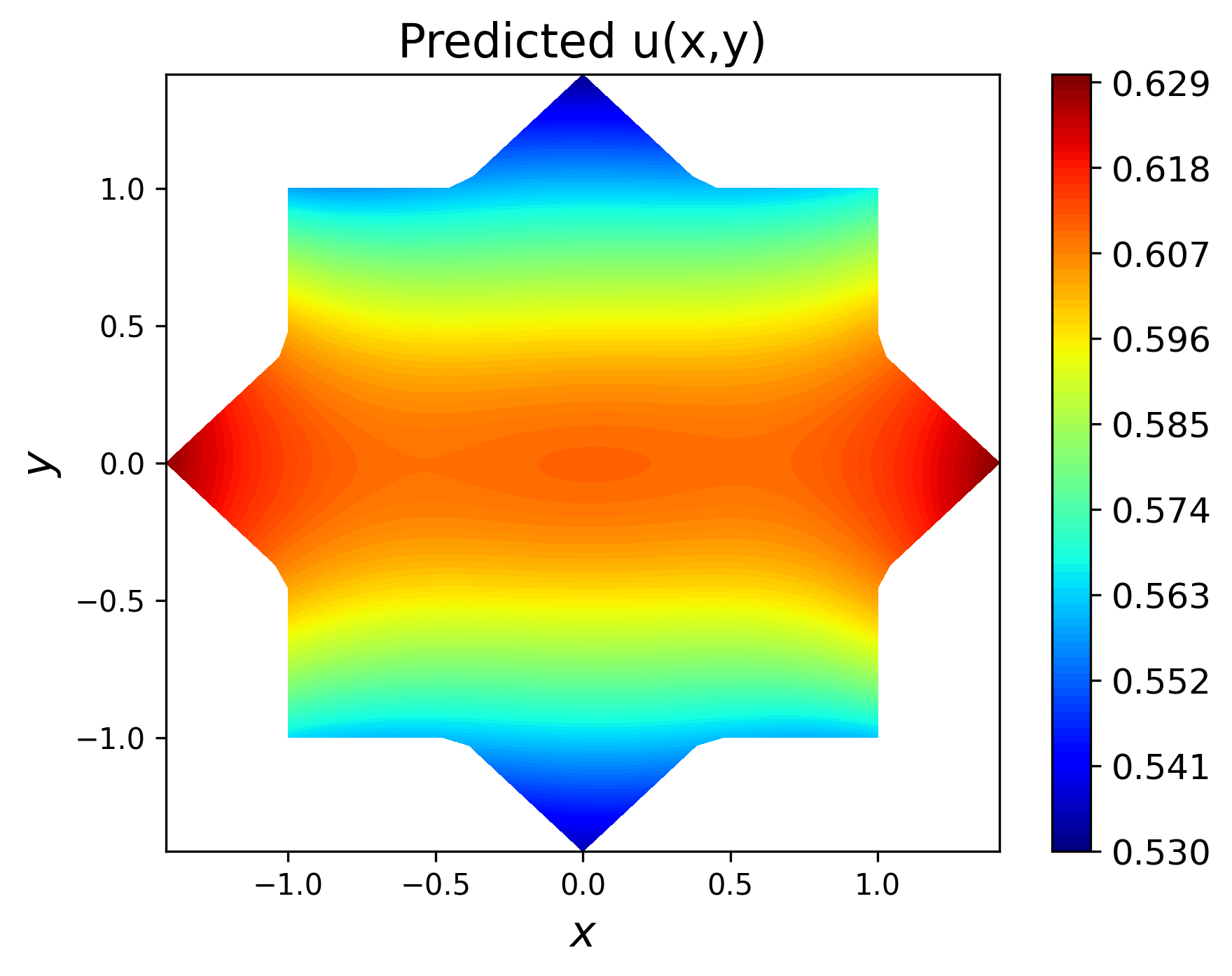}
	}%
	\caption{Two-dimensional Poisson Equation on irregular domain: (a) The reference solution (b) The predicted solution for the AD-PINN.}
	\label{OcADPINNpred}
\end{figure}

\begin{figure}[htbp]
	\centering
	\subfigure[HFD-PINN-sdf]{
		\includegraphics[scale=0.35]{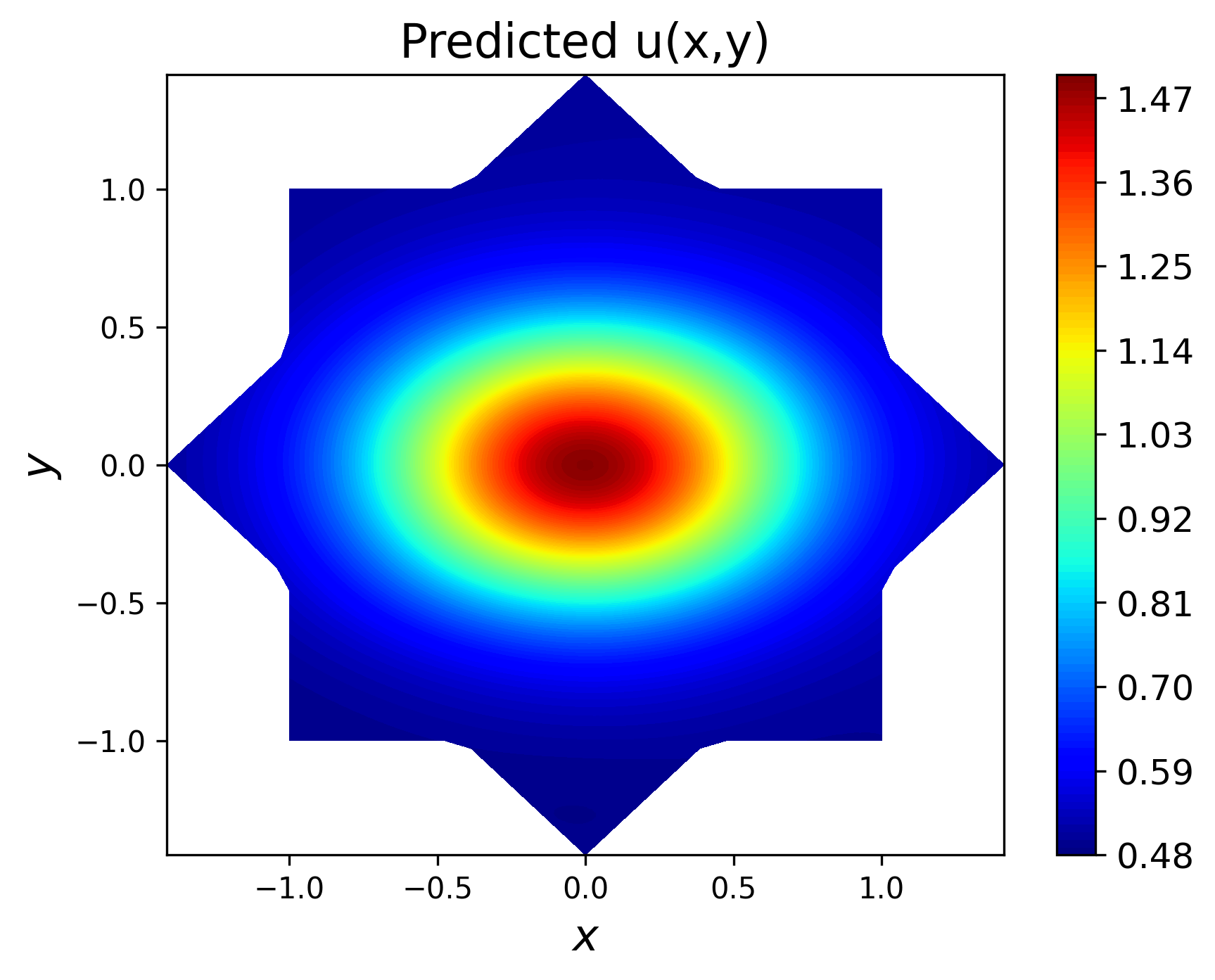}
	}%
	\subfigure[Error]{
		\includegraphics[scale=0.35]{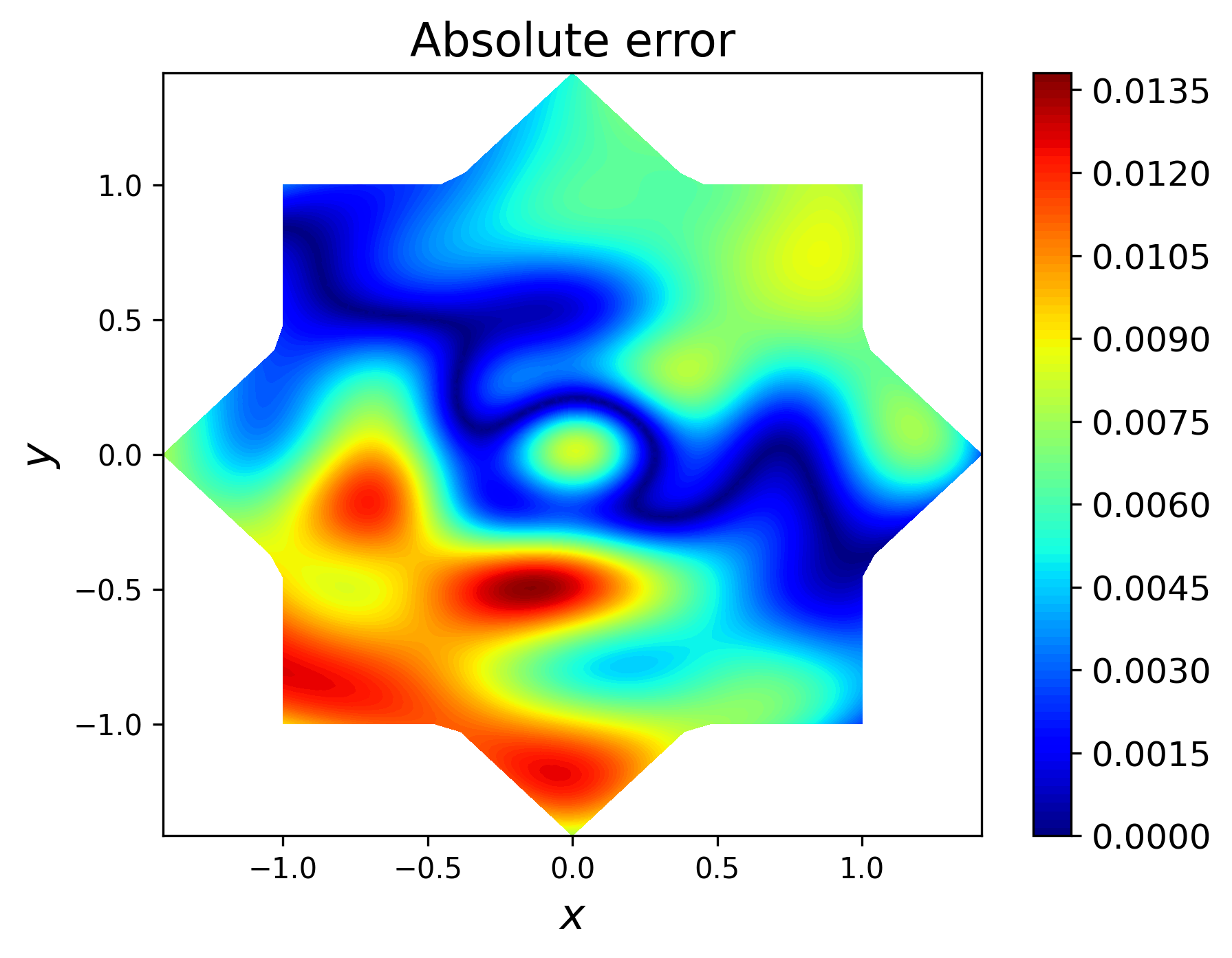}
	}%
	\centering
	\caption{Two-dimensional Poisson Equation on irregular domain: Solution and error for the HFD-PINN-sdf.}
	\label{OcFDPINNpred}
\end{figure}

\begin{figure}[htbp]
	\centering
	\subfigure{
		\includegraphics[scale=0.45]{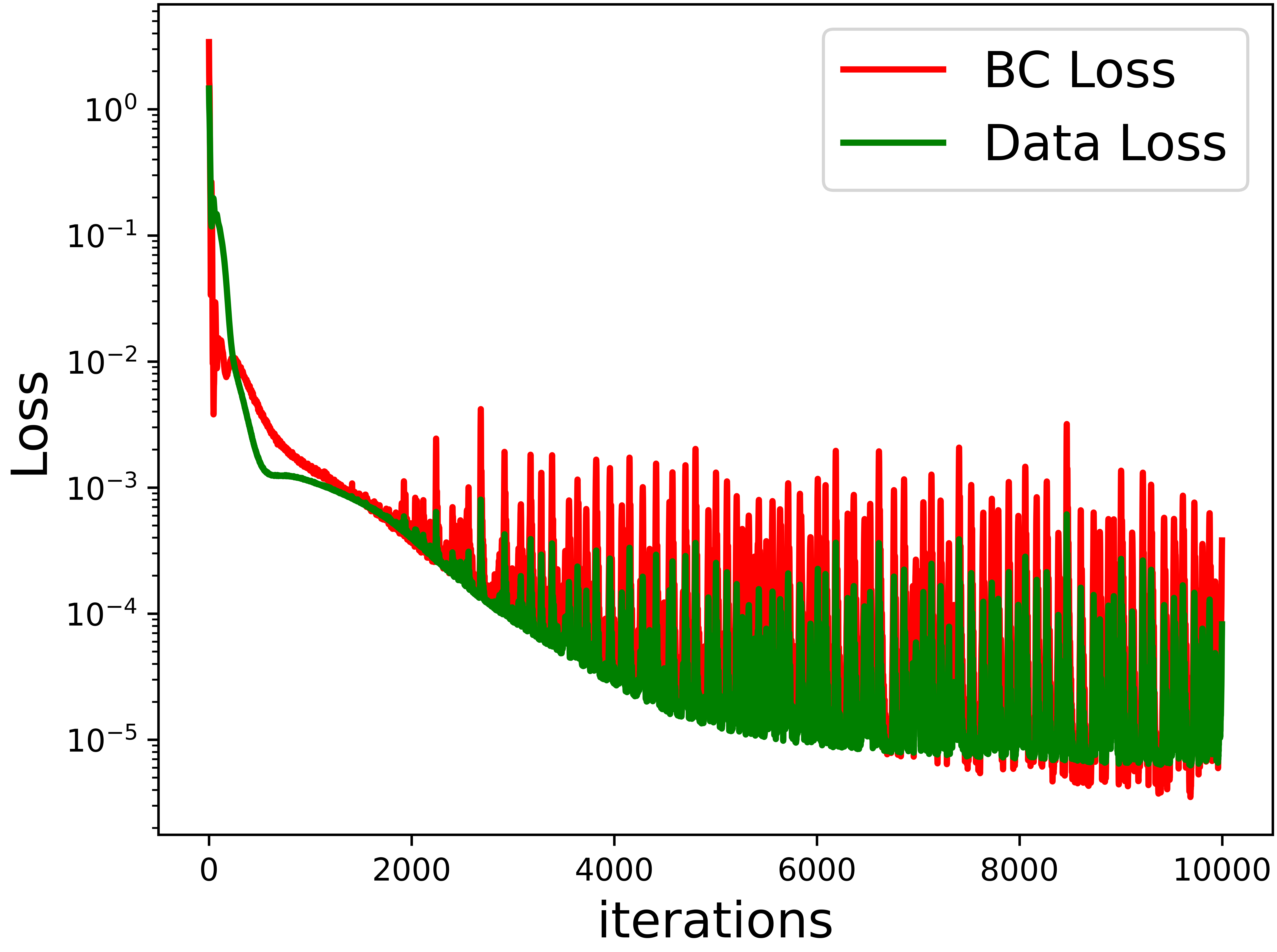}
	}%
	\centering
	\caption{Two-dimensional Poisson Equation on irregular domain: Evolution of the loss function along with the training of the HFD-PINN-sdf.}
	\label{OcFDPINNloss}
\end{figure}

The AD-PINN cannot obtain more accurate predictions shown in Fig. \ref{Ocpoints}. In Fig. \ref{OcADPINNpred} and \ref{OcFDPINNpred}, comparing the difference between the exact solution and predicted solution, we could see from the absolute error plot that the HFD-PINN-sdf can do the job in adapting to the sharper irregular areas and boundaries, which results in a relative L2 prediction error of 0.013. The variation of the loss with different iterations of HFD-PINN-sdf is shown in Fig. \ref{OcFDPINNloss}, where Data Loss represents the test error, and BC Loss represents the error on the boundary. Compared with the AD-PINN, we can see by Fig. \ref{OcFDPINNpred}(b) that the absolute error on the boundary and the area are effectively reduced. The prediction accuracy of the HFD-PINN-sdf is more than 40 times better than that of AD-PINN.

\section{Conclusions}
Despite recent successes in some applications, the PINN often has difficulty efficiently approximating the solutions of PDEs. In the original PINN approach named AD-PINN, the residual form of PDEs is lumped as a regularization loss term through the automatic differentiation. In this paper, we propose the hybrid finite difference with the physics-informed neural network to use the finite difference method (FDM) locally instead of AD in the framework of PINN. To take advantage of the proposed method, we deal with the derivatives in the regular domain with FDM, while the AD is suitable for calculating the derivatives at complex boundaries. It can be challenging to generate background mesh for complicated geometries. To circumvent this issue, we propose a Self-adaptive HFD-PINN with a signed distance function, which locally uses the finite difference scheme at random points. In addition, we use the signed distance function to define the difference interval at each collocation point. We take the Poisson equation, Burgers equation as examples to verify the performance of the HFD-PINN and HFD-PINN-sdf with three difference schemes such as compact finite difference method and crank-nicolson method. We have also showcased the flexibility and robustness of our framework for learning from the different number of collocation points. In addition, we have demonstrated the efficacy of our framework by solving heat transfer problems in the center hole and corner hole. We further solve the Poisson equation on the irregular domain with HFD-PINN-sdf. Therefore, HFD-PINN and HFD-PINN-sdf are more instructive and efficient, significantly when solving PDEs in complex geometries. In summary, our study provides new insights into the development of PINN and continuously improves their prediction accuracy.

Despite some recent progress, we still need future endeavors to shed light on challenging questions such as: 1) How does the loss of the neural network change with the finite difference? 2) How can we effectively reduce these gradient fluctuations of loss functions? 3) How can we combine traditional PDE solving methods to improve the generalization and prediction accuracy of PINN? These interesting discussions will be further explored in future work.



\section*{Acknowledgments}
This work was supported by the National Natural Science Foundation of China (No.11725211, 52005505, and 62001502).

%
%


\bibliographystyle{model1-num-names}
\bibliography{reference}

%

\end{document}